\documentclass{aa}
\usepackage{graphicx}
\usepackage{subcaption}
\usepackage{amssymb}
\usepackage{amsmath}
\usepackage{txfonts}
\usepackage{gensymb}
\usepackage{hyperref}
\bibpunct{(}{)}{;}{a}{}{,}

\def\phn{\phantom{0}}

\usepackage{color}
\definecolor{darkblue}{rgb}{0,0,0.6}
\newcommand{\kibitz}[2]{\ifnum\Comments=1\textcolor{#1}{#2}\fi}

\authorrunning{Magee et al.}
\titlerunning{Determining the $^{56}$Ni distribution}

\begin{document} 
            
    \title{Determining the $^{56}$Ni distribution of type Ia supernovae from observations within days of explosion} 
	\author{M. R. Magee \inst{1,2}
	\and 
	K. Maguire \inst{1}
	\and
	R. Kotak \inst{3}
	\and
	S. A. Sim \inst{2}
	\and 
	J. H. Gillanders \inst{2}
	\and
	S. J. Prentice \inst{1}
	\and
	K. Skillen \inst{1}
            }

	\institute{School of Physics, Trinity College Dublin, The University of Dublin, Dublin 2, Ireland \\  (\email{mrmagee.astro@gmail.com} \label{inst1})
	\and Astrophysics Research Centre, School of Mathematics and Physics, Queen's University Belfast, Belfast, BT7 1NN, UK \label{inst2}
	\and 	Tuorla Observatory, Department of Physics and Astronomy, FI-20014 University of Turku, Finland \label{inst3}
		}

   \date{Received -	- -; accepted - - - }

 
  \abstract{Recent studies have shown how the distribution of $^{56}$Ni within the ejected material of type Ia supernovae can have profound consequences on the observed light curves. Observations at early times can therefore provide important details on the explosion physics in thermonuclear supernovae, which are poorly constrained. To this end, we present a series of radiative transfer calculations that explore variations in the $^{56}$Ni distribution. Our models also show the importance of the density profile in shaping the light curve, which is often neglected in the literature. Using our model set, we investigate the observations that are necessary to determine the $^{56}$Ni distribution as robustly as possible within the current model set. Additionally, we find that this includes observations beginning at least $\sim$14 days before $B$-band maximum, extending to approximately maximum light with a relatively high ($\lesssim$3 day) cadence, and in at least one blue and one red band are required (such as $B$ and $R$, or $g$ and $r$). We compare a number of well-observed type Ia supernovae that meet these criteria to our models and find that the light curves of $\sim$70-80\% of objects in our sample are consistent with being produced solely by variations in the $^{56}$Ni distributions. The remaining supernovae show an excess of flux at early times, indicating missing physics that is not accounted for within our model set, such as an interaction or the presence of short-lived radioactive isotopes. Comparing our model light curves and spectra to observations and delayed detonation models demonstrates that while a somewhat extended $^{56}$Ni distribution is necessary to reproduce the observed light curve shape, this does not negatively affect the spectra at maximum light. Investigating current explosion models shows that observations typically require a shallower decrease in the $^{56}$Ni mass towards the outer ejecta than is produced for models of a given $^{56}$Ni mass. Future models that test differences in the explosion physics and detonation criteria should be explored to determine the conditions necessary to reproduce the $^{56}$Ni distributions found here.
  }
   \maketitle
%
   
\section{Introduction}
\label{sect:intro}

Within the modern era of optical, all-sky surveys, the diversity of type Ia supernovae (SNe~Ia) light curves has become evermore apparent and subsequently investigated in greater detail. Although it is thought that SNe~Ia result from thermonuclear explosions of white dwarfs \cite[see e.g. ][]{hillebrandt--2013}, these discoveries at increasingly early times demonstrate new and poorly understood phenomena that present challenges for our current understanding of progenitor scenarios and explosion mechanisms. Therefore, these early epochs are particularly important as they provide additional constraints that are not available at later times.

\par

\cite{jiang--18} demonstrate the variety of light curve behaviours that have been observed at early times through comparisons between a large number of objects. A variety of shapes in the light curves of SNe~Ia have been observed during the rise to maximum light \cite[e.g. ][]{firth--sneia--rise, papadogiannakis--2019}. In addition, certain SNe~Ia have shown short-lived, faint bumps that last only a few days following explosion and reach approximately 10\% of the peak luminosity. Early semi-analytical work by \cite{kasen--10} explores how bumps can be produced through the interaction between the expanding SN ejecta and companion star, with the strength of the bump dependent on the exact nature of the companion. The early light curves of SNe~Ia, therefore, could provide additional constraints on the companion star in the single-degenerate channel; however, such observational signatures are highly sensitive to viewing angle effects and are likely only visible under specific conditions. Following from the work of \cite{kasen--10}, \cite{liu--2015c} show how the luminosity distribution of early bumps expected from binary population synthesis calculations can be used to constrain the contribution of the single-degenerate channel to the overall SNe~Ia rate. Subsequent observational studies employing a large ($\textgreater$80) number of SNe~Ia \cite[e.g. ][]{hayden--10, bianco--11} or individual objects (e.g. SN~2011fe, \citealt{11fe--nature}; SN~2009ig, \citealt{foley--2012--09ig}; SN~2013dy, \citealt{zheng--13}; KSN~2011b, KSN~2012a, \citealt{olling--15}) have searched for the signatures of interaction predicted by \cite{kasen--10} for red giant companions. No direct evidence for such companions has been observed, however limits have been placed on their contribution to the single degenerate channel. More recent work has shown that bumps may also be produced through interaction with circum-stellar material \citep{piro-16}, which is similar to the production of bumps in some core-collapse SNe \cite[e.g. ][]{arcavi--17, nyholm--17}. The decay of short-lived radioactive isotopes resulting from the detonation of a thin helium shell \citep{noebauer-17} can also produce bumps in the early light curve; however, such a scenario may not be compatible with the majority of SNe~Ia. In addition to this, an off-centre $^{56}$Ni distribution in which $^{56}$Ni is concentrated in two physically distinct locations, with a large abundance of $^{56}$Ni in the outer ejecta could also produce an early light curve bump \citep{dimitriadis--19}.

\par

The $^{56}$Ni distribution of SNe~Ia is not well constrained by explosion models and recently, particular attention has been paid to exploring exactly the effects of the $^{56}$Ni distribution on observations \cite[e.g. ][]{piro--12, piro-nakar-2013, dessart-11fe}. 
Theoretical work has shown that the often assumed power-law rise of $L\propto t^2$ is not an intrinsic property of SNe~Ia but that a variety of different shapes can be produced depending on the assumed distribution of $^{56}$Ni within the ejecta \citep{piro-nakar-2014, piro-16, noebauer-17, magee--18}. Additionally, \cite{noebauer-17} show that a break in the power-law rise can also be produced if the $^{56}$Ni distribution is strongly stratified, in other words confined to the inner ejecta. Observational studies have indeed demonstrated significant variation in the rising light curve shapes for samples of SNe~Ia \citep{firth--sneia--rise, olling--15, papadogiannakis--2019}. Aside from the overall light curve shape, the colours are also highly sensitive to the $^{56}$Ni distribution. \cite{piro-16} show that those models with more extended $^{56}$Ni distributions are also bluer at early times, although this work assumes grey opacities and bolometric corrections. Recently, \cite{magee--18} presented models calculated with non-grey opacities and also demonstrated that more extended $^{56}$Ni distributions should produce bluer colours. Furthermore, \cite{magee--18} showed that the full colour light curve of at least one object could be explained by a model in which $^{56}$Ni is present throughout the ejecta.

\par

Building upon previous work by \cite{magee--18}, in this study we perform the first direct comparison of colour light curves from models invoking different $^{56}$Ni distributions to a large sample of objects (35 SNe~Ia in total). We investigate which objects can be explained by extending the $^{56}$Ni abundances into the outer layers of the ejecta and which objects require an excess of flux (i.e. a bump) at early times. In Sect.~\ref{sect:models} we discuss the construction of our model ejecta profiles and present the light curves in Sect.~\ref{sect:model_lightcurves}. In section~\ref{sect:robust} we present tests of what observations are required in order to robustly determine the $^{56}$Ni distribution within the current model set. In Sect.~\ref{sect:comparisons}, we use these requirements to create a sample of SNe and compare them to our models. Finally, we discuss our results in Sect.~\ref{sect:discussion} and present our conclusions in Sect.~\ref{sect:conclusions}.

%

\section{Models}
\label{sect:models}
All of our model light curves presented in this work have been calculated using the radiative transfer code, TURTLS\textbf{\footnote{The Use of Radiative Transfer for Light curves of Supernovae}}, which is described in detail by \cite{magee--18}. We briefly outline the operation of the code and construction of the models presented here. Light curves for all of our models are available on GitHub\footnote{\href{https://github.com/MarkMageeAstro/TURTLS-Light-curves}{https://github.com/MarkMageeAstro/TURTLS-Light-curves}}.

\par

\begin{table}
\centering
\caption{Ejecta density parameters}\tabularnewline
\label{tab:model-params}\tabularnewline
\resizebox{\columnwidth}{!}{
\begin{tabular}{ccccc}
\hline
\multicolumn{5}{c}{Double power law density profiles} \tabularnewline
\hline
\tabularnewline[-0.25cm]
Model   &	Transition                  & Inner     & Outer & Kinetic 	\tabularnewline
        &	velocity  	                & slope     & slope & energy 	\tabularnewline
        &	v$_{\rm{t}}$ (km~s$^{-1}$)  & $\delta$  & n     & (erg) 	\tabularnewline
\hline
\hline
\tabularnewline[-0.2cm]
DPL\_KE0.50	&	\phn7\,500 	& 1  & 12 		& 	$5.04\times10^{50}$	\tabularnewline 
DPL\_KE0.60	&	\phn7\,500 	& 0  & 12 		& 	$6.04\times10^{50}$	\tabularnewline 
DPL\_KE0.65	&	\phn7\,500 	& 1  & \phn8	& 	$6.53\times10^{50}$	\tabularnewline 
DPL\_KE0.78	&	\phn7\,500 	& 0  & \phn8	& 	$7.84\times10^{50}$	\tabularnewline 
DPL\_KE1.40	&	12\,500 	& 1  & 12 		& 	$1.40\times10^{51}$	\tabularnewline 
DPL\_KE1.68	&	12\,500 	& 0  & 12 		& 	$1.68\times10^{51}$	\tabularnewline 
DPL\_KE1.81	&	12\,500 	& 1  & \phn8	& 	$1.81\times10^{51}$	\tabularnewline 
DPL\_KE2.18	&	12\,500 	& 0  & \phn8	& 	$2.18\times10^{51}$	\tabularnewline 

\hline
\hline
\multicolumn{5}{c}{Exponential density profiles} \tabularnewline
\hline
\tabularnewline[-0.25cm]
Model & \multicolumn{2}{c}{Velocity scale                }     & \multicolumn{2}{c}{ Kinetic energy}	\tabularnewline
      & \multicolumn{2}{c}{v$_{\rm{e}}$ (km~s$^{-1}$)   }     & \multicolumn{2}{c}{(erg)}	\tabularnewline
\hline
\hline
EXP\_KE0.50	&	\multicolumn{2}{c}{1735.91} & 	\multicolumn{2}{c}{$5.04\times10^{50}$}	 \tabularnewline 
EXP\_KE0.60	&	\multicolumn{2}{c}{1901.60} & 	\multicolumn{2}{c}{$6.04\times10^{50}$}	 \tabularnewline 
EXP\_KE0.65	&	\multicolumn{2}{c}{1976.42} & 	\multicolumn{2}{c}{$6.53\times10^{50}$}	 \tabularnewline 
EXP\_KE0.78	&	\multicolumn{2}{c}{2165.06} & 	\multicolumn{2}{c}{$7.84\times10^{50}$} \tabularnewline 
EXP\_KE1.10 &   \multicolumn{2}{c}{2565.20} & 	\multicolumn{2}{c}{$1.10\times10^{51}$}	 \tabularnewline 
EXP\_KE1.40	&	\multicolumn{2}{c}{2893.19} & 	\multicolumn{2}{c}{$1.40\times10^{51}$}	 \tabularnewline 
EXP\_KE1.68	&	\multicolumn{2}{c}{3169.33} & 	\multicolumn{2}{c}{$1.68\times10^{51}$}	 \tabularnewline 
EXP\_KE1.81	&	\multicolumn{2}{c}{3294.04} & 	\multicolumn{2}{c}{$1.81\times10^{51}$}	 \tabularnewline 
EXP\_KE2.18	&	\multicolumn{2}{c}{3608.44} & 	\multicolumn{2}{c}{$2.18\times10^{51}$}	 \tabularnewline 
\hline
\hline
\end{tabular}
}
\tablefoot{Parameters of the artificial density profiles. }
\end{table}

TURTLS is a one-dimensional Monte Carlo radiative transfer code for modelling the early (up to approximately maximum) light curves of thermonuclear SNe. The input structure of the SN ejecta (density, composition, and velocity range) is freely defined by the user. Monte Carlo packets are injected into the model tracing the radioactive decay of $^{56}$Ni and subsequently $^{56}$Co. Packets are initially injected representing bundles of $\gamma$-ray photons and their propagation throughout the ejecta is followed assuming a grey $\gamma$-ray opacity of $\kappa / \rho$ = 0.03~cm$^{2}$~g$^{-1}$. After  interacting with the SN ejecta, $\gamma$-packets are converted into optical radiation packets (r-packets). To follow the propagation of these r-packets, we use TARDIS \citep{tardis,tardis_v2} to calculate non-grey expansion opacities for the appropriate density, temperature, and composition during each time step. We also include the effects of Thomson scattering by free electrons. Packets that escape the model region are binned and convolved with the desired set of filter functions to construct multi-colour light curves. TURTLS assumes local thermodynamic equilibrium and therefore, we only follow the light curve evolution until approximately 25 days post-explosion. 

\par
\begin{figure*}
\centering
\includegraphics[width=\textwidth]{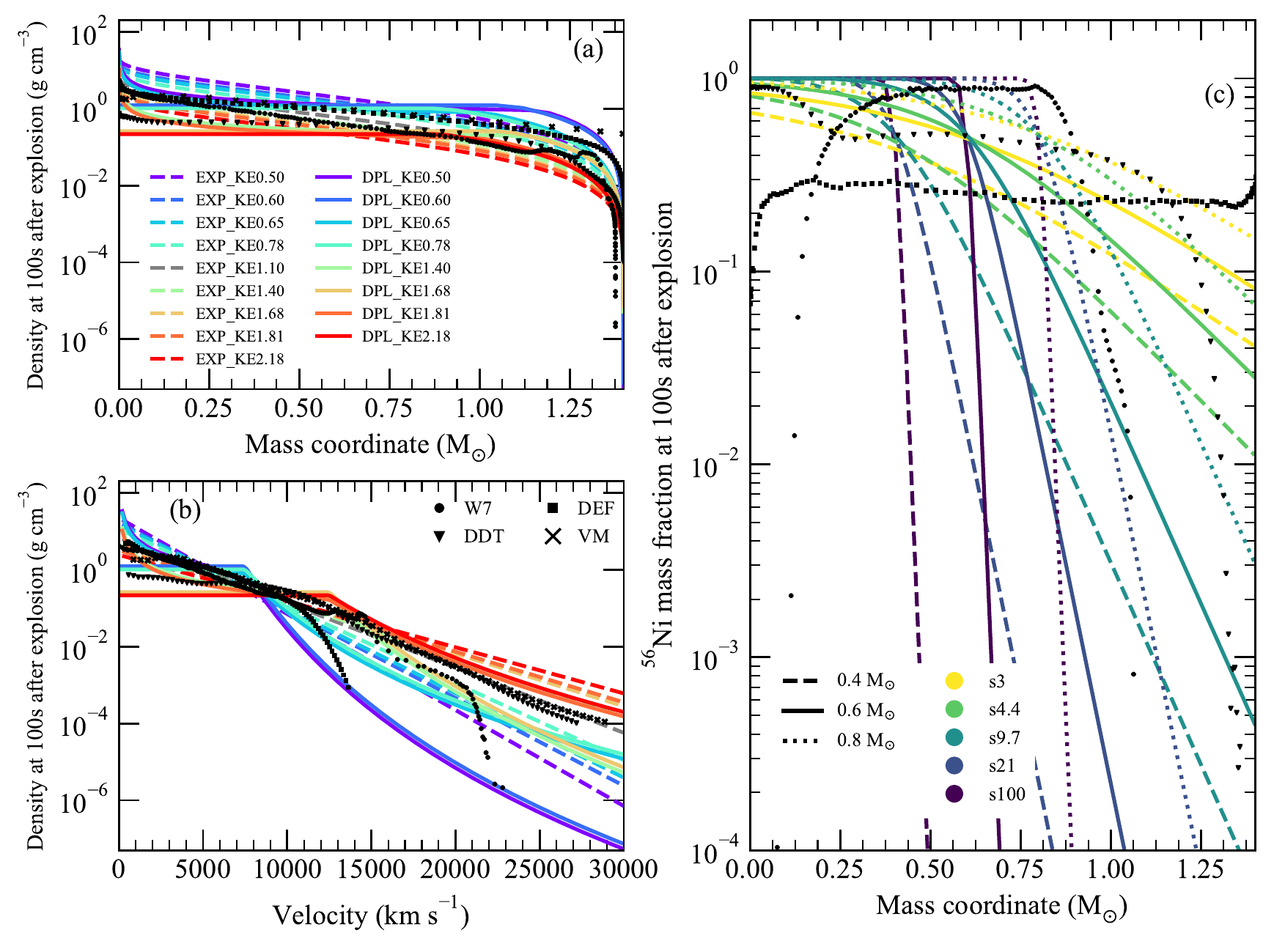}
\caption{Ejecta profiles for the models used as part of this work. Density profiles are shown in mass- {\it (Panel a)} and velocity-space {\it (Panel b)}. {\it Panel c:} $^{56}$Ni distributions are shown for the EXP\_KE1.10 density profiles with 0.4, 0.6, and 0.8~$M_{\rm{\odot}}$ of $^{56}$Ni. Density profiles and $^{56}$Ni distributions are shown for a selection of explosion models: W7 \cite[pure deflagration; ][]{nomoto-w7}, DDT N100 \cite[deflagration-to-detonation transition; ][]{seitenzahl--13}, DEF N1600 \cite[pure deflagration; ][]{fink-2014}, VM \cite[violent merger; ][]{pakmor-2012}.
}
\label{fig:densities}
\centering
\end{figure*}

Following \cite{botyanszki-2017}, in \cite{magee--18} we constructed a parameterised set of ejecta density profiles that broadly mimic those predicted by thermonuclear explosion models. These density profiles have double power law structures; the density at velocity $v$ is given as:
\begin{equation}
\rho(v) =\left\{
	\begin{array}{lr}
     \rho_0~~\left(v/v_{\rm{t}}\right)^{-\delta} & ~v \leq v_{\rm{t}}~ \\  
     \rho_0~~\left(v/v_{\rm{t}}\right)^{-n} & ~v > v_{\rm{t}}, \\
	\end{array} 
\right.
\end{equation}
where $v_{\rm{t}}$ gives the velocity boundary between the inner and outer regions, $\delta$ gives the slope of the inner region, $n$ gives the slope of the outer region, and the reference density, $\rho_0$, is given by:
\begin{equation}
\rho_0 = \frac{M_{\rm{ej}}}{4 \pi \left(v_{\rm{t}} t_{\rm{exp}}\right)^3} \left[\frac{1}{3 - \delta} + \frac{1}{n - 3}\right]^{-1},
\end{equation}
where $\delta\textless3$, $n\textgreater3$, $M_{\rm{ej}}$ is the ejecta mass, and $t_{\rm{exp}}$ is the time since explosion. In \cite{magee--18}, we presented models calculated using two values each for $v_{\rm{t}}$, $\delta$, and $n$, which equalled eight density profiles and broadly covered the parameter space predicted by explosion models. In order to further explore the effect of the density profile shape, we also include here a set of models with exponential density profiles. As in \cite{botyanszki-2017}, for each double power law density profile we calculate the kinetic energy of the ejecta as:
\begin{equation}
E_{\rm{k}} = \frac{1}{2}  M_{\rm{ej}} v_{\rm{t}}^2 \left[\frac{1}{5 - \delta} + \frac{1}{n - 5}\right] \left[\frac{1}{3 - \delta} + \frac{1}{n - 3}\right]^{-1},
\end{equation}
We then choose an exponential density profile with the same kinetic energy. Following \cite{kasen--06a}, the density at velocity $v$ is given by: 
\begin{equation}
\rho \left(v \right) = \frac{M_{\rm{ej}}}{8 \pi \left(v_{\rm{e}} t_{\rm{exp}}\right)^3} \rm{e}^{\left(-v/v_{\rm{e}} \right)},
\end{equation}
where $v_{\rm e} = \sqrt{ E_{\rm{k}}/6 M_{\rm {ej}} }$ is the velocity scale. 
Finally, we include an additional exponential density profile model with $E_{\rm{k}} = 1.1 \times 10^{51}$~erg to bridge the gap between our $v_{\rm{t}}$ = 7\,500~km~s$^{-1}$ and 12\,500~km~s$^{-1}$ models. The density profiles presented as part of this work use a fixed ejecta mass of $M_{\rm{ej}} = 1.4 M_{\odot}$ and maximum velocity of 30\,000~km~s$^{-1}$. The parameters of all the model density profiles are listed in Table~\ref{tab:model-params} and are shown in Fig.~\ref{fig:densities}.

\par

For each density profile, we have constructed a series of $^{56}$Ni distributions, such that the $^{56}$Ni abundance decreases monotonically towards larger radii. The $^{56}$Ni mass fraction at mass coordinate $m$ is given by:
\begin{equation}
\label{eqn:ni_dist}
^{56}{\rm Ni}\left(m\right) = \frac{1}{\exp\left(s\left[m - M_{\rm{Ni}}\right]/M_{\rm{\odot}}\right) + 1 },
\end{equation}
where $M_{\rm{Ni}}$ is the total $^{56}$Ni mass in $M_{\odot}$ and $s$ is the scaling parameter. The scaling parameter controls how quickly the ejecta transitions from $^{56}$Ni-rich to $^{56}$Ni-poor, with larger numbers corresponding to a sharper transition between these two regions. In \cite{magee--18}, we presented one $^{56}$Ni mass (0.6~M$_{\odot}$) and three scaling parameters (3, 9.7, and 100) for each density profile. Here, we extend our previous work to include additional $^{56}$Ni masses of 0.4 and 0.8~M$_{\odot}$ (broadly covering the range expected for SNe~Ia) and scaling parameters of 4.4 and 21 (intermediate to the scaling parameters presented in \cite{magee--18}). In total, this brings our model set to three $^{56}$Ni masses, 17 density profiles, and five $^{56}$Ni distributions, which equates to 255 models. We note that our model set is designed to explore a large parameter space and does not represent sampling of an underlying probability distribution for parameters, therefore the exact number of models is not overly meaningful and is simply given for reference.

\par

\begin{figure*}
\centering
\includegraphics[width=\textwidth]{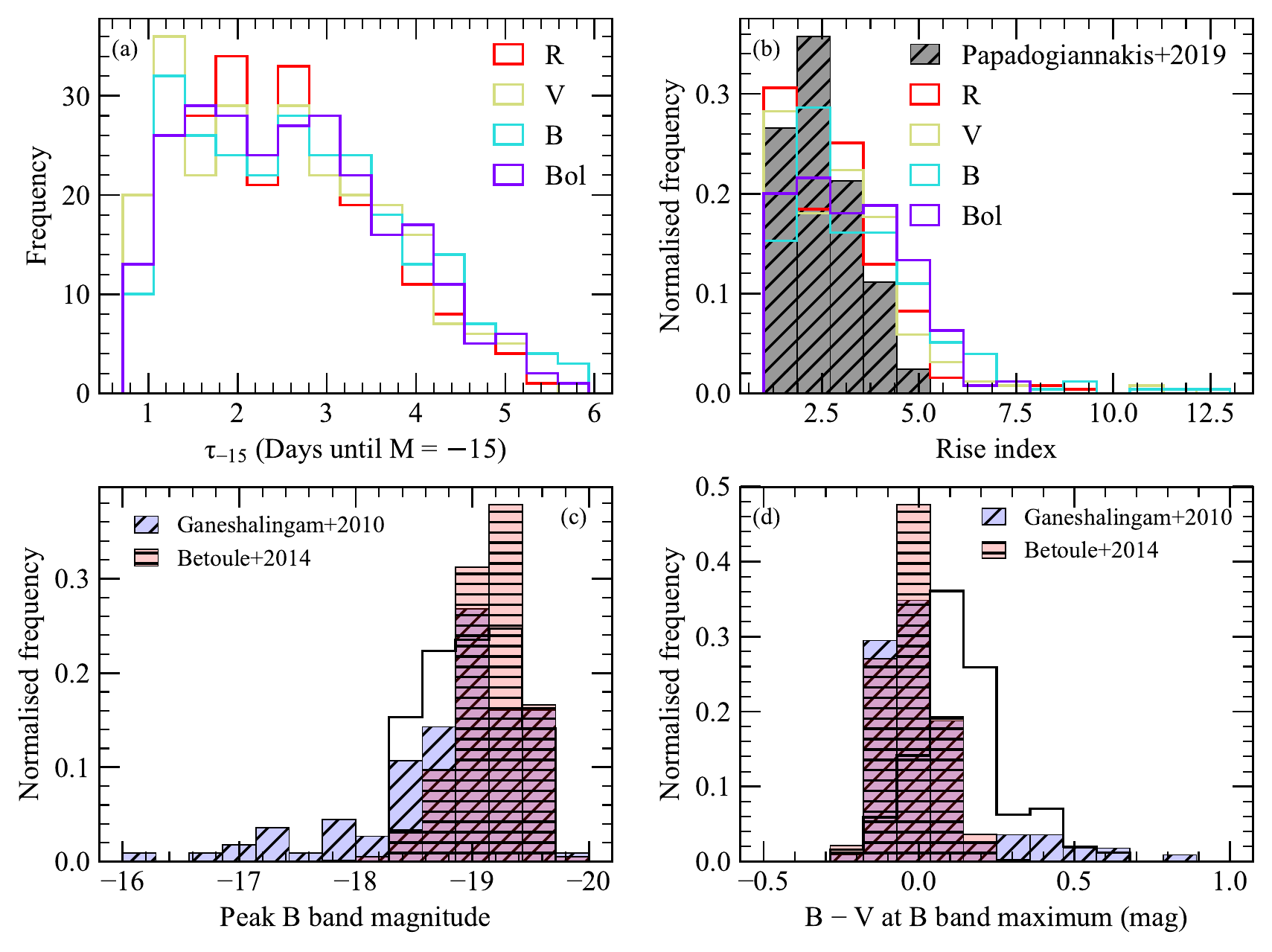}
\caption{Distribution of light curves properties for model sequence. {\it Panel a:} $\tau_{-15}$ is calculated as the time it takes to reach an absolute magnitude of $M = -15$ in each band. {\it Panel b:} Distribution of rise indices in each band. {\it Panel c:} Histogram of peak absolute $B$-band magnitudes for our models. {\it Panel d:} $B-V$ colour at $B$-band maximum. 
}
\label{fig:observables}
\centering
\end{figure*}

As in \cite{magee--18}, we wish to test only the effects of the $^{56}$Ni distribution on the observed light curve. We therefore use a simplified ejecta composition comprised of three zones \cite[see Fig.~4b of][]{magee--18}. The inner zone is dominated by iron group elements (IGE), where we have assumed that $^{56}$Ni comprises 100\% of the IGE mass fraction immediately following explosion, the outer $\sim0.1$~M$_{\odot}$ of the ejecta is dominated by carbon and oxygen, and the remaining material is composed of intermediate mass elements. Relative abundance ratios of elements are determined from the W7 model \citep{nomoto-w7}, however we have also explored abundance ratios determined from other explosion models. Specifically, we have determined compositions using the N100 model of \cite{seitenzahl--13}, the N1600 model of \cite{fink-2014}, and the violent merger model of \cite{pakmor-2012}. We find that these changes to the abundance ratios have little effect on the overall light curve properties. We stress that the models presented in this work form a simple parameter study of the $^{56}$Ni distribution. Such a simplified ejecta structured is unlikely to be realised in nature, but allows us to efficiently explore the effects of the $^{56}$Ni distribution alone. As we have assumed that the mass of IGE is composed solely of radioactive $^{56}$Ni, with no contribution from stable material, our models are typically brighter and bluer than models that are otherwise identical, but contain a larger contribution from stable IGE. Comparisons with observations should therefore be taken as representative of lower limits on the amount of $^{56}$Ni present at higher velocities. In addition, the assumption of LTE limits the applicability of our models to later times and in the near-infrared \cite[e.g. ][]{artis}.

%

\section{Model light curves}
\label{sect:model_lightcurves}

\subsection{Light curve properties}
\label{sect:lc_props}

In the following section, we demonstrate that our models span the parameter space of interest for SNe~Ia. The general light curve properties (absolute peak magnitude, colour, rise index) of our models are shown in Fig.~\ref{fig:observables}. In Fig.~\ref{fig:observables}(a), we show the distribution of $\tau_{-15}$, which we define as simply the time it takes for the model to reach an absolute magnitude of $-15$ in a given band. The choice of an absolute magnitude of $-15$ ensures that our results are not affected by noise due either to the Monte Carlo simulation or packet binning. In addition, at $M = -15$, the simulation has exited the dark phase. \cite{piro-nakar-2013} define the dark phase as the time between explosion and the emergence of the first photons resulting from the decay of $^{56}$Ni. At early times, there is also be some contribution from shock-breakout. Using semi-analytical methods, \cite{piro--2010} find that the bolometric magnitude of shock-breakout in SNe~Ia is fainter than $\sim$ $-$13, while in the $V$-band the shock-breakout peaks at $\sim$ $-9$ -- $-10$. Therefore regardless of the band of observation, once the absolute magnitude has reached $-15$, the luminosity is dominated by photons resulting from $^{56}$Ni. Hence, although our models do not contain any contribution from shock-heating, at $\tau_{-15}$ all models have exited the dark phase. As shown in Fig.~\ref{fig:observables}(a), most of our models reach $-15$ after a few days, however approximately one third take more than three days to reach $M_{B} = -15$. Only $\lesssim$5\% of our models have $\tau_{-15}(B)$ lasting more than five days or less than one day. All bands show a broadly similar spread in values. A number of SNe Ia have been claimed to show dark phases ranging from zero to a few days \citep{piro-nakar-2014, hsiao--2015, jiang--2017}, however this range is uncertain as the explosion epoch is unknown.

\par

The length of $\tau_{-15}$ for each model is well correlated with how quickly the early light curve increases in brightness, parameterised as the rise index. We calculate the rise index by fitting the first five days after explosion using:
\begin{equation}
    f(t) = \alpha t^n,
\end{equation}
where $f$ is flux, $\alpha$ is a normalising constant, $t$ is the time since explosion, and $n$ is the rise index. Those models with longer $\tau_{-15}$ typically have higher rise indices and show much sharper rises after a few days. As previously shown by \cite{magee--18}, our models demonstrate that a range of rise indices (above and below the commonly used $n$ = 2) can be obtained simply by varying the distribution of $^{56}$Ni. With the inclusion of a lower $^{56}$Ni mass (0.4~$M_{\odot}$) and exponential density profiles, our models show that very sharp rises ($n \textgreater 5$) may be achieved. Currently, no known SN Ia has shown a similarly sharp rise, however if such an object is found it would have $^{56}$Ni concentrated only deep within the ejecta and show a sudden and drastic transition between $^{56}$Ni-rich and -poor regions. Figure~\ref{fig:observables}(b) shows that the range of rise indices calculated for our models is consistent with the observed range for SNe Ia, as shown by the sample of 207 objects presented by \cite{papadogiannakis--2019}.

\par

In Fig.~\ref{fig:observables}(c), we show the distribution of peak $B$-band magnitudes for our models. We find a relatively uniform distribution between $-19.7 \lesssim M_{B} \lesssim -18.3$ -- again we note that there is no underlying probability distribution to the model parameters presented in this work. For comparison, we show the range of peak magnitudes obtained from the large samples of SNe Ia published by \cite{ganeshalingam--2010} and \cite{betoule--2014}, excluding any highly reddened SNe. The data have been corrected for Milky Way extinction and distance moduli were calculated from redshifts assuming $H_0 = 70$~km~s$^{-1}$~Mpc$^{-1}$, $\Omega_{\rm{M}} = 0.3$, and $\Omega_{\lambda} = 0.7$. The distribution of observed peak magnitudes is more concentrated around $M_{B} \sim -19.2$ than the relatively flat distribution of our models, however the overall range of peak magnitudes produced by our models is broadly consistent with those observed. In addition, \cite{ganeshalingam--2010} also include a number of fainter objects, with peak magnitudes $M_{B} \textgreater -18$. The aim of this paper is to study normal SNe Ia and therefore these faint events are not covered by the model parameter space presented here. Future work will focus on exploring the parameter space for fainter SNe Ia. 

\par

In Fig.~\ref{fig:observables}(d) we compare the colours of our models at $B$-band maximum to those obtained by \cite{ganeshalingam--2010} and \cite{betoule--2014}. We note that \cite{betoule--2014} only include those objects for which the colour at maximum light is $-0.3$ \textless ~$(B - V)_{\rm{max}}$ \textless ~$0.3$. Although our models cover a similar range to normal SNe Ia the distribution of model colours tends to peak towards somewhat redder values. As previously mentioned, assuming a more realistic composition with additional IGE would shift each model to redder colours, as well as fainter luminosities shortly after explosion. In addition to the colours at maximum light, our models also generally cover the range in $g-r$ colour evolution observed within the first approximately one week following explosion (Bulla et al., in prep.).

\par

\begin{figure}
\centering
\includegraphics[width=\columnwidth]{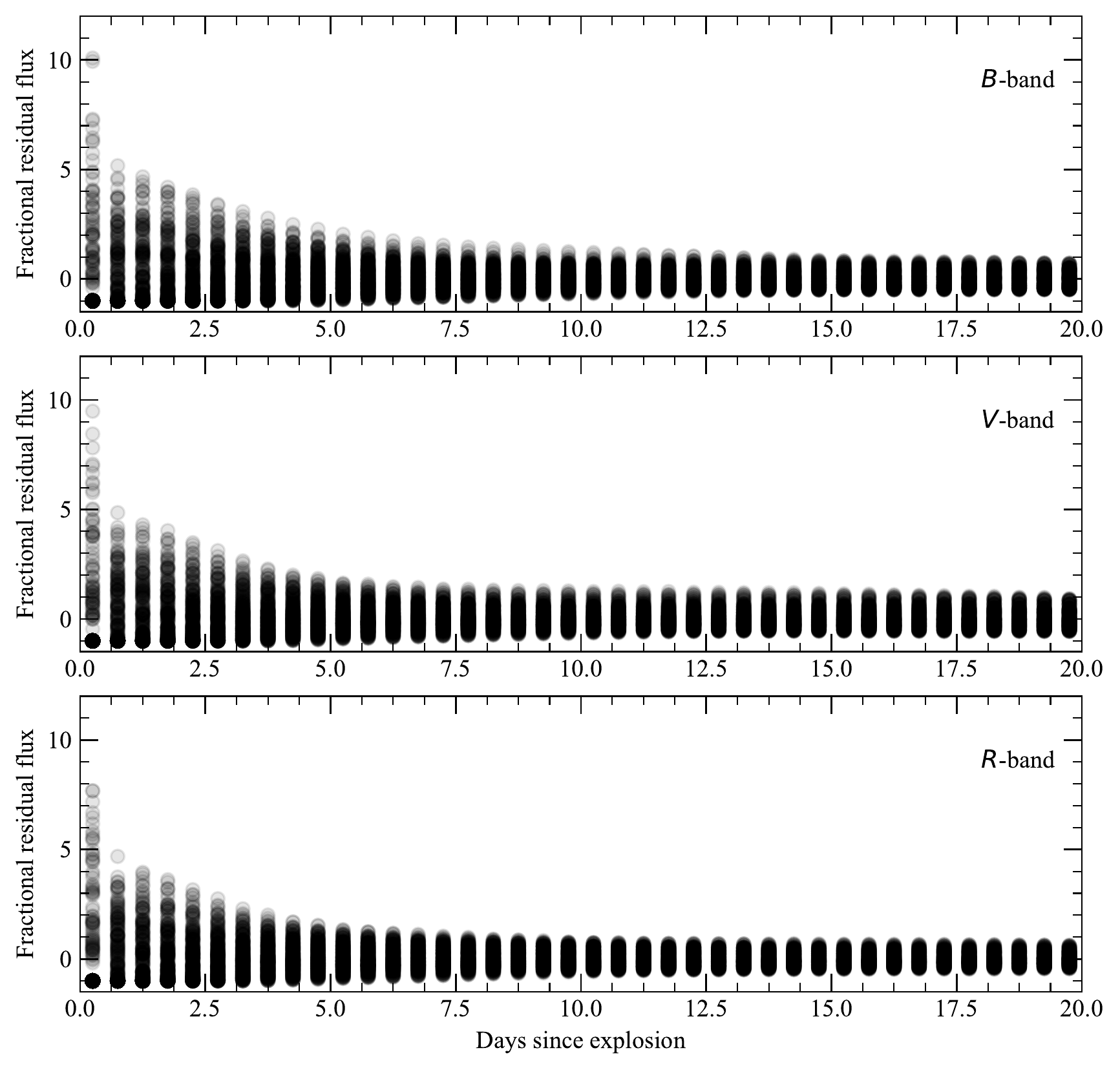}
\caption{Fractional residual flux of each model in the $B$-, $V$-, and $R$-bands with respect to the mean across all model is shown as a function of days since explosion. }
\label{fig:mean_diff}
\centering
\end{figure}

To further demonstrate the diversity of our models, we have calculated an average model for each band. This is simply the mean flux per epoch across our entire model set, including all density profiles and $^{56}$Ni masses. Figure~\ref{fig:mean_diff} shows the fractional difference between this average model and each individual model for the $B$-, $V$-, and $R$-bands. It is clear that the models are most diverse within approximately five days of explosion before becoming more similar -- therefore these times provide the tightest constraints on the ejecta properties of observed SNe. In Section~\ref{sect:robust}, we discuss more quantitatively the requirements on early data for producing well-constrained model fits.

%

\subsection{Importance of the density profile}

\begin{figure*}
\centering
\includegraphics[height=15cm]{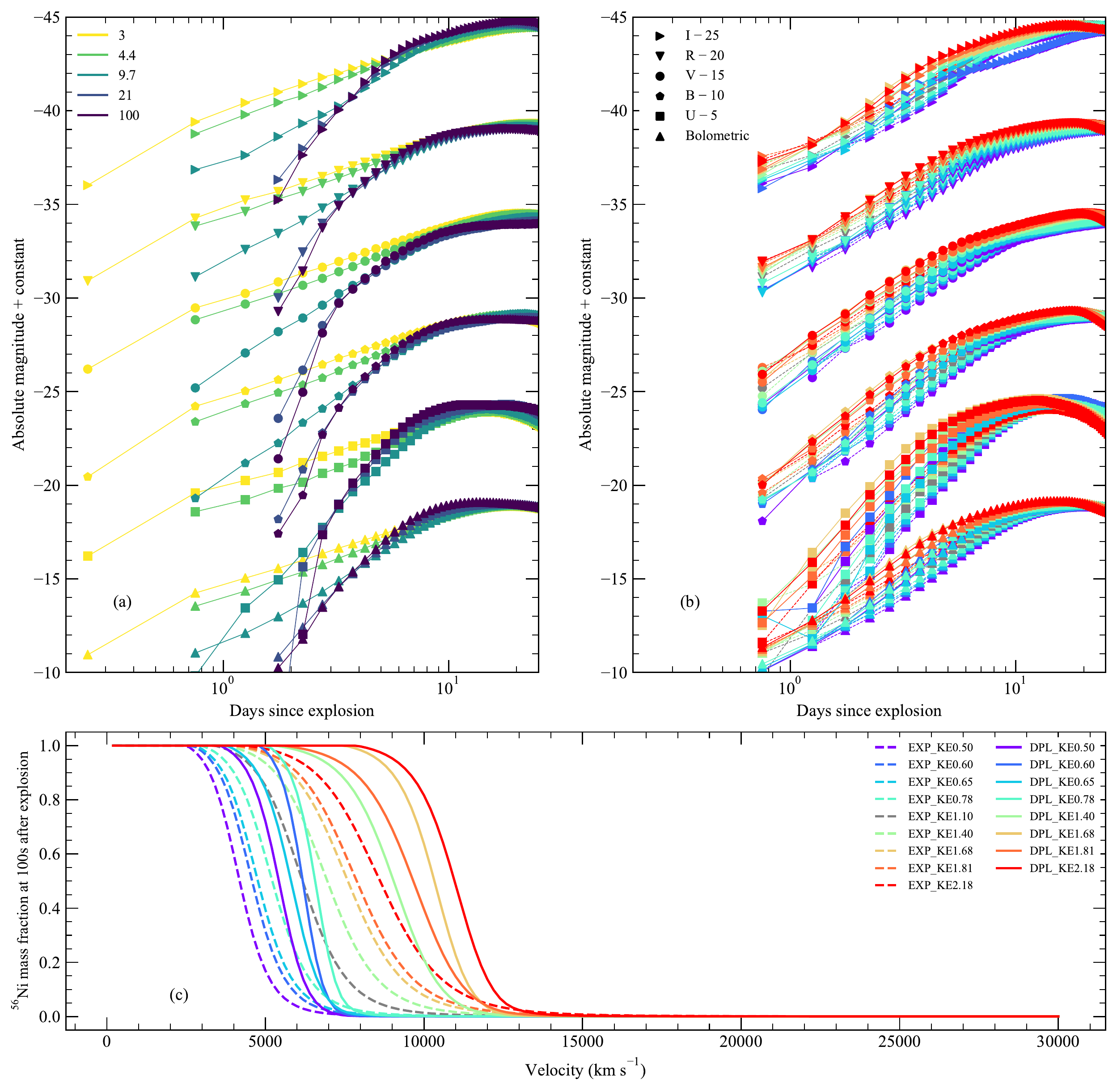}
\caption{Diversity of light curves for different $^{56}$Ni distributions and density profiles. {\it Panel a:} Light curves are shown for our EXP\_Ni0.6\_KE1.10 model with different scaling parameters and $^{56}$Ni distributions calculated from the Eqn.~\ref{eqn:ni_dist}. Symbols show different bands of observations, which have been arbitrarily offset. The legend for symbols in {\it Panels a} and {\it b}, is given in {\it Panel b}. Models with smaller scale parameters (i.e. more extended $^{56}$Ni distributions) show brighter light curves at early times. {\it Panel b:} Light curves calculated with different density profiles are shown for the case of models with $0.6~M_{\rm{\odot}}$ of $^{56}$Ni and a scale parameter of $s = 9.7$. Colours in {\it Panel b} are as in {\it Panel c}. {\it Panel c:} Density profiles for light curves shown in {\it Panel b}. All models have a maximum velocity of 30\,000~km~s$^{-1}$. 
}
\label{fig:compare_models}
\centering
\end{figure*}

In Fig.~\ref{fig:compare_models}(a) we show the variation in the observed light curves produced by the different $^{56}$Ni distributions given in Fig.~\ref{fig:densities}(c). We focus here on the representative case of the EXP\_Ni0.6\_KE1.10 models, which have exponential density profiles, 0.6~$M_{\odot}$ of $^{56}$Ni, and kinetic energies of $1.10\times10^{51}$ erg. The trends observed however, are broadly applicable to other density profiles and $^{56}$Ni masses. As shown previously by \cite{magee--18} and other studies \citep{piro-16, noebauer-17}, models with extended $^{56}$Ni distributions ($s$ = 3, 4.4) produce light curves that are brighter and bluer at earlier times than more compact $^{56}$Ni distributions ($s$ = 21, 100). The extended models contain a significant amount of $^{56}$Ni at high velocities and so the ensuing photons are able to escape more easily, as they experience fewer interactions. The bluer colours result from the outer ejecta experiencing local heating due to the $^{56}$Ni decay. This is particularly important at early times, before thermal energy from the inner ejecta is able to diffuse outwards. 

\par

In Fig.~\ref{fig:compare_models}(b), we clearly demonstrate that although the $^{56}$Ni distribution in mass-space may be the same for certain models, their light curves can show significant differences. As a representative case, the models shown in Fig.~\ref{fig:compare_models}(b) contain 0.6~$M_{\odot}$ of $^{56}$Ni and have a scale parameter of $s = 9.7$. Following from Eqn.~\ref{eqn:ni_dist}, the distribution of $^{56}$Ni is therefore the same for all models (as shown by Fig.~\ref{fig:densities}(c)) in mass coordinate. Differences in the density profiles however, result in different $^{56}$Ni distributions in velocity-space, as shown in Fig.~\ref{fig:compare_models}(c). We find that those models with higher kinetic energies (i.e. more $^{56}$Ni at higher velocities) typically show brighter light curves at early times. Up to approximately one week post explosion, the difference in $B$-band magnitude between the highest and lowest kinetic energies can be $\gtrsim$2 magnitudes. At later times, the difference in magnitude decreases, and all models reach similar peak $B$-band magnitudes (as this is driven primarily by the total $^{56}$Ni mass). The rise times also show significant variation and can range from $\sim$17 -- 24 days to reach $B$-band maximum.

\par

Previous studies have investigated the effects of $^{56}$Ni distributions in the ejecta on the light curves by varying the mass fraction of $^{56}$Ni at various mass-depths below the ejecta surface \citep{piro-nakar-2014, contreras--2018, shappee--2019}. Our models show that this mass-distribution of $^{56}$Ni is not the sole factor driving the shape of the light curves and that the density profile cannot be ignored (Fig.~\ref{fig:compare_models}). As shown above, different density profiles and velocity-distributions of $^{56}$Ni can result in significant variation in the light curves. The $^{56}$Ni mass fraction at specific mass-depths cannot therefore be uniquely identified from observed SN Ia light curves. 

\subsection{Sensitivity to $^{56}$Ni distribution}

\begin{figure*}
\centering
\includegraphics[width=\textwidth]{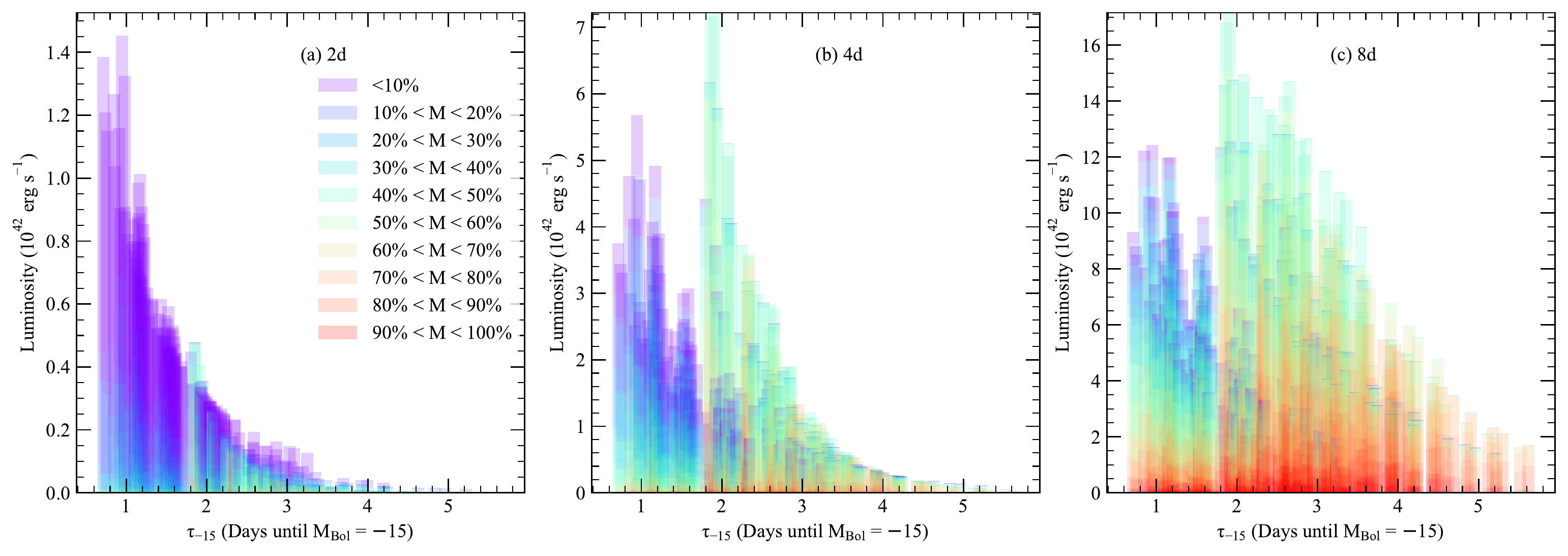}
\caption{Contribution of ejecta zones to luminosity at various times. Vertical bars represent individual models and are sorted by the bolometric $\tau_{-15}$. Panels show the total luminosity emitted by models at various times after explosion. Models are discretised into ten zones, each containing 10\% of the ejecta mass, beginning from the outer ejecta. Each zone is coloured to show the contribution towards the total luminosity emitted by the model at each time.
}
\label{fig:zone_lums}
\centering
\end{figure*}

\par

In Fig.~\ref{fig:zone_lums}, we show how the $^{56}$Ni distribution throughout the entire ejecta contributes to the bolometric luminosity at various times. We look at demonstrative times of 2, 4, and 8 days after explosion. For each model we discretise the ejecta into ten zones, each containing 10\% of the total ejecta mass, beginning at the outermost regions and moving progressively inwards. For each time, we bin escaping packets depending on the zone from which they are first injected, allowing us to determine the contribution of each zone to the total luminosity at that time. Models in Fig.~\ref{fig:zone_lums} are ordered based on the length of their bolometric $\tau_{-15}$ (see Sect.~\ref{sect:lc_props}).

\par

Figure~\ref{fig:zone_lums}(a) shows -- as expected -- that at two days after explosion, very little luminosity is escaping for those models with $\tau_{-15}$ $\textgreater$2 days and the luminosity at two days generally decreases with increasing $\tau_{-15}$. At these times, the outer 10\% of the ejecta mass contributes as much as 80\% of the luminosity for models with $^{56}$Ni present throughout the entire ejecta. For longer $\tau_{-15}$ models, the luminosity is dominated by the inner regions -- hence, these models are fainter as less light is able to escape from deep inside the ejecta. 

\par

Four days after explosion, the relative importance of the outer 10\% has decreased, however this zone still contributes a significant fraction of the total luminosity emitted for models with short $\tau_{-15}$. Figure~\ref{fig:zone_lums}(b) reveals an interesting property of our model set: it shows a bimodal distribution. Beginning with the shortest $\tau_{-15}$, there is a general trend towards lower luminosities until a $\tau_{-15}$ of $\sim$1.8~d, at which point the models show a sudden increase in brightness before again decreasing with increasing $\tau_{-15}$. Those models with $\tau_{-15}$ $\gtrsim$1.8~d also show a larger fraction of the luminosity originating from deeper inside the ejecta. This break in model behaviour results from the different $^{56}$Ni distributions and ejecta compositions used in this study. All models with $\tau_{-15}$ $\lesssim1.8$~d have $^{56}$Ni present throughout the entire ejecta (i.e. s = 3 or 4.4), while no models with more stratified ejecta structures (i.e. s = 9.7, 21, or 100) produce such short $\tau_{-15}$. The sudden increase in brightness shown in Fig.~\ref{fig:zone_lums}(b) results from these more stratified models. The rise index and $\tau_{-15}$ are highly correlated and models with longer $\tau_{-15}$ show significantly faster rises. By four days after explosion, the models that are somewhat more stratified (s = 9.7) rapidly increased in brightness over models with $^{56}$Ni present throughout the entire ejecta. This is also likely related to one of the limitations of our models, which is that the only IGEs present are due to the radioactive decay chain of $^{56}$Ni. Once the inner regions become sufficiently optically thin, light is able to escape very quickly -- giving the sudden increase in brightness. Therefore this bimodal behaviour is more likely attributable to the specific arrangement of the models presented in this work. The inclusion of additional opacity from other IGEs at high velocities would likely slow the release of light and will be explored in future work.

\par

By eight days post-explosion (Fig.~\ref{fig:zone_lums}(c)), the luminosity is now dominated by the inner ejecta regions for all models, although those models with high velocity $^{56}$Ni still show some contribution from the outer layers. This trend of decreasing sensitivity to the outer layers continues at later times. Figure~\ref{fig:zone_lums} demonstrates the complexity of determining the effects of the $^{56}$Ni distribution. In order to obtain any information about the outermost ejecta, observations within the first approximately four days following explosion are necessary. Observations at later times show a significantly reduced sensitivity to these layers and therefore are not useful for providing robust constraints on the outer $^{56}$Ni distribution.

%

\section{Observational requirements for determining the $^{56}$Ni distribution of SNe~Ia}
\label{sect:robust}

\begin{figure*}
\centering
\includegraphics[width=\textwidth]{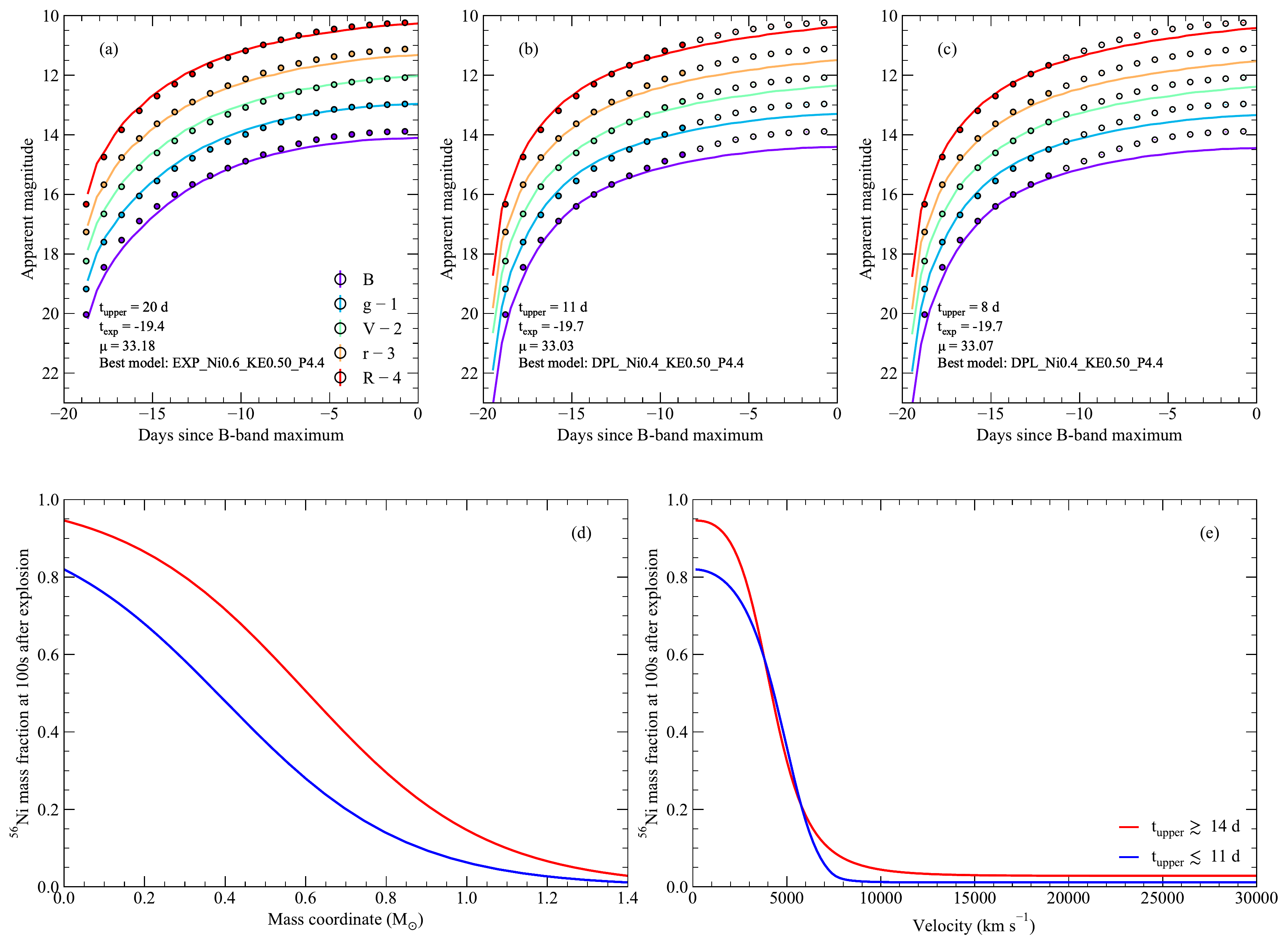}
\caption{Comparison of the H07 template and the best fitting models, assuming different upper cut-offs to the light curve. In each case, we assume first detection occurs $\sim$1\,d after explosion. Points included in the $\chi^2$ analysis are shown as solid circles. Unfilled circles show the remainder of the light curve for reference. {\it Panel a:} Best matching model parameters are given when fitting the full light curve of the H07 template, up to close to maximum light. {\it Panel b:} Best matching model parameters when fitting only less than the first two weeks of the light curve. {\it Panel c:} Only the first week of the light curve is included in the $\chi^2$ analysis, again showing how the data included in the comparison can affect the results. {\it Panels d} and {\it e} show the $^{56}$Ni distributions for the best fitting models in mass- and velocity-space, assuming different upper cut-offs to the template light curve.
}
\label{fig:upper_cut_off}
\centering
\end{figure*}

We wish to explore how the distribution of $^{56}$Ni varies among SNe~Ia. The number of bands and length of time over which a SN~Ia is observed however, typically varies from object to object. In addition, the cadence of observations also varies. It is therefore important to understand how changes in the completeness of the light curve (i.e.  wavelength coverage, temporal coverage, and cadence) affects the inferred distribution of $^{56}$Ni. Here we explore quantitatively how variations in the epoch of last detection, epoch of first detection, and number of bands affects the inferred explosion parameters.

\par

We have opted to perform these tests against existing templates for SNe Ia rather than to individual objects in order to ensure that variations due to differences in observational bands, epoch of first detection, or cadence can be carefully controlled. We have created synthetic light curves for the \cite{hsiao--07} (hereafter H07) spectroscopic template using SNCosmo \citep{sncosmo} in the $BVRgr$ bands. We have arbitrarily chosen a redshift of $z = 0.01$ (hence we assume a distance modulus of $\mu = 33.18$) and peak absolute $B$-band magnitude of $M_{B}$ = $-19.2$. We measure rise indices in the $BVRgr$ bands of $\sim$1.96. The choice of $M_{B}$ = $-19.2$ therefore produces a light curve with a peak absolute $B$-band magnitude and rise indices that are coincident with the maxima of the model distributions shown in Fig.~\ref{fig:observables}. This allows us to produce matches on either side of the distribution peak, covering a broad range of models. We note however that the exact values for peak magnitude and distance modulus are unimportant as we are not necessarily interested in which model produces the best match in this case, but rather the conditions under which the best match changes. Additionally, we have created templates using our EXP\_Ni0.6\_KE1.10\_P9.7, EXP\_Ni0.4\_KE2.18\_P3, and DPL\_Ni0.8\_KE0.60\_P4.4 models, which represent a range of light curve behaviours. The benefit of using templates made from our own model set is that the true model properties are exactly known, and we are able to test whether they are recovered during the fitting process. Conversely, comparisons to our H07 template more closely mimic comparisons with real observations as it is unlikely that our models will produce a perfect match. For each of our model templates we introduce scatter into the observations for each band assuming a Gaussian distribution of random offsets with a standard deviation of 0.05 mag., which is comparable to typical photometric uncertainties. We also set the uncertainty of each point equal to 0.05 mag. 

\par

To determine the best matching models for a template, we compare the template light curve against every model light curve in our set using a $\chi^2$ analysis in flux space. Each template is matched to the models by assuming that the explosion epoch can occur at any point from 30 days before maximum light up to the date of the first 5$\sigma$ detection, in steps of 0.1 days. We also include an uncertainty on the distance modulus for each template of $\delta\mu = 0.15$~mag., which is typical of the distance modulus uncertainty for SNe. To account for potential underestimates of the photometric uncertainties and uncertainties in the model setup \cite[see ][]{magee--18}, we have included a scaling factor to adjust the best matching model for each template to have a reduced-$\chi^2$ of $\chi^2_{\nu} = 1$. From this, we determined the number of models within 3$\sigma$ of $\chi^2_{\nu} = 1$.

\par

When finding the best match to the H07 template, we infer an explosion date of $t_{\rm{exp}} = -19.6$\,d relative to $B$-band maximum. For templates created from our own models, the explosion date occurs exactly at $t_{\rm{exp}} = 0$. We use these fiducial values to explore how various cuts in the light curve affect our results. For each template, we also test observing cadences of one and three days.


\subsection{Effect of fitting varying time ranges}
\label{sect:varying_times}

Here we explore the importance of obtaining a well-sampled light curve. To do this, we perform separate tests in which we progressively omit observations either at later (closer to maximum light) or earlier (closer to explosion) times. 

\subsubsection{Neglecting later points} 
\label{sect:later_times}
The first test we perform is on the overall temporal extent of the light curve -- in other words, the length of time after explosion that the observations continue. We test temporal ranges extending from 5 to 20 days after the first detection, with cadences of one and three days. In each case, we assume first detection occurs approximately one day after explosion.

\begin{figure*}
\centering
\includegraphics[width=\textwidth]{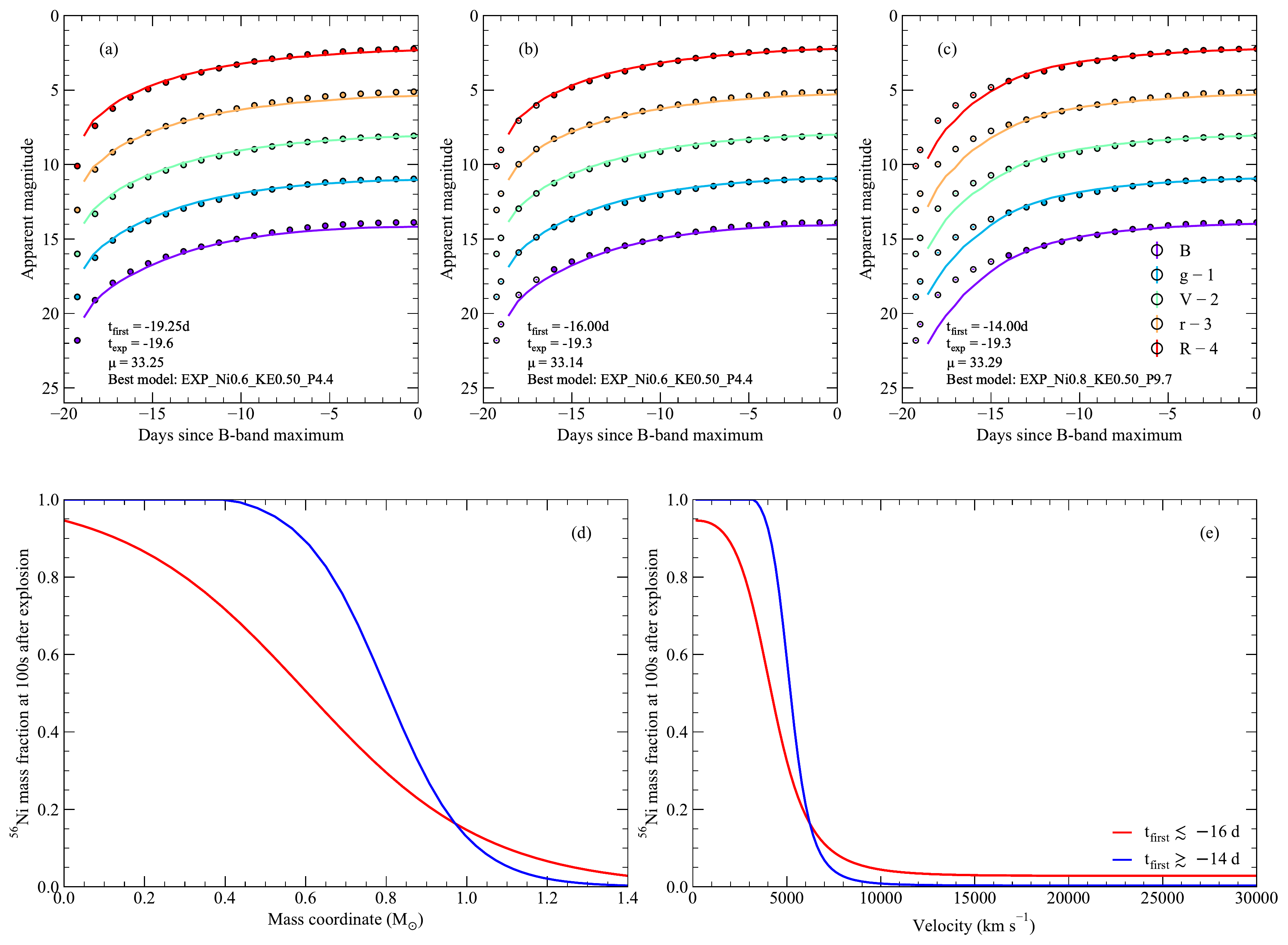}
\caption{Comparison of the H07 template and the best fitting models, assuming different epochs of first detection ($t_{\rm{first}}$) relative to $B$-band maximum. Points included in the $\chi^2$ analysis are shown as solid circles. Unfilled circles show the remainder of the light curve for reference. {\it Panel a:} Best matching model parameters are given when fitting the full light curve of the H07 template. In this case, first detection occurs only hours after our estimated date of explosion. {\it Panel b:} Best matching model parameters for a first detection approximately 3 days after explosion. {\it Panel c:} Best matching model parameters for a first detection approximately 5 days after explosion.  {\it Panels d} and {\it e} show the $^{56}$Ni distributions for the best fitting models in mass- and velocity-space, assuming different first detection dates for the template light curve.}
\label{fig:lower_cut_off}
\centering
\end{figure*}

\par

The best matching $^{56}$Ni distributions of the H07 template show dramatic changes with the date of last detection. In Fig.~\ref{fig:upper_cut_off} we show the best fitting models assuming different upper cut-offs to the light curve. When analysing the full 20\,d light curve, we find that we consistently favour models with $0.6~M_{\rm{\odot}}$ of $^{56}$Ni, low kinetic energies (0.50, 0.60), and extended $^{56}$Ni distributions. Figure~\ref{fig:upper_cut_off}(a) shows that our EXP\_Ni0.6\_KE0.50\_P4.4 model is able to broadly match the H07 template. Some differences in colour are apparent, however we again stress that we have taken a rather simplistic approach to the composition and it is likely that modifications could be made to improve agreement. If the light curve extends to $\gtrsim$2 weeks after explosion, we find little change in the best matching models parameters. Including only 11\,d after the first detection however, we find that we also obtain good agreement in $\sim$20\% of cases with models containing a lower $^{56}$Ni mass of $0.4~M_{\rm{\odot}}$, as shown in Fig.~\ref{fig:upper_cut_off}(b). The distance modulus also shifts to lower values, while the explosion date remains largely unchanged. We note however that the best matching model (i.e. the model producing the lowest $\chi^2$) also remains unchanged in this case. Limiting the light curve to only approximately the first week (Fig.~\ref{fig:upper_cut_off}(c)), we now find that a lower $^{56}$Ni mass is the preferred model, and accounts for 75\% of matches within the 3$\sigma$ range. A similar trend is obtained when using lower cadence light curves, however here there is a larger spread in the kinetic energies. 

\par

Conversely, for templates created from our own model set we see little change in the best matching models, explosion times, and distance moduli regardless of the date of last detection, assuming the light curve has a 1\,d cadence. For a 3\,d cadence, we find much greater diversity in the $^{56}$Ni distribution for some templates. When comparing to the full light curve of our DPL\_Ni0.8\_KE0.60\_P4.4 template with a 3\,d cadence, approximately 33\% of the matches contain only $0.6~M_{\rm{\odot}}$ of $^{56}$Ni, which is clearly incorrect. This fraction increases (to $\sim$50\%) as observations closer to maximum light are excluded.

\par

The first of our robustness tests demonstrates how limiting the light curve could change the inferred $^{56}$Ni distribution. For our H07 template, we find that models with 0.6~$M_{\rm{\odot}}$ of $^{56}$Ni produce the best agreement to the full light curve. Figure~\ref{fig:upper_cut_off}(c) shows however, that a lower $^{56}$Ni mass model does indeed provide a good match to only the first week of the light curve, while it is clearly discrepant at later times. It is possible that a combination of the $^{56}$Ni distributions shown in Fig.~\ref{fig:upper_cut_off}(d) and (e) could produce the best agreement overall. In this way, the light curve could be used to map out a more detailed profile for the $^{56}$Ni distribution, rather than those assumed here. This would necessitate however, fixing the explosion date and distance modulus. Limiting the light curve comparison to only the first approximately two weeks following explosion could also lead to an entirely incorrect result, as discussed previously in the case of our DPL\_Ni0.8\_KE0.60\_P4.4 template. Nevertheless, we typically find a convergence towards a $^{56}$Ni distribution as more of the light curve is included. We therefore suggest that the light curve must cover at least a two week time span, with a cadence of $\lesssim$3\,d. Fitting only the first approximately one week after explosion can result in significantly different model parameters. Future work will continue to explore more complex $^{56}$Ni distributions than the functional form given by Eqn.~\ref{eqn:ni_dist}, which could provide better matches to observations in some cases.


\subsubsection{Neglecting early points}
Our model light curves show the most significant variations at early times, as demonstrated by Fig.~\ref{fig:mean_diff}. Therefore these epochs provide important constraints on the $^{56}$Ni distributions. To determine how early observations need to be obtained to effectively constrain the $^{56}$Ni distribution and explosion properties, we artificially trim the early light curves by excluding all points before a certain cut-off phase relative to maximum light. This allows us to simulate the first detection ($t_{\rm{first}}$) of an object at various times after explosion. In all cases, we include observations up to twenty days post-explosion.

\par

For the H07 template, we find significant differences in the $^{56}$Ni distribution the later the first detection occurs after explosion. As mentioned in Sect.~\ref{sect:later_times}, when fitting the full light curve, we find the EXP\_Ni0.6\_KE0.50\_P4.4 model to produce the best match. If first detection occurs at $\sim$ $-$15\,d relative to $B$-band maximum however (equivalent to $\sim$4.5\,d after explosion), we find that $\sim$25\% of models within the 3$\sigma$ range invoke a higher $^{56}$Ni mass of $0.8~M_{\rm{\odot}}$. For a first detection epoch of $\sim$ $-$14\,d, models with $0.8~M_{\rm{\odot}}$ of $^{56}$Ni account for $\sim$33\% of matches and also provide the best agreement (lowest $\chi^2$), as shown in Fig.~\ref{fig:lower_cut_off}. These models also typically have more compact $^{56}$Ni distributions. Although models containing $0.8~M_{\rm{\odot}}$ of $^{56}$Ni provide a good match to the later light curve, they are clearly significantly brighter than the H07 template at early times. When comparing to models created from our own templates, we consistently find that the fitting process returns the correct $^{56}$Ni mass, however there remains a significant spread in the kinetic energies of the models. Again, we find this spread also increases as first detections occurs increasingly later after explosion and if the cadence is lowered to 3\,d.

\begin{figure}
\centering
\includegraphics[width=\columnwidth]{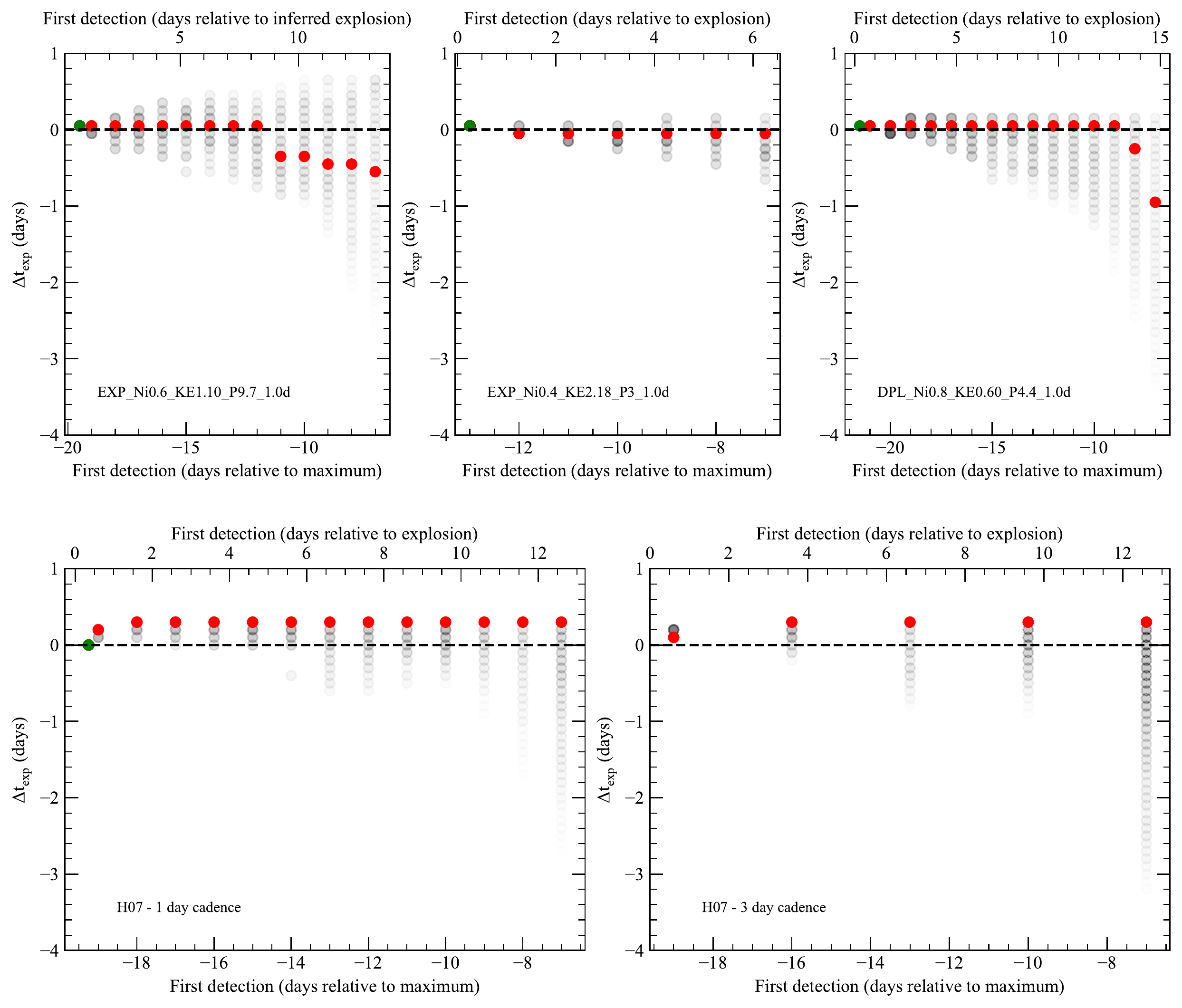}
\caption{ Difference between explosion dates for various epochs of first detections relative to fitting the full light curve ($\Delta$t$_{\rm{\exp}}$). Each panel represents a different template light curve. The H07 template is shown with both a 1\,d and 3\,d cadence, while all other templates are shown with a 1\,d cadence. Dates of first detection are given relative to explosion and maximum light in each case. Comparisons including the full light curve are shown as green points and used to set the fiducial explosion date for each template. Best fitting explosion dates for each epoch of first detection are shown in red, while the $3\sigma$ spread in explosion dates is given by shaded circles. Darker shades indicate how frequently a given explosion date occurs within the 3$\sigma$ range.
}
\label{fig:early_cut_off_exp}
\centering
\end{figure}

Aside from the $^{56}$Ni distribution itself, we find a significant increase in the explosion date uncertainty as first detection occurs at later times. In Fig.~\ref{fig:early_cut_off_exp}, we show how the explosion date varies depending on the epoch of first detection. Unsurprisingly, we find that first detections occurring later result in a larger scatter in the best fitting explosion times. For first detections occurring $\textless$4\,d after explosion, the typical spread in the explosion time is $\lesssim$0.5\,d. If the first detection occurs approximately one week after explosion, this increases further to \textgreater1\,d. For our H07 template, we also find that excluding the first detection at $\sim$0.35\,d after explosion results in a systematic shift to an explosion date that is 0.2\,d later than the fiducial value and to a larger distance modulus by 0.07~mag, although the best fitting model remains the same. If observations begin $\textgreater$1\,d after explosion, the best fitting explosion date is shifted later by a further 0.1\,d. Lowering the cadence of observations to 3\,d, we again find that the spread in model explosion times typically increases for all epochs of first detection. Those objects with lower cadence light curves would need to be discovered earlier in order to provide tighter constraints on their explosion properties -- however, discovering such objects at the required early times would likely prove challenging if the cadence were so low.

\par

In general, we find that increasing the time between explosion and first detection results in a greater diversity of the best fitting ejecta kinetic energies, $^{56}$Ni distributions, distance moduli, and explosion times. Lower cadence templates show greater diversity in all properties within the best matching 3$\sigma$ range. To provide the tightest constraints on the explosion date and $^{56}$Ni distribution (here, we define `tightest constraints' as picking the minimum number of models or a single model), observations beginning $\lesssim$3\,d after explosion are necessary, with a cadence of $\textless$3\,d. Based on rise time estimates for samples of SNe Ia \cite[e.g. ][]{hayden--10, ganeshalingam--2011} this corresponds to $\lesssim-14$ relative to $B$-band maximum.


\subsection{Effect of multiple filters}
\label{sect:multiple bands}

We now look at whether excluding specific bands changes which model is determined to be best. We have performed our fitting procedure including only the $B$-band, only the $R$-band, both the $B$- and $R$-bands, and the full $BVRgr$ light curve. In each case, we assume first detection occurs approximately one day after explosion and include observations up to three weeks after this point.

\begin{figure*}
\centering
\includegraphics[width=\textwidth]{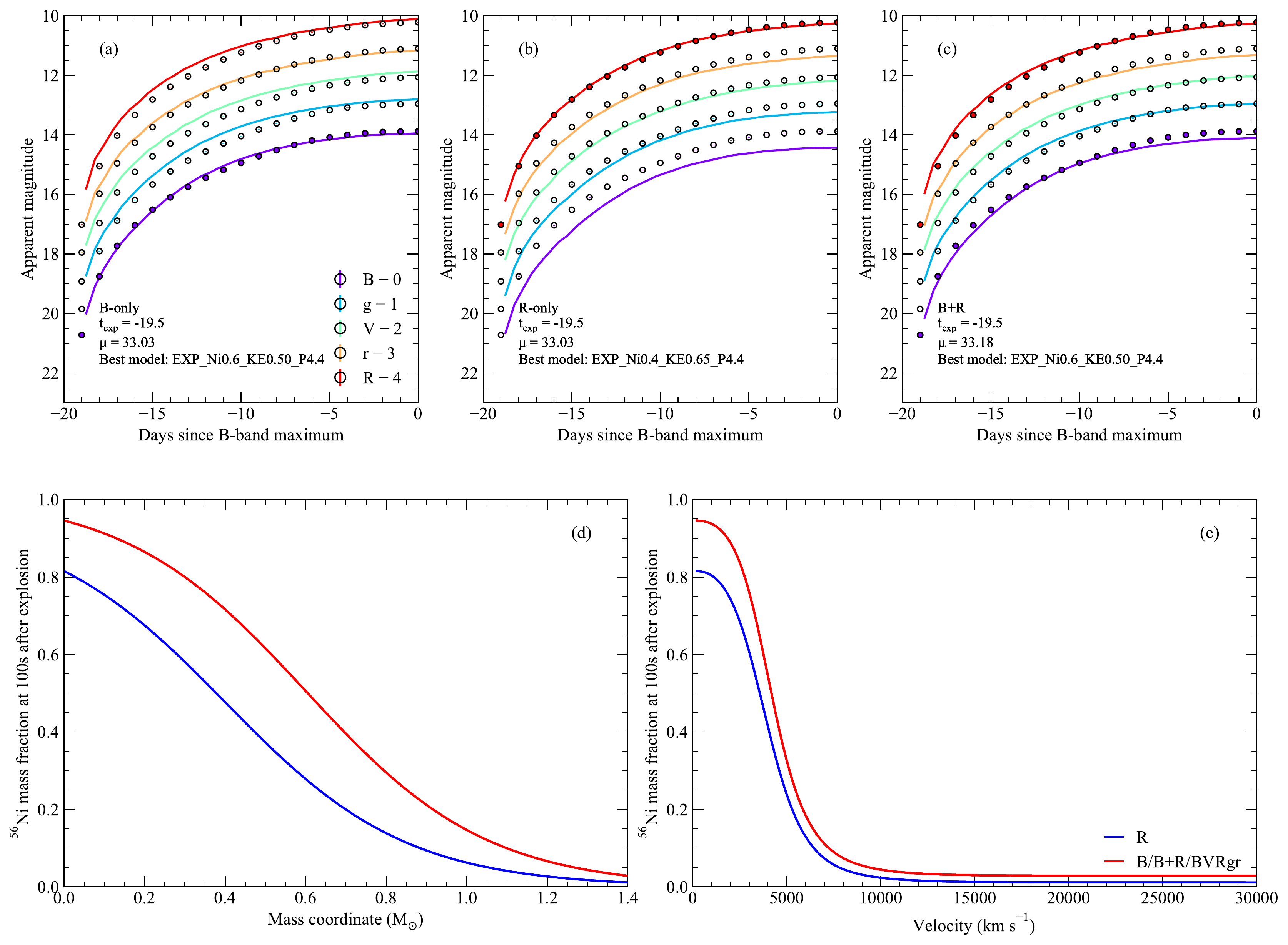}
\caption{Comparison of the H07 template and the best fitting models when including only certain bands. Points included in the $\chi^2$ analysis are shown as solid circles. Unfilled circles show the remainder of the light curve for reference. In each case, the template is first detected one day after explosion and extends to approximately three weeks later. {\it Panel a:} Best matching model parameters are given when fitting only the $B$-band light curve of the H07 template. {\it Panel b:} Fitting only the $R$-band, we find the best model has changed. {\it Panel c:} Best matching model when including both the $B$- and $R$-bands. {\it Panels d} and {\it e} show the $^{56}$Ni distributions for the best fitting models in mass- and velocity-space, assuming the noted filters are observed.}
\label{fig:filters}
\centering
\end{figure*}
   
For our H07 template, we find that including only the $B$-band does not result in a change in the best matching models, however there is generally a decrease in the distance modulus (by $\sim$0.1~mag). Figure~\ref{fig:filters}(a) shows that including only the $B$-band results in better agreement with the $B$-band, at the expense of agreement in redder bands. Conversely, including only the $R$-band results in an entirely different best matching model. We find the $R$-band is best reproduced with a model containing a lower $^{56}$Ni mass of 0.4~$M_{\rm{\odot}}$, however this model clearly underestimates the flux in bluer bands. Indeed, this model (EXP\_Ni0.4\_KE0.65\_P4.4) is the only one that appears within the 3$\sigma$ range for the $R$-band only matches, and does not appear within the 3$\sigma$ range when comparing to the full light curve.

\par

For our H07 template, the $B$-band alone is sufficient to determine the $^{56}$Ni distribution, however this is not the case for all of our other templates. Approximately 33\% of matches to our DPL\_Ni0.8\_KE0.60\_P4.4 template contain 0.6~$M_{\rm{\odot}}$ of $^{56}$Ni  (an incorrect result) when limiting the comparison to either the $B$- or $R$-bands. Across all of our templates however, in general we find that including only the $R$-band results in a larger uncertainty in (or even incorrect) $^{56}$Ni distribution, explosion time, and distance modulus compared to including only the $B$-band. For lower cadences, the spread in best matching parameters increases significantly. In the case of templates created from our own models with a 3\,d cadence, including only one band generally results in an incorrect $^{56}$Ni distribution as the best match. 
   
\par

Although, the wavelength coverage has a significant effect on the inferred $^{56}$Ni distributions, the explosion date is generally insensitive to the band of observation. This would indicate that the shape of the light curve is the overall strongest constraint on the explosion epoch. For example, our models show that $\tau_{-15}$ in different bands is highly correlated, and therefore the luminosity generally increases in all bands around the same time -- although this increase is typically more extreme in the bluer bands.

\subsection{Summary of observational requirements}

Our robustness tests provide a prescription for determining the $^{56}$Ni distribution of an object as accurately as possible within the model set explored here. We find that high cadence ($\lesssim$3\,d) observations beginning within $\lesssim$3\,d after explosion, extending to at least two weeks after explosion, and in at least one blue and one red band (e.g. $B$- and $R$-band) are necessary to provide robust results. Additional filters can reduce the uncertainty on best fitting parameters further, however a minimum of one blue and one red band is necessary. Additional scatter within the best fitting models for individual objects can likely be attributed to the coarseness of the parameter space explored. Future work will continue to expand the parameter space of models and determine $^{56}$Ni distributions in greater detail. 

\par

The tests conducted here also highlight the importance of modelling observations in a consistent manner. For example, including only observations within the first few days of explosion can produce results that are in direct contradiction to those at maximum light. Therefore, not only should observations be obtained with the correct coverage, but full exploitation of such observations is neccessary to avoid poor or contradictory results.

%

 %
\section{Comparisons with observations of SNe Ia}
\label{sect:comparisons}

\begin{table*}
\centering
\caption{Best fitting models per object }
\tabularnewline
\label{tab:object_fits}\tabularnewline
\resizebox{\textwidth}{!}{
\begin{tabular}{lllllllll}\hline
\hline
\multicolumn{5}{c}{Supernova properties} \vline & \multicolumn{4}{c}{Best fitting model properties} \tabularnewline
\hline
Name & Extinction  & Host   & Filters & Reference & Best match & Explosion date  & Matches & Unique   \tabularnewline
     & ($A_{\rm{V}}$) &  redshift &  &  &  & (MJD) &  & matches   \tabularnewline

\hline
\multicolumn{9}{c}{Consistent with $^{56}$Ni distributions} \tabularnewline
\hline
PTF09dsy		        		&	0.35    &  0.013 	 & $R$          		& 9 	        		&   EXP\_Ni0.6\_KE0.50\_P9.7	&	55052.91(0.52)	&	36	&	14	 \tabularnewline
\textbf{SN~2009ig}		&	0.09    &  0.009 	 & $BVRr$       		& 12	        		&   EXP\_Ni0.8\_KE1.10\_P9.7	&	55061.20(0.54)	&	52	&	16	 \tabularnewline
PTF10duz		        		&	0.07    &  0.064 	 & $R$          		& 9 	        		&   EXP\_Ni0.6\_KE1.10\_P4.4	&	55266.98(0.42)	&	755	&	48	 \tabularnewline
PTF10hml		        		&	0.04    &  0.053 	 & $R$          		& 9 	        		&   DPL\_Ni0.6\_KE1.68\_P3	    &	55334.42(0.66)	&	987	&	57	 \tabularnewline
PTF10iyc		        		&	0.04    &  0.059 	 & $R$          		& 9 	        		&   EXP\_Ni0.8\_KE1.68\_P4.4	&	55343.35(0.32)	&	31	&	11	 \tabularnewline
\textbf{SN~2010jn}		&	1.21    &  0.025 	 & $Rgr$        		& 13	        		&   EXP\_Ni0.8\_KE2.18\_P4.4	&	55478.42(0.51)	&	770	&	38	 \tabularnewline
PTF10accd		        &	0.35    &  0.035 	 & $R$          		& 9 	        		&   EXP\_Ni0.6\_KE1.10\_P3	    &	55537.10(0.32)	&	33	&	10	 \tabularnewline
PTF11hub		        		&	0.03    &  0.029 	 & $R$          		& 9 	        		&   DPL\_Ni0.4\_KE1.81\_P9.7	&	55753.23(0.29)	&	17	&	5	 \tabularnewline
\textbf{SN~2011fe*}		&	0.04    &  0.001     	 & $BVRgr$ 	   	& 14, 15, 16	&   EXP\_Ni0.6\_KE0.78\_P4.4	&	55796.79(0.57)	&	41	&	8	 \tabularnewline
PTF12emp		        &	0.03    &  0.037 	 & $R$          		& 9 	        		&   EXP\_Ni0.4\_KE0.78\_P4.4	&	56061.09(1.04)	&	108	&	30	 \tabularnewline
\textbf{SN~2012cg*}		&	0.62    &  0.002     	 & $BVRr$ 	    	& 17        		&   EXP\_Ni0.6\_KE0.78\_P4.4	&	56062.75(0.10)	&	12	&	3	 \tabularnewline
PTF12gdq		        		&	0.10    &  0.033 	 & $R$          		& 9 	        		&   EXP\_Ni0.4\_KE1.10\_P9.7	&	56097.42(0.72)	&	247	&	22	 \tabularnewline
KSN2012a		        &	0.00    &  0.086 	 & K2           		& 8	            	&   DPL\_Ni0.4\_KE2.18\_P3	    &	56161.11(0.59)	&	23	&	3	 \tabularnewline
\textbf{SN~2012fr}		&	0.06    &  0.005     	 & $BVgr$ LSQ 	& 18, 19    	&   EXP\_Ni0.6\_KE1.40\_P9.7	&	56224.23(0.13)  &	30	&	3	 \tabularnewline
\textbf{LSQ12fxd}		&	0.07    &  0.031 	 & $BVr$ LSQ    	& 9, 10	        &   DPL\_Ni0.8\_KE1.40\_P4.4	&	56228.05(0.59)	&	147	&	20	 \tabularnewline
\textbf{SN~2012ht}		&	0.08    &  0.004     	 & $BVRgr$ 	    	& 20        		&   EXP\_Ni0.4\_KE0.50\_P9.7	&	56275.80(0.53)	&	56	&	14	 \tabularnewline
\textbf{LSQ13ry}	    	&	0.12    &  0.030 	 & $BVgr$ LSQ   	& 9, 10	        &   EXP\_Ni0.6\_KE0.78\_P9.7	&	56376.83(0.75)	&	152	&	20	 \tabularnewline
\textbf{iPTF13asv*}		&	0.14    &  0.036 	 & $BVRr$       		& 3	            	&   EXP\_Ni0.8\_KE1.68\_P9.7	&	56411.48(0.25)	&	93	&	6	 \tabularnewline
\textbf{SN~2013dy}		&	1.07    &  0.004 	 & $BVR$ Unf.   	& 21, 22    	&   EXP\_Ni0.4\_KE0.50\_P9.7	&	56479.55(0.84)	&	60	&	14	 \tabularnewline
\textbf{iPTF13dge*}		&	0.34    &  0.016 	 & $BVgr$       		& 4	            	&   EXP\_Ni0.4\_KE1.40\_P9.7	&	56536.30(0.78)	&	535	&	24	 \tabularnewline
LSQ13cpk		        		&	0.06    &  0.033 	 & LSQ          		& 9	            	&   EXP\_Ni0.6\_KE1.68\_P4.4	&	56572.34(0.54)	&	35	&	12	 \tabularnewline
\textbf{iPTF13ebh*}		&	0.37    &  0.013 	 & $BVgr$       		& 5	            	&   EXP\_Ni0.4\_KE1.10\_P9.7	&	56605.58(0.41)	&	70	&	10	 \tabularnewline
\textbf{SN~2013gy}		&	0.48    &  0.014     	 & $BVgr$ 	    	& 23        		&   EXP\_Ni0.6\_KE1.10\_P4.4	&	56630.48(0.23)	&	12	&	2	 \tabularnewline
\textbf{iPTF14bdn*}		&	0.03    &  0.016 	 & $BVR$        		& 6	            	&   EXP\_Ni0.6\_KE1.40\_P9.7	&	56801.17(0.47)	&	232	&	19	 \tabularnewline
\textbf{SN~2015F*}   	&	0.80    &  0.005 	 & $BVRgr$      		& 24, 25    	&   EXP\_Ni0.6\_KE0.65\_P9.7	&	57087.57(0.26)	&	29	&	5	 \tabularnewline
\textbf{iPTF16abc*}		&	0.25    &  0.023 	 & $BVgr$       		& 7	            	&   EXP\_Ni0.6\_KE1.40\_P4.4	&	57480.22(0.17)  &	17	&	2	 \tabularnewline
\textbf{SN~2017erp}		&	0.60    &  0.006     	 & $BVRgr$ 	    	& 29        		&   EXP\_Ni0.6\_KE0.78\_P4.4	&	57915.42(0.18)	&	12	&	3	 \tabularnewline

\hline
\multicolumn{9}{c}{Require additional early flux} \tabularnewline
\hline
LSQ12gpw		        &	0.19    &  0.058 	& LSQ          	& 9	            	&   EXP\_Ni0.8\_KE0.78\_P4.4		&	56246.71(0.42)	&	583	&	33	 \tabularnewline
\textbf{LSQ12hzj}		&	0.18    &  0.029 	& $BVr$ LSQ    & 9, 10	        &   EXP\_Ni0.8\_KE0.78\_P100		&	56284.34(0.49)	&	131	&	6	 \tabularnewline
\textbf{SN~2016jhr}		&	0.08    &  0.117 	 	& $gr$         	& 11	        		&   EXP\_Ni0.4\_KE0.65\_P3	    	&	57481.13(1.30)  &	3185&	69	 \tabularnewline
\textbf{SN~2016coj}		&	0.06    &  0.005 	& $BVR$ Unf.   & 26, 27    	&   DPL\_Ni0.4\_KE1.40\_P100		&	57531.30(0.55)	&	232	&	19	 \tabularnewline
\textbf{SN~2017cbv}		&	0.46    &  0.004     	& $BVgr$ 	    	& 28         		&   EXP\_Ni0.8\_KE0.65\_P4.4		&	57820.39(0.21)	&	16	&	6	 \tabularnewline
\textbf{SN~2018oh}		&	0.11    &  0.011	 	& $Vg$ K2      	& 2	            	&   EXP\_Ni0.6\_KE1.68\_P4.4		&	58144.57(0.05)	&	2	&	1	 \tabularnewline
\hline
\multicolumn{9}{c}{Ambiguous} \tabularnewline
\hline
KSN2011b		&	0.00    &  0.052 	 & K2           & 8	            &   EXP\_Ni0.4\_KE2.18\_P4.4    &   55828.58(0.05)	&	2	&	1	 \tabularnewline
\textbf{ASASSN-14lp}		&	1.09    &  0.005 	 & $Vgr$ Unf.   & 1	            &   DPL\_Ni0.8\_KE2.18\_P9.7	&	56997.88(0.48)	&	19	&	4	 \tabularnewline

\hline
\end{tabular}
}
\tablefoot{Best fitting model parameters for each SN. Where appropriate, host galaxy extinction is included. The filters used to determine the best matching models are given. Explosion dates are given by the best matching model, while the uncertainty is calculated as the standard deviation of explosion dates from models within the 3$\sigma$ range. Objects included in our gold sample are shown in bold. Within the group `consistent with $^{56}$Ni distributions', those SNe denoted with an asterisk represent objects for which we believe additional tuning of our models is necessary to produce improved agreement. References: (1) \cite{shappee--16}; (2) \cite{shappee--2019}; (3) \cite{cao--16}; (4) \cite{ferretti--16}; (5) \cite{hsiao--2015}; (6) \cite{smitka--15}; (7) \cite{miller--18}; (8) \cite{olling--15}; (9) \cite{firth--sneia--rise}; (10) \cite{walker--15}; (11) \cite{jiang--2017}; (12) \cite{foley--2012--09ig}; (13) \cite{hachinger--13}; (14) \cite{11fe--nature}; (15) \cite{2011fe}; (16) \cite{vinko--12}; (17) \cite{marion--16}; (18) \cite{zhang--14}; (19) \cite{contreras--2018}; (20) \cite{yamanaka--14}; (21) \cite{zheng--13}; (22) \cite{pan--15}; (23) \cite{holmbo--18}; (24) \cite{cartier--15}; (25) \cite{im--15}; (26) \cite{richmond--17}; (27) \cite{zheng--17a}; (28) \cite{hosseinzadeh--17}; (29) \cite{brown--19}
}
\end{table*}

We have compiled a sample of well studied SNe~Ia from the literature with early light curve data for which we can determine the best matching models from our parameter space using the $\chi^2$ analysis discussed in Sect.~\ref{sect:robust}. For the reasons previously discussed (see Sect.~\ref{sect:models}) we do not include near-UV (such as $U$ or $u$) bands or near-IR (such as $I$ or $i$) bands in our analysis. We also note that, for those objects with unfiltered observations, we compared directly to the $R$-band, as unfiltered observations are often calibrated to the $R$-band \cite[e.g][]{zheng--17a, shappee--18}

\par

Following from Sect.~\ref{sect:robust}, we select those objects that have at least one 3$\sigma$ detection $\sim$14 days before $B$-band maximum. Those SNe with observations in more than one optical band form our gold sample of 23 objects. An additional 12 objects have observations in only one band, bringing our full sample to 35 SNe. In Table~\ref{tab:object_fits}, we list the details of objects studied here. The explosion date is given by the best matching models (i.e. lowest $\chi^2$), while the uncertainty shows the spread of values obtained within the 3$\sigma$ range. Table~\ref{tab:object_fits} also lists the number of models within our set that produce a match to each object within the 3$\sigma$ range. A single model may produce multiple matches, due to changes in explosion date and/or distance modulus, therefore we also list the number of unique matches (out of a possible 255). We again stress that the model set presented as part of this work does not represent a statistical sampling of the most likely outcomes for thermonuclear explosions -- our model set is simply designed to explore a large parameter space. Therefore, while the relative numbers of models within the 3$\sigma$ range are useful for comparisons among objects, the absolute numbers are not overly meaningful.   

\par

\begin{figure*}
\centering
\includegraphics[width=\textwidth]{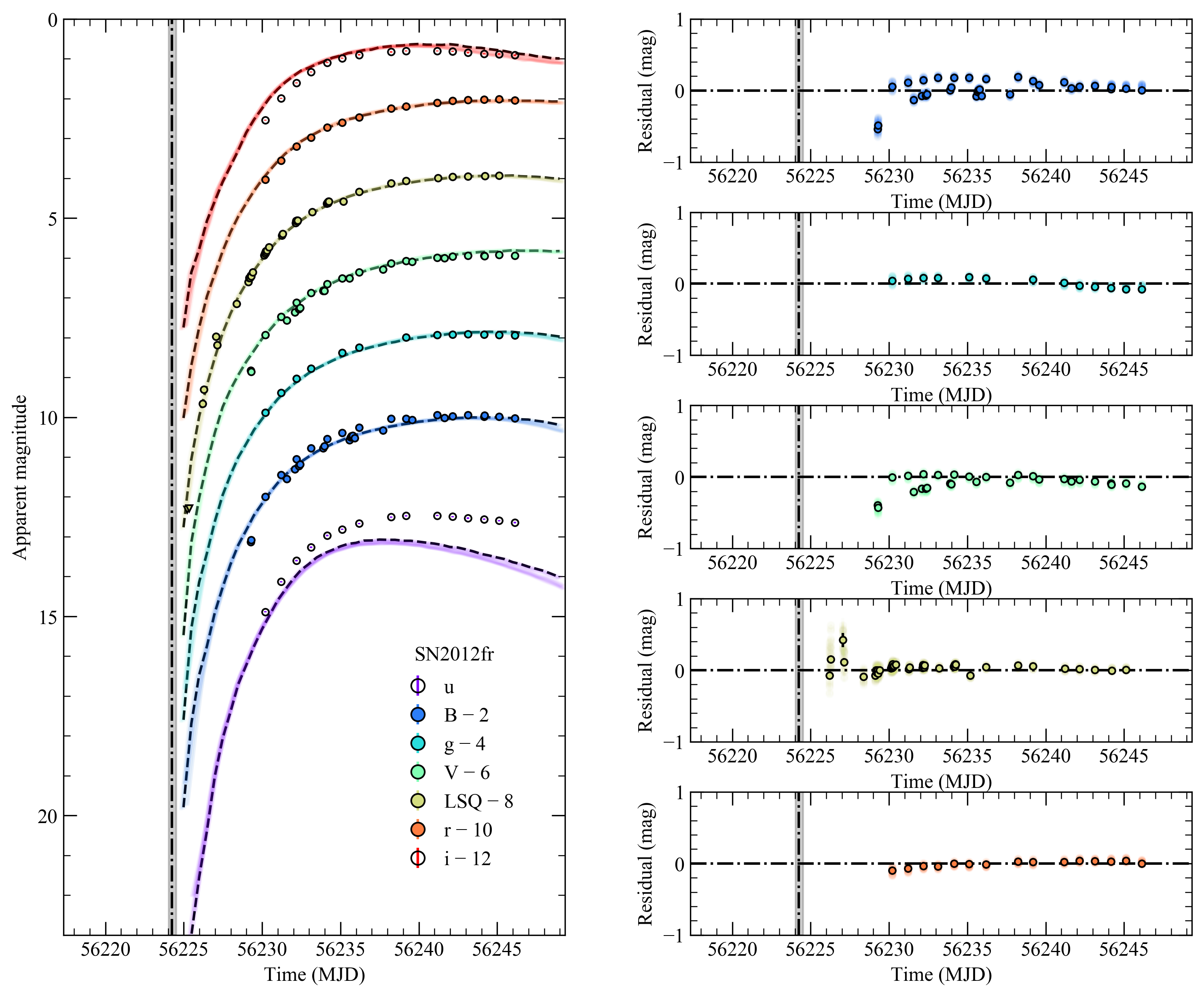}
\caption{Light curves and residuals of the best matching models for SN~2012fr within the 3$\sigma$ range. {\it Left:} Observations used in calculating the $\chi^2$ value are shown as filled circles, while those not included in the fit are shown as unfilled circles. Any epochs of $\textless3\sigma$ detections are considered upper limits and shown as triangles. The lowest $\chi^2$ model is shown as a dashed line, while the shaded colour regions show the models within the 3$\sigma$ range. The range of possible explosion dates is given by the shaded grey region with the best match explosion date denoted by a dash-dot line. {\it Right:} Residuals of the best matching models within the 3$\sigma$ range compared to observations of SN~2012fr. The lowest $\chi^2$ model is denoted by points with a black outline. }
\label{fig:SN2012fr_best}
\centering
\end{figure*}

In the following section, we divide the sample into three separate groups: i) SNe~Ia that are consistent with the models presented here, or would require minor adjustments to models to produce better agreement, ii) SNe~Ia requiring additional flux at early times (e.g. an early `bump'), and iii) ambiguous cases that may fall under either of the previous cases. For each object, we calculate the residuals between observations and models within the 3$\sigma$ range to place it into one of the three groups. Figure~\ref{fig:SN2012fr_best} demonstrates the residuals obtained for SN~2012fr.


\subsection{SNe Ia that are consistent with model $^{56}$Ni distributions}

For this first group of SNe~Ia, we include any event for which the residual has a maximum value of \textless$\pm1$ mag. in each band and the mean residual is approximately zero ($\lesssim0.1$~mag.). We assume that these objects are well matched by one of the models in our studied parameter space, however some objects do show slightly larger residuals than the mean, particularly at early times. Given the relatively coarse nature of our model set however, we suggest that somewhat modified $^{56}$Ni distributions could provide improved agreement in these cases. Therefore, this group can be further divided into those objects for which the models do or do not require additional tuning. We find that 17 out of 23 ($\sim$74\%) objects in our gold sample fall into this category, which are split approximately evenly between those that require additional tuning and those that do not. For our full sample, this fraction increases slightly to 27 out of 35 ($\sim$77\%) objects showing good agreement with our models. The remaining objects are discussed further in Sects.~\ref{subsect:excess} and \ref{subsect:ambig}.

\subsubsection{Examples of good matches}   

As an example of the typical agreement between models and data, in Fig.~\ref{fig:SN2012fr_best} we show the best matching models for SN~2012fr. Additional objects are shown in Appendix.~\ref{sect:apdx:good}. Our model with 0.6~$M_{\rm{\odot}}$ of $^{56}$Ni, a relatively intermediate kinetic energy (1.40), and an intermediate $^{56}$Ni distribution ($s$ = 9.7) shows very good agreement with the light curves of SN~2012fr. Some small discrepancies are apparent, however we again stress that we have taken a relatively simple approach to the ejecta composition.

\par

SN~2012fr has been extensively studied in the literature. Compared to our best match, \cite{contreras--2018} found good agreement with a $^{56}$Ni distribution containing a higher mass fraction in the outer ejecta (at larger mass coordinates). \cite{contreras--2018} also find an explosion time that is 1.5 days later than our best match of MJD = 56\,224.23, however this neglects the effects of a dark phase. The lower $^{56}$Ni fraction in the outer ejecta of our models is consistent with the inferred earlier explosion time due to the dark phase. The first significant flux begins to emerge at $\sim$1 day after explosion, at which point the $B$-band quickly reaches M$_B$ $\sim$ $-10$. We also note that, due to this dark phase, our model is fully consistent with the limits observed on MJD = 56\,225 occurring after explosion and the subsequent \textgreater2.5 magnitude rise in $\lesssim$1 day, as can be seen in Fig.~\ref{fig:SN2012fr_best}. SN~2012fr clearly shows the importance of the dark phase and that the explosion date can not simply be estimated as occurring between the first detection and previous non-detection -- the explosion may have already occurred at the epochs of non-detections. Accurate estimation of the explosion date requires comparisons with model light curves.

\par

For some objects, we find that the model with the lowest $\chi^2$ does not necessarily produce the best match at early times. An example of this is shown in Fig.~\ref{fig:SN2013gy_best} for SN~2013dy. Although the lowest $\chi^2$ is somewhat fainter than the $g$-band observations at very early times, we do find models within the 3$\sigma$ range that can perfectly reproduce these points at the expense of slightly worse agreement at later times. This is likely related to the fact that no weighting has been applied to observations at earlier times, which are also generally less numerous than at later times.

\begin{figure*}
\centering
\includegraphics[width=\textwidth]{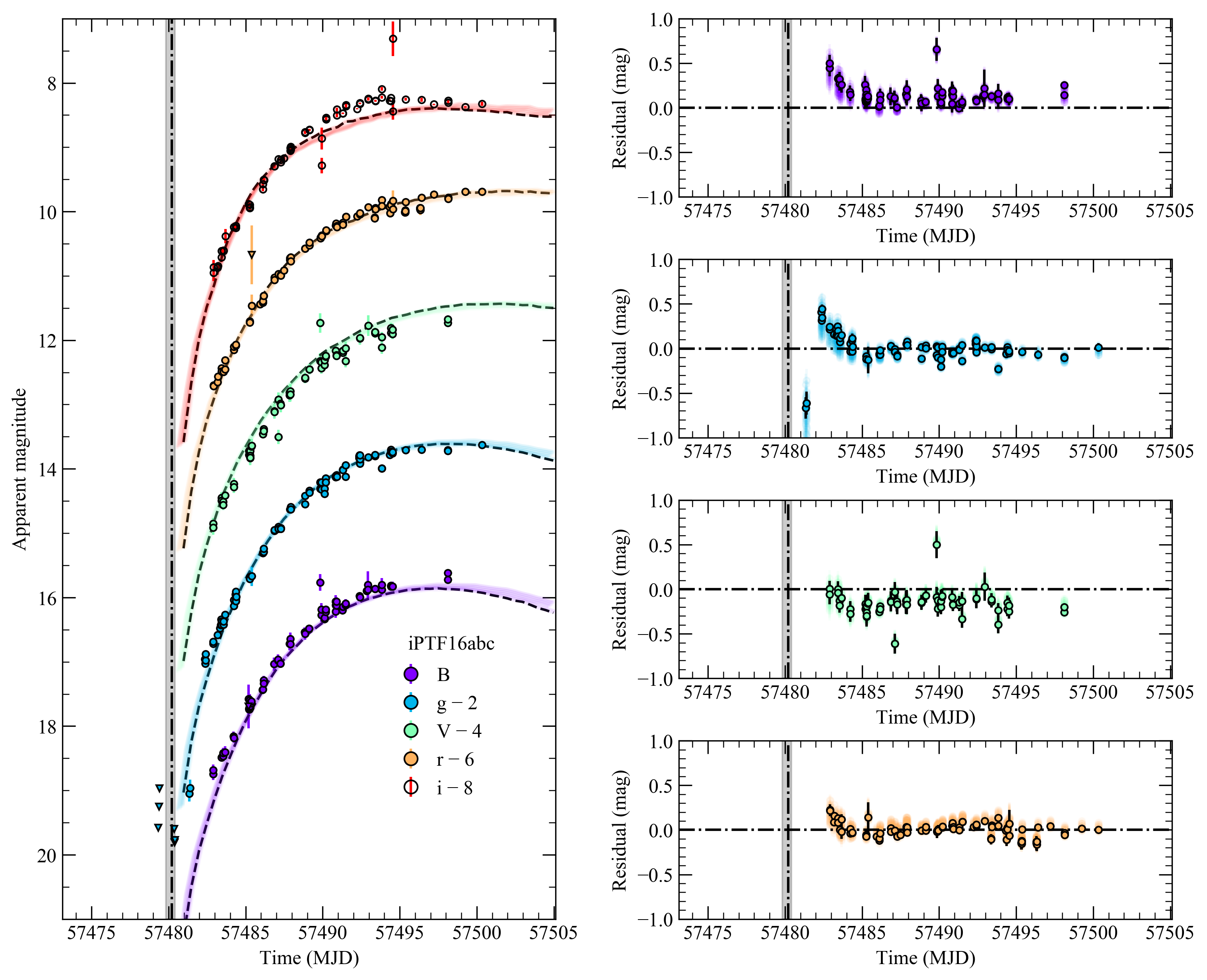}
\caption{Same as in Fig.~\ref{fig:SN2012fr_best}, but for iPTF16abc. }
\label{fig:iPTF16abc_best}
\centering
\end{figure*}

\begin{figure*}
\centering
\includegraphics[width=\textwidth]{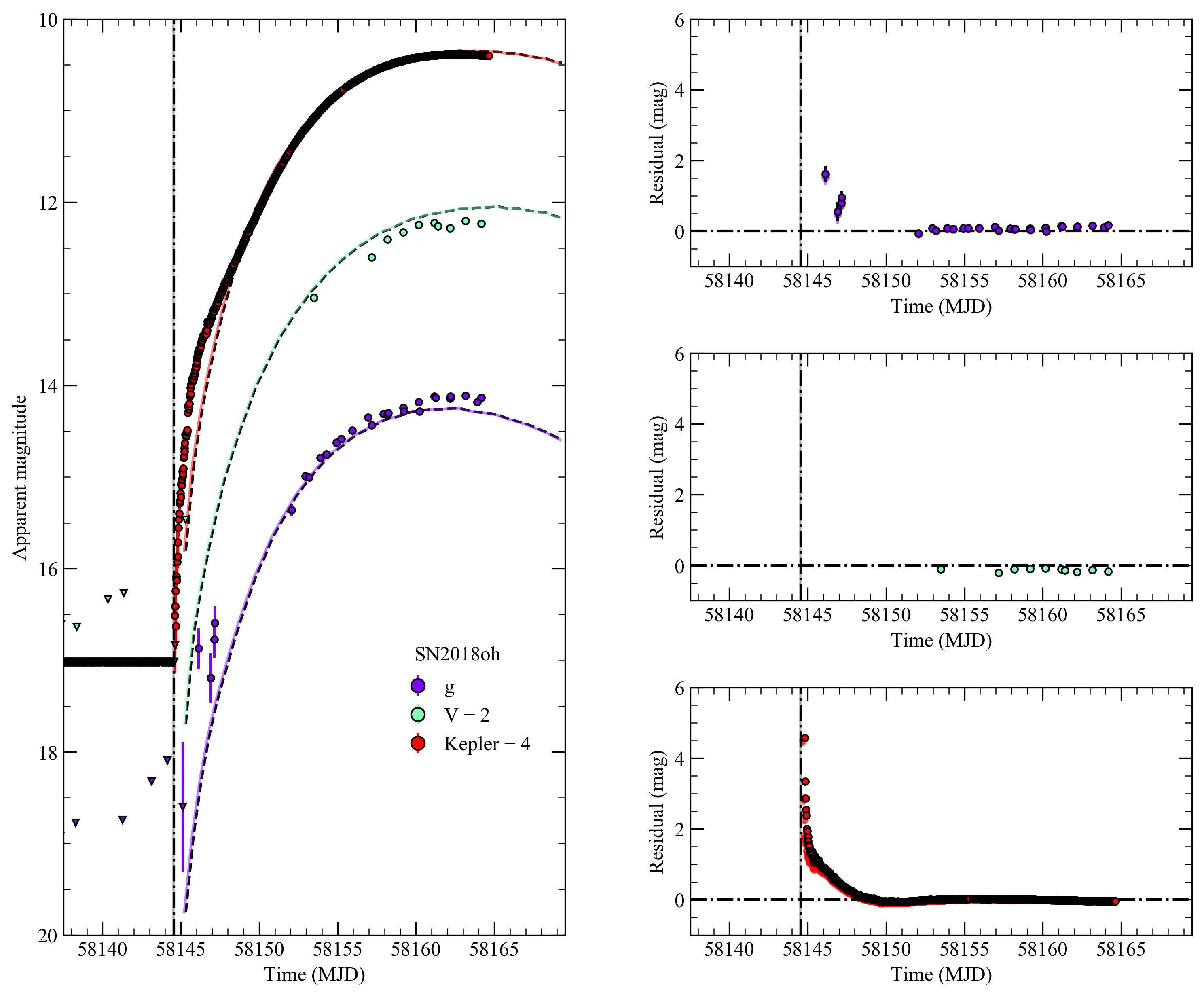}
\caption{Same as in Fig.~\ref{fig:SN2012fr_best}, but for SN~2018oh. }
\label{fig:ASASSN-18bt_best}
\centering
\end{figure*}

\subsubsection{Models that require additional tuning}

For some objects within our sample, we find generally good agreement with the models, however some larger discrepancies are apparent at certain times. Comparisons can therefore be used to inform what changes are necessary to the models in order to produce improved agreement, and provide indications of more complicated $^{56}$Ni distributions than the simple functional form assumed here. An example of this is shown in Fig.~\ref{fig:iPTF16abc_best} for the case of iPTF16abc. Additional objects are shown in Appendix.~\ref{sect:apdx:good_but}.

\par

Figure~\ref{fig:iPTF16abc_best} shows that we are generally able to reproduce the light curve of iPTF16abc beginning a few days after explosion, however the earliest points show a much sharper rise than our models. A similar issue was observed by \cite{miller--18} using the models of \cite{piro-16}. This would indicate that although iPTF16abc is well matched by a model containing an extended $^{56}$Ni distribution, a sharper drop-off of the $^{56}$Ni fraction in the very outermost ejecta is necessary to reproduce the sharp rise observed.

\par

Similar to iPTF16abc, we find that SN~2012cg can be reproduced using an extended $^{56}$Ni distribution, but models likely require a sharper drop-off of $^{56}$Ni in the outer ejecta to be fully consistent with the observations. \cite{marion--16}, and later \cite{jiang--18}, suggested that SN~2012cg shows signs of an excess of flux in the bluer bands compared to a power-law rise -- possibly indicating signs of interaction. Figure~\ref{fig:SN2012cg_best} shows that the optical light curve shape of SN~2012cg is fully consistent with being produced solely by $^{56}$Ni. As stated previously we do not included UV bands, however given that our models are able to reproduce the shape of the $B$- and $V$-band light curves without invoking interaction, it is possible that the excess relative to a power-law rise observed in UV bands could also be reproduced. Furthermore, \cite{shappee--18} have also argued against interaction with a non-degenerate companion after re-analysis of the light curve of SN~2012cg (including UV bands) and the inclusion of additional constraints from X-ray limits and nebular spectroscopy. We note that our best fitting explosion time would indicate that the detection reported by \cite{lipunov--12} occurs only hours after explosion. We caution over-interpretation of this however, as this detection is based on an unfiltered image and reported without uncertainties -- as previously mentioned we compare unfiltered observations to the $R$-band. 

\par

\cite{im--15} argued for the possible detection of shock-cooling from interaction with a companion in the light curve of SN~2015F, based on 2$\sigma$ detections at 2 and 3 days before the first unambiguous detection around MJD $\sim$57\,089. \cite{cartier--15} included additional 3$\sigma$ non-detections shortly before explosion and argued against signs of shock-cooling. In Fig.~\ref{fig:SN2015F_best}, we show that SN~2015F is consistent with our models containing an intermediate $^{56}$Ni distribution -- a slightly increased $^{56}$Ni fraction in the outer ejecta could further improve agreement. Figure~\ref{fig:SN2015F_best} also shows that SN~2015F is consistent with post-explosion non-detections -- again demonstrating the importance of the dark phase in determining the explosion date.

\par

The good matches between $\sim$74\% of SNe Ia in our gold sample and our models show that simply altering the distribution of $^{56}$Ni within the ejecta can reproduce matches to the light curves of numerous objects. We find that none of the objects within our sample produce satisfactory matches with our most compact $^{56}$Ni distributions ($s = 100$), while satisfactory matches for our next most compact $^{56}$Ni distribution ($s = 21$) are produced for only LSQ12fxd. As shown in Table~\ref{tab:object_fits}, our intermediate $^{56}$Ni distributions ($s = 9.7$) account for 10 out of 17 ($\sim$59\%) of the best matches for those gold sample SNe~Ia consistent with our models. Our analysis would therefore indicate that most objects in our sample do not show such a sharp drop-off in the $^{56}$Ni mass fraction as in the most compact $^{56}$Ni distribution models. Instead, models showing a more gradual decrease in $^{56}$Ni fraction or containing $^{56}$Ni throughout the ejecta are preferred -- there must exist some transition region from $^{56}$Ni-rich to -poor ejecta. Some objects however, could potentially be reproduced by extended $^{56}$Ni distributions that show a sharp decrease in the $^{56}$Ni fraction in the outer $^{56}$Ni-poor ejecta only.


\subsection{SNe Ia requiring an early flux excess}
\label{subsect:excess}

\begin{figure*}
\centering
\includegraphics[width=\textwidth]{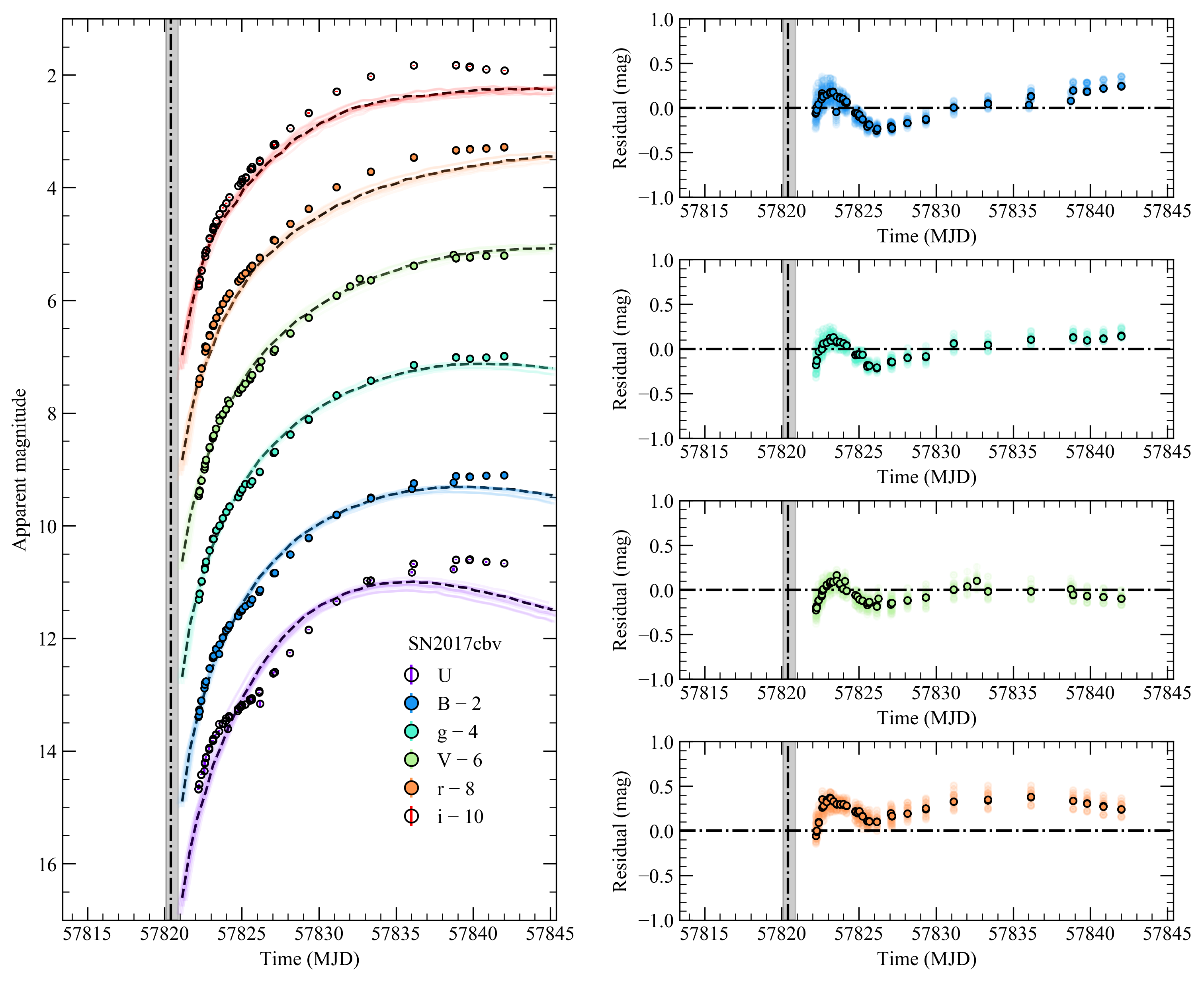}
\caption{Same as in Fig.~\ref{fig:SN2012fr_best}, but for SN~2017cbv. }
\label{fig:SN2017cbv_best}
\centering
\end{figure*}

Within our gold sample, there are 5 out of 23 ($\sim$22\%) objects for which our models indicate that an excess of flux is necessary at early times. The distinction between whether or not an object requires a flux excess is based on two criteria: either a residual of \textgreater1 magnitude in any band or a quasi-sinusoidal residual pattern. The first of these criteria is self-explanatory -- if there is a significantly large difference between the model and SN fluxes, it is likely that significant changes to the model would be necessary or the model lacks some additional physics, such as an interaction. For some SNe~Ia, a bump is clearly observed in the light curve, however the residual is always $\lesssim$$\pm1$~mag. and instead shows a quasi-sinusoidal pattern in which the zero line is crossed twice within approximately three days. None of the models in our set reproduce similar bumps, therefore this likely results from the $\chi^2$ analysis and the fact that these SNe tend to have more points at early times -- in other words, no models can match the bump and the lowest $\chi^2$ values are obtained for models that go through the bump, hence the sinusoidal pattern. Comparisons with SNe falling into this category are shown in Appendix~\ref{sect:apdx:excess}.

\par

A comparison between SN~2018oh and our models is given in Fig.~\ref{fig:ASASSN-18bt_best}. It is clear that the distinctive early bump of SN~2018oh observed in the Kepler band is not matched -- SN~2018oh is an example of a SN for which there is a large flux excess (\textgreater1 mag.) at early times, relative to our models. Beginning approximately four days after explosion however, we find that the light curve is reasonably well matched by our models containing $^{56}$Ni throughout the ejecta (i.e. one of our more extended $^{56}$Ni distributions). This would indicate that while an extended $^{56}$Ni distribution can provide good agreement with the light curve closer to maximum light, an additional source of luminosity not accounted for in our models is required to produce the early bump. Indeed, the early bump observed in SN~2018oh has been extensively studied and favoured interpretations invoke additional sources of flux beyond the smooth, monotonically decreasing $^{56}$Ni distributions presented here. Scenarios investigated include interaction with a companion star, interaction with circum-stellar material, a double detonation explosion, and an off-centre $^{56}$Ni distribution in which an additional clump of $^{56}$Ni is placed near the surface of the ejecta, however the exact nature of the bump is currently unclear \citep{shappee--2019, dimitriadis--19}. SN~2017cbv similarly shows a bump at early times that is not reproduced in our models, which \cite{hosseinzadeh--17} argue results from interaction. Figure~\ref{fig:SN2017cbv_best} demonstrates that the bump in the light curve is most prominent in the bluer bands, however the residuals highlight that the sinusoidal pattern, and hence early bump, can be observed across all bands. The models presented here could therefore be used to identify weak bumps in future objects and provide a more physical point of comparison than a simple $t^n$ rise in luminosity. 

\par

Although we find that $\sim$22\% of objects in our sample require an excess of flux at early times, our models do not provide arguments in favour of any one origin for the flux excess. All models presented here show a smooth rise in the early phase, and therefore such bumps cannot be produced by simply having a large $^{56}$Ni fraction in the outer ejecta that monotonically increases towards the inner regions. If such features are the result of $^{56}$Ni decay, then the $^{56}$Ni distribution must be clumpy or irregular in some way, however exactly the shape of the $^{56}$Ni distribution required is not clear and warrants further investigation.


\subsection{SNe Ia with ambiguous results}
\label{subsect:ambig}
    
Here we discuss the two objects (ASASSN-14lp and KSN2011b) for which our model fits are potentially inconclusive. In both cases, we find that our $\chi^2$ analysis produces matches for the full light curve and for a flux excess.

\par

In Fig.~\ref{fig:ASASSN-14lp_and_KSN2011b} we show our best matches for ASASSN-14lp and KSN2011b. For ASASSN-14lp, we find that our lowest $\chi^2$ match under-predicts the flux at early times in the $V$-band by $\sim$1.6 mag, which could indicate an excess of flux is necessary. The two earliest unfiltered observations at approximately 2 days after explosion also show a flatter rise than indicated by our best match. We also find however, that if we assume explosion occurred $\sim$1 day earlier, a model with a lower kinetic energy (1.81 or 1.68 compared to 2.18) can reproduce the early $V$-band observations. While this also reproduces the flux in most other bands, the unfiltered observations are now fainter than the model predictions at 3 days after explosion. \cite{shappee--16} present extensive observations of ASASSN-14lp and see no evidence for a flux excess at early times, but suggest the possibility of a broken power law rise. For KSN~2011b, we find only two matches within the 3$\sigma$ range -- one with a maximum residual \textless1 mag. and one with a maximum residual \textgreater1 mag. As shown in Fig.~\ref{fig:KSN2011b_best} neither model reproduces the earliest points. Based on analysis of the K2 light curve, \cite{olling--15} have argued against signs of interaction in KSN~2011b.

\par

For both ASASSN-14lp and KSN~2011b, we argue that they likely could be explained solely through an irregular distribution of $^{56}$Ni. This may include significantly increasing the $^{56}$Ni mass fraction in the outer ejecta, while still maintaining the monotonically decreasing structure. Based on the current model set and criteria described in this section however, we cannot rule out the possibility of an early excess. Additional work exploring different $^{56}$Ni distributions will help to better constrain the $^{56}$Ni distributions in these SNe.



\section{Discussion}
\label{sect:discussion}

\begin{figure*}
\centering
\includegraphics[width=\textwidth]{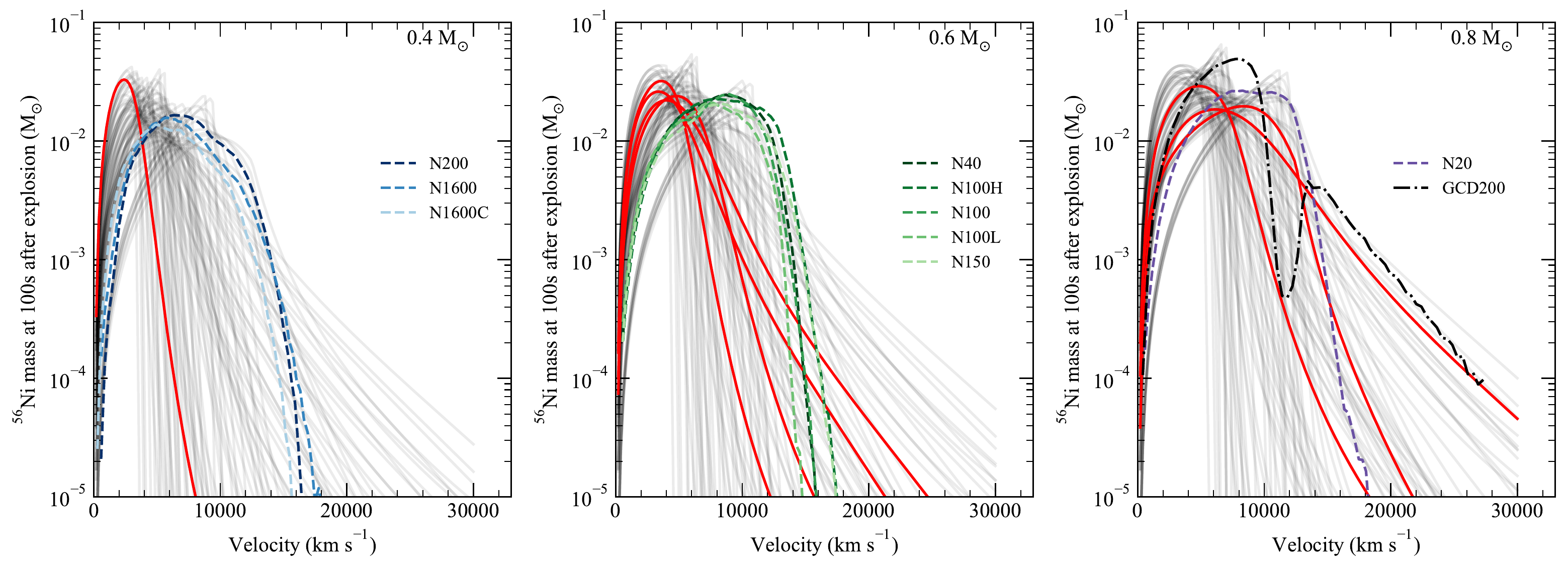}
\caption{$^{56}$Ni distributions for the best matching models to SNe in our gold sample that do not require additional tuning (red). Distributions are given in mass- and velocity-space and separated in different $^{56}$Ni masses. Grey lines show the $^{56}$Ni distributions for our full model set. $^{56}$Ni distributions are also shown for the DDT explosion models of \cite{seitenzahl--13}, with similar $^{56}$Ni masses, as dashed lines. The GCD model of \cite{seitenzahl--16} is shown as a dash-dot line.}
\label{fig:exp_comp}
\centering
\end{figure*}
  
\subsection{Chandrasekhar-mass explosions and extended $^{56}$Ni distributions}

Based on comparisons with observations, our models show that the early light curves of many objects ($\sim$74\% of our gold sample) can be explained by differences in the $^{56}$Ni distributions and that a wide range of $^{56}$Ni distributions is necessary to match the diversity among SNe~Ia. In Fig.~\ref{fig:exp_comp}, we show the best matching models to SNe~Ia within our gold sample that require no additional tuning. Figure~\ref{fig:exp_comp} demonstrates that, in general, a relatively gradual decrease in $^{56}$Ni mass is required in the outer ejecta. Models with extreme drop-offs ($s = 100$) in the $^{56}$Ni mass fraction close to the inner ejecta rise much more sharply than any yet observed SNe~Ia and do not show good agreement with observations. If a similarly sharp decrease in the $^{56}$Ni mass fraction occurs only in the very outermost regions of the ejecta, this could potentially explain some objects in our sample that show steeper rises than our models (such as iPTF16abc).

\par

In Fig.~\ref{fig:exp_comp} we also show the angle-averaged $^{56}$Ni distributions predicted by multi-dimensional explosion simulations invoking two different near-Chandrasekhar mass explosion scenarios. We focus on near-Chandrasekhar mass explosions as these are most similar to the models explored in this work. The first of these scenarios is the deflagration-to-detonation transition \cite[DDT; ][]{khokhlov--91a}. Here, following ignition of the core, nuclear burning initially progresses sub-sonically before transitioning to a super-sonic detonation. The exact conditions under which a transition to detonation occurs are unknown, but are often treated as a free parameter. It is generally believed that a strongly turbulent deflagration front is needed to drive mixing of burned and unburned material. A detonation is typically then enforced if this mixing occurs over sufficient length- and time-scales \citep{woosley--07,roepke--07b,woosley--09,seitenzahl--13}. 

\par

\begin{figure}
\centering
\includegraphics[width=\columnwidth]{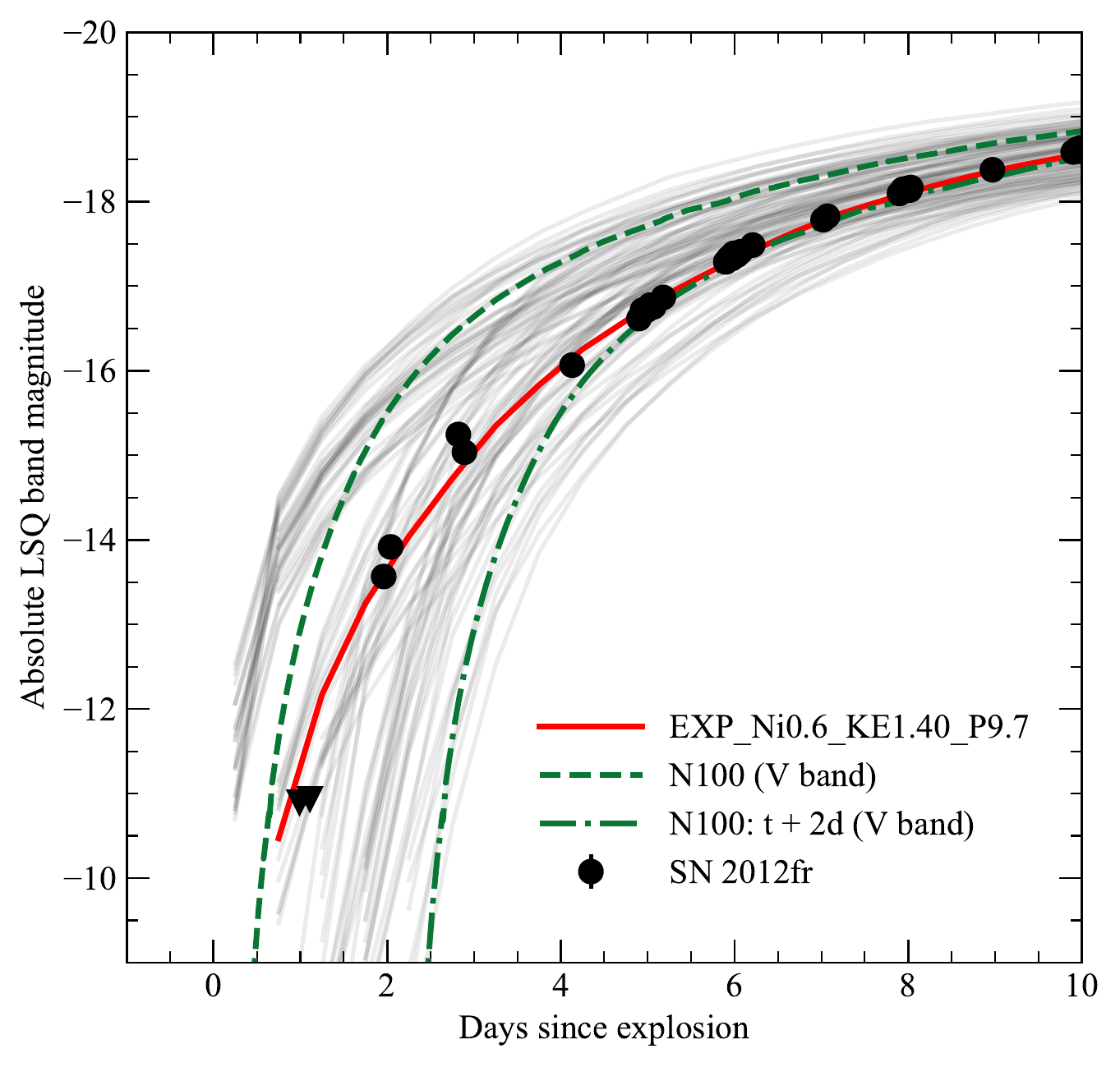}
\caption{Light curve comparison of SN~2012fr (black), our best matching model (red), and the N100 DDT model \cite[green;][]{noebauer-17}. Light curves for our model and SN~2012fr are shown in the LSQ band, while the DDT model is shown in the $V$-band, which our modelling shows is extremely similar to LSQ. The N100 model shows a much sharper rise than is exhibited by SN~2012fr and our best matching model. Offsetting the N100 model by 2\,d, we find that is does produce favourable agreement beginning approximately 4\,d after explosion. }
\label{fig:12fr_exp_comp_lc}
\centering
\end{figure}

The angle-averaged DDT models of \cite{seitenzahl--13} are shown in Fig.~\ref{fig:exp_comp} as dashed lines. We find in general that the DDT models produce a more abrupt drop-off in the $^{56}$Ni mass than is favoured by our models. Such a sudden change in $^{56}$Ni abundance is a natural consequence of the lower density in the outer regions, where full burning in nuclear statistical equilibrium (NSE) cannot be achieved, however this typically results in a sharper rise to maximum. In Fig.~\ref{fig:12fr_exp_comp_lc}, we show the light curve of SN~2012fr compared to our best matching model and the N100 DDT model \citep{noebauer-17}. It is clear that the sudden decrease in $^{56}$Ni mass exhibited by the N100 model produces a rise that is too sharp to match SN~2012fr. Conversely, in our model with an extended $^{56}$Ni distribution, the more gradual decrease in $^{56}$Ni mass towards higher velocities produces a broader light curve that is closer to what is observed. Producing favourable agreement with the N100 model would require shifting the explosion date of SN~2012fr by two days (Fig.~\ref{fig:12fr_exp_comp_lc}; dast-dot line), however in this case agreement is only for observations beginning approximately four days after explosion. At earlier times, SN~2012fr is much brighter than the N100 models. Our results would therefore indicate that some amount of $^{56}$Ni in the outer ejecta is indeed a requirement. 

\par

One possibility to produce the $^{56}$Ni distributions favoured by our modelling within the DDT scenario is to delay somewhat the transition from turbulent deflagration to detonation, relative to the models of \cite{seitenzahl--13}. In such a case, the deflagration ash (including $^{56}$Ni) would continue to expand and mix into the outer ejecta after NSE burning has ceased. This could potentially produce a smoother transition region between the $^{56}$Ni-rich inner ejecta and the (relatively) $^{56}$Ni-poor outer ejecta. The transition from deflagration to detonation is highly uncertain, but as previously mentioned some amount of mixing between hot burned ash and colder fuel is believed to be necessary. \cite{seitenzahl--13} assume mixing occurs over at least a half eddy turnover time. In addition, \cite{seitenzahl--09} explore conditions for detonations and demonstrate the sensitivity to the temperature profile, density, and composition -- none of which are known precisely. The temperature profile in particular is not resolved in the \cite{seitenzahl--13} simulations. Therefore there is considerable scope for variation in the time delay between deflagration and detonation, which could potentially allow for sufficient mixing. It is not clear however, the length of delay that is required to ensure sufficient outward mixing of $^{56}$Ni. In addition, the subsequent detonation would then necessarily occur in lower densities, impacting the overall composition of the ejecta. Given the uncertainties inherent in the triggering of a detonation, the scenario outlined here does not necessarily require meticulous fine-tuning -- future simulations should explore different detonation conditions and the effects on synthetic observables.

\par

The scenario described above (delaying the detonation and allowing the deflagration ash to expand and mix further) is broadly similar to (although not as extreme as) gravitationally confined detonations \cite[GCD; ][]{plewa--04}. The GCD scenario initially progresses in much the same way as some of the weaker deflagration DDT models -- the sub-sonic deflagration has burned material up to the white dwarf surface before the detonation begins. In this case, the deflagration plume remains bound to and accelerates along the surface of the exploding white dwarf, before eventually converging on the point opposite that of the initial breakout. This causes compression of unburned fuel and may provide the spark for subsequent detonation. 

\par

In Fig.~\ref{fig:exp_comp}, we show the GCD200 model of \cite{seitenzahl--16}. The distribution of $^{56}$Ni mass in the outer layers of the GCD200 model is qualitatively similar to our favoured models. Figure~\ref{fig:exp_comp} shows the angle-averaged $^{56}$Ni distribution, however the true ejecta structure is highly asymmetric, which results from the detonation occurring at a single point on the stellar surface. Therefore, although the $^{56}$Ni fraction shown in Fig.~\ref{fig:exp_comp} appears similar in the outer regions to our models, lower $^{56}$Ni fractions are obtained along certain viewing angles \cite[see ][]{seitenzahl--16}. The various distributions of $^{56}$Ni observed could therefore potentially be explained by simple viewing angle dependencies, however polarimetry observations of SNe~Ia indicate a low level of continuum polarisation and approximately symmetric geometry \cite[e.g.][]{wang--wheeler--08}.

\par

We also note that the total $^{56}$Ni mass produced by the GCD scenario may be too high to explain all SNe~Ia. Indeed, the two-dimensional explosion simulations of \cite{meakin--09} consistently produce $\sim$1.1~$M_{\rm{\odot}}$ of $^{56}$Ni. \cite{seitenzahl--16} obtain the lower $^{56}$Ni mass of 0.74~$M_{\rm{\odot}}$ for GCD200 by assuming the initial deflagration is ignited far from the centre of the white dwarf. Larger offsets could potentially produce lower $^{56}$Ni masses, however this may be unrealistic \cite[e.g. ][]{nonaka--12}. Futhermore, \cite{roepke--off-centre} have shown that the conditions believed to be necessary for detonation may not be robustly achieved in three-dimensional simulations of GCD explosions -- these conditions were relaxed by \cite{seitenzahl--16} in order to achieve detonation. In addition, \cite{maguire--18} argue that the GCD model of \cite{seitenzahl--16} can be ruled out for the SNe~Ia in their sample due to the presence of stable Fe and Ni at high-velocities, rather than the low-velocities that are observed in the late-time spectra. The applicability of the GCD scenario to SNe~Ia is therefore unclear, however whether fainter explosions could be achieved and produce $^{56}$Ni distributions similar to those found here, warrants further investigation.

\subsection{Early flux excess}
\label{sect:discuss_excess}

For 5 out of 23 ($\sim$22\%) SNe in our gold sample, we find that our models are not able to reproduce the observations within the first few days of explosion. This would suggest that our models are not exploring the appropriate parameter space for these objects or additional physics is required. 

\par 

One potential source of an early flux excess is the interaction between the expanding supernova ejecta and companion star. During the interaction, the ejecta is shock heated and produces a prompt burst of X-rays. The ejecta continues to expand and cool, resulting in a shift of emission to longer UV and optical wavelengths. \cite{kasen--10} argue that signs of interaction are only visible under favourable viewing angles, regardless of the companion type, and therefore should only be observed in $\sim$10\% of SNe~Ia. The origin of the excess in the five objects from our gold sample is unclear, however multiple scenarios have been proposed and it is likely the origin is not the same for all objects. Therefore, we suggest our result is broadly comparable to the previous estimates of \cite{kasen--10}, however this is based on a small number of objects. Interaction with circum-stellar material can also produce an excess of flux, as shown by \cite{piro-16}, depending on the mass and radius of the circum-stellar material. For those models with more extended $^{56}$Ni distributions, the signature of circum-stellar interaction is masked by the broader light curve and therefore is much weaker. Hence, it is still possible that the SNe included in our sample that do match our model light curves also experienced some degree of interaction, but the signatures are too weak to observe.   

\par

Another potential source for an early flux excess is the detonation of a helium shell on top of a white dwarf, which has been proposed for SN~2016jhr \cite[MUSSES1604D;][]{jiang--2017}. In this so-called double detonation scenario, a helium shell is accumulated on the surface of the white dwarf through accretion. If the shell is able to grow to a certain critical mass, it ignites. \cite{fink--07} have shown that once the helium shell is ignited, a second detonation within the core is all but inevitable due to the convergence of the initial shock. Previous studies have shown that, depending on the size of the helium shell, short-lived isotopes such as $^{52}$Fe and $^{48}$Cr are produced in varying amounts and it is the decay of these isotopes that produces the pronounced early bump \citep[e.g. ][]{noebauer-17}. Compared to observations however, the ash produced in the helium shell leads to colours that are too red \citep{kromer--10}. Strong spectral features of titanium and chromium are also produced in the models, but not observed in spectra of normal SNe~Ia. For SN~2016jhr, \cite{polin--19} show good agreement to the light curve for a double detonation model with a 1.0~$M_{\rm{\odot}}$ white dwarf and 0.05~$M_{\rm{\odot}}$ helium shell. The overall light curve width is too narrow however, and even in this case the early bump produced by the model is less pronounced than observed in the earliest $g$-band points.

\par

Recent work has also shown that variations on this double detonation scenario, with smaller helium shells, could produce agreement with a range of SNe~Ia luminosities, particularly if the helium shell has been polluted with some additional carbon, oxygen, and nitrogen \citep{shen--18, townsley--19}. This begs the question of whether sub-Chandrasekhar mass explosions could be consistent with all or the majority of SNe~Ia, and the fact that some show a bump is simply due to the size and/or composition of the helium shell. The models presented by \cite{shen--18} show good agreement with the light curve of SN~2011fe around maximum light, however it is clear that the light curves are much narrower and rise much more sharply than what is observed (this could be related to the assumption of LTE producing a faster decline after maximum light). The $^{56}$Ni distributions show a flattening towards the outer ejecta, however they are generally at lower mass fractions and do not extend as far into the outer ejecta as the models proposed here. Future simulations should test whether it is possible to achieve increased $^{56}$Ni abundances in the outer ejecta both with and without the production of additional short-lived isotopes.

\subsection{Effects of extended $^{56}$Ni distributions on maximum light spectra}
\label{sect:discuss_max}

Our models indicate that some amount of $^{56}$Ni in the outer ejecta is necessary to reproduce the light curve shapes for a number of objects. The presence of this additional source of opacity at high velocities naturally raises the question of whether the spectra at maximum light are negatively affected. To test this, we bin escaping Monte Carlo packets into spectra using 1000 log-separated frequency bins between 10$^{14}$ -- 10$^{16}$~Hz and sum over their luminosities. We stress that our model spectra have not been modified in anyway and the simplified composition used means that spectra likely do not perfectly reproduce all observed features in SNe~Ia, in particular for intermediate mass elements. They may still be used however, for investigating whether or not notable Ni or Fe features are produced and to what extent these disagree with observations.

\par

\begin{figure}
\centering
\includegraphics[width=\columnwidth]{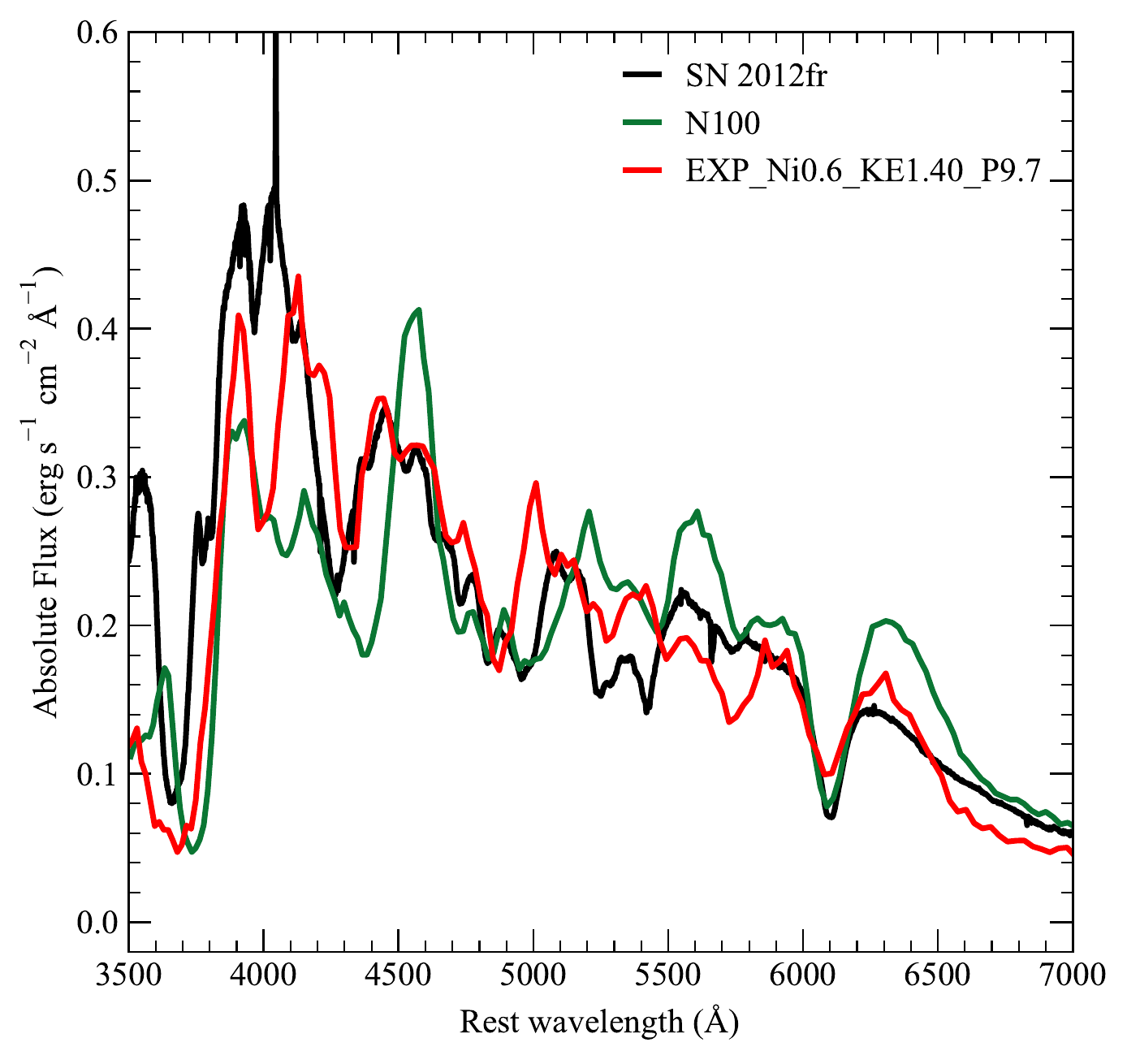}
\caption{Spectral comparison of SN~2012fr (black), our best matching model (red), and the N100 DDT model \citep{seitenzahl--13} at 18 days after explosion. Spectra are shown on an absolute flux scale.}
\label{fig:12fr_exp_comp_spec}
\centering
\end{figure}

In Fig.~\ref{fig:12fr_exp_comp_spec}, we show the spectrum of SN~2012fr at +18\,d relative to explosion, along with spectra of our best matching model (EXP\_Ni0.6\_KE1.40\_P9.7) and the N100 model at a similar phase. Figure~\ref{fig:12fr_exp_comp_spec} shows that our model spectrum produces generally favourable agreement and reproduces many of the optical features (including the \ion{Si}{ii}~$\lambda$6355 feature), although not necessarily with the appropriate strength (again this is likely due to the simplified composition). In addition, our model is able to match many of the features in the blue part of the spectrum between $\sim$3\,500 -- $\sim$5\,000~$\AA$, where the spectrum is highly sensitive to the presence of high velocity IGEs. Figure~\ref{fig:12fr_exp_comp_spec} also shows that the level of agreement between our model and SN~2012fr is comparable, or perhaps better in some ways, to that between N100 and SN~2012fr. This clearly demonstrates that the $^{56}$Ni distributions favoured by our model do not produce spectra that are in disagreement with observations.

\par

For models with less extended $^{56}$Ni distributions than our best matching model, the overall luminosity of the P21 is generally slightly fainter than the P9.7 model at this epoch and many of the features show similar strengths, however in particular those around $\sim$4\,500~$\AA$ show worse agreement than in the P9.7 model. In the case of the more extended $^{56}$Ni distribution model, P4.4, the presence of a larger fraction of high velocity $^{56}$Ni results in stronger line blanketing and slight discrepancies in the relative strengths of features around $\sim$4\,000~$\AA$.

\par

We have also compared our model spectra to SN~2012fr at +4\,d relative to explosion. At these epochs, comparisons between spectral features are even more sensitive to the simplifications of our model due to the smaller region above the photosphere being interrogated. At this epoch, we find that the EXP\_Ni0.6\_KE1.40\_P9.7 does produce features similar to SN~2012fr, although the strength of features due to IMEs (and the \ion{Si}{ii} feature in particular) are noticeably weaker than what is observed. In addition, we find that the velocities in our model are systematically lower than in SN~2012fr. Shifting the model spectrum by a few thousand km~s$^{-1}$ does produce favourable agreement. We speculate that changes to the density structure potentially could reproduce such features, however this is beyond the scope of the parameter space exploration presented here. Overall, the generally favourable agreement between our model and the maximum light spectrum of SN~2012fr further strengthens the claim that at least some SNe~Ia require extended $^{56}$Ni distributions to match the light curve shapes. Future work will explore more complicated composition structures and their effects on the light curve shapes and spectra.

%

\section{Conclusions}
\label{sect:conclusions}

Using the Monte Carlo radiative transfer code of \cite{magee--18}, we have performed an extensive parameter study of the $^{56}$Ni distribution in SNe~Ia. Building upon previous work, we explore additional $^{56}$Ni masses and distributions, and density profile shapes. Consistent with previous studies \citep{dessart-11fe,piro-16, noebauer-17, magee--18}, we show significantly different light curves can be achieved for various $^{56}$Ni distributions. Our models also highlight that similarly diverse light curves can be obtained by altering the density profile and that this must be taken into account when comparing against observations. This is particularly important as the date of explosion is not known exactly and our models show some degeneracy between the density profile, explosion date, and absolute flux level.

\par

Using a series of template light curves, we investigated the observations that are necessary in order to constrain the distribution of $^{56}$Ni for SNe~Ia within the current set of models. We find that if first detection occurs $\gtrsim$3 days after explosion or $\lesssim$14 days before $B$-band maximum, the $^{56}$Ni distribution and explosion date are not well constrained. If observations are limited to only one band, it is possible that many models may be shifted in such a way as to produce favourable agreement -- due to the inherent uncertainties in explosion epochs and distances. This could lead to an incorrect determination of the $^{56}$Ni distribution and caution must be used. Fitting only the first approximately one week after explosion can lead to results that are in contradiction to those at maximum light. In all cases, if the cadence is low ($\gtrsim$3\,d) the $^{56}$Ni distribution and explosion epoch show much larger uncertainties. We therefore suggest the following requirements for observations of SNe~Ia in order to provide robust constraints: observations should begin at least two weeks before $B$-band maximum, extend to at least approximately $B$-band maximum with a relatively high ($\lesssim$3\,d) cadence, and in at least one blue and one red band (such as the $B$- and $R$-bands, or $g$- and $r$-bands). Following these tests, we also highlight the need to properly exploit such data by modelling colour light curves up to maximum light in a consistent manner. 

\par

Based on the criteria listed above, we selected a sample of well-observed SNe~Ia and performed comparisons to our models. Of the 23 objects included in our gold sample (i.e. matching all of the criteria listed above), we find $\sim$74\% are consistent with our models. Additional tuning could be made to certain models to produce improved agreement, however in general we find that the full colour light curves of most objects are well reproduced. Our models would therefore indicate that a variety of $^{56}$Ni distributions could be produced in nature -- even for a given $^{56}$Ni mass. Through our comparisons, we also show the importance of the dark phase. The explosion date for a SN~Ia is often estimated based on non-detections occurring shortly before the first unambiguous detection, however our models show that some SNe are consistent with relatively deep post-explosion limits and accurate determination of the explosion epoch requires comparisons with models. Finally, direct comparison of light curves and spectra from our models and existing explosion models to SN~2012fr shows that while an extended $^{56}$Ni distribution is necessary to produce the light curve shape, this does not negatively impact the spectra observed at maximum light.

\par

By comparing our favoured $^{56}$Ni distributions to existing explosion models, we show that observations of SNe~Ia typically require a more gradual decrease in the $^{56}$Ni mass towards the outer ejecta. Within the Chandrasekhar-mass delayed detonation scenario, we speculate that such $^{56}$Ni distributions could be achieved if there was some delay between the initial deflagration and subsequent detonation (a `delayed' delayed detonation model) -- such a scenario could potentially allow for sufficient outward expansion and mixing of the deflagration ash to produce distributions similar to those favoured by observations. Qualitatively, such expansion of deflagration ash could be comparable to that produced by gravitationally confined detonations, which produce a similar distribution to our models for at least one case. We suggest future explosion simulations testing the limits of the detonation ignition warrant further investigation.

%

\begin{acknowledgements}
We thank F. R\"opke and I. Seitenzahl for useful discussions. This work made use of the High Performance Computing cluster at Queen's University Belfast, and the Heidelberg Supernova Model Archive (HESMA), https://hesma.h-its.org. This work was supported by TCHPC (Research IT, Trinity College Dublin). Calculations were performed on the Kelvin cluster maintained by the Trinity Centre for High Performance Computing. This cluster was funded through grants from the Higher Education Authority, through its PRTLI program. MM, KM, and SP are funded by the EU H2020 ERC grant no. 758638. KM and SS acknowledge support from STFC through grant ST/P000312/1.
\end{acknowledgements}

\bibliographystyle{aa}
\bibliography{Magee}

\begin{appendix}

\onecolumn
\section{SNe~Ia requiring no additional tuning to $^{56}$Ni distribution}
\label{sect:apdx:good}
Here we present light curves and residuals for the best matching models and additional objects included in our sample. Figures are as in Fig.~\ref{fig:SN2012fr_best}. Objects included in this section are those for which no additional tuning is required to produce good agreement with our models.

\begin{figure}[h!]
    \centering
    \begin{subfigure}[b]{0.49\textwidth}
        \includegraphics[width=\textwidth]{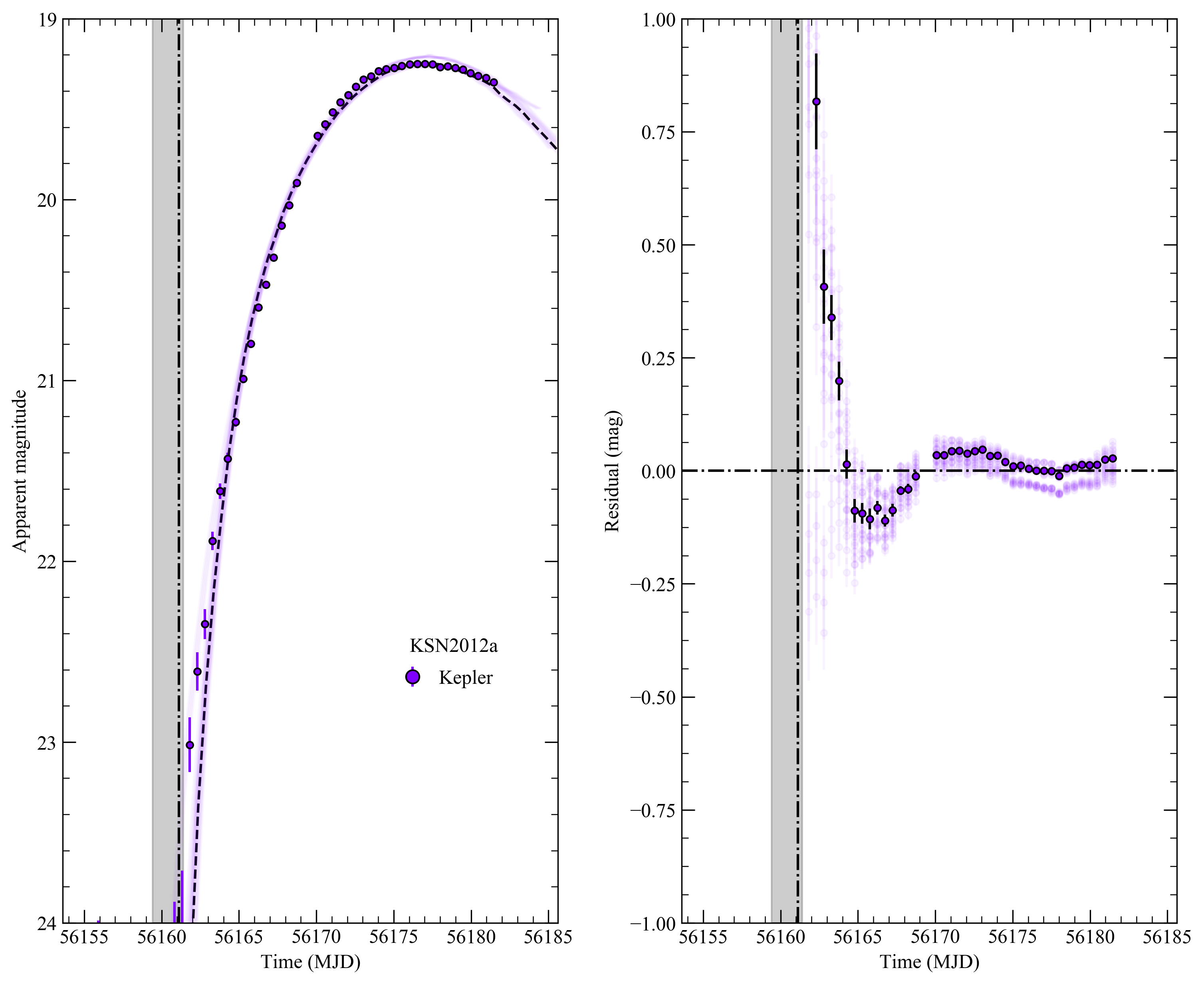}
        \caption{KSN2012a}
        \label{fig:KSN2012a_best}
    \end{subfigure}
    ~ 
    \begin{subfigure}[b]{0.49\textwidth}
        \includegraphics[width=\textwidth]{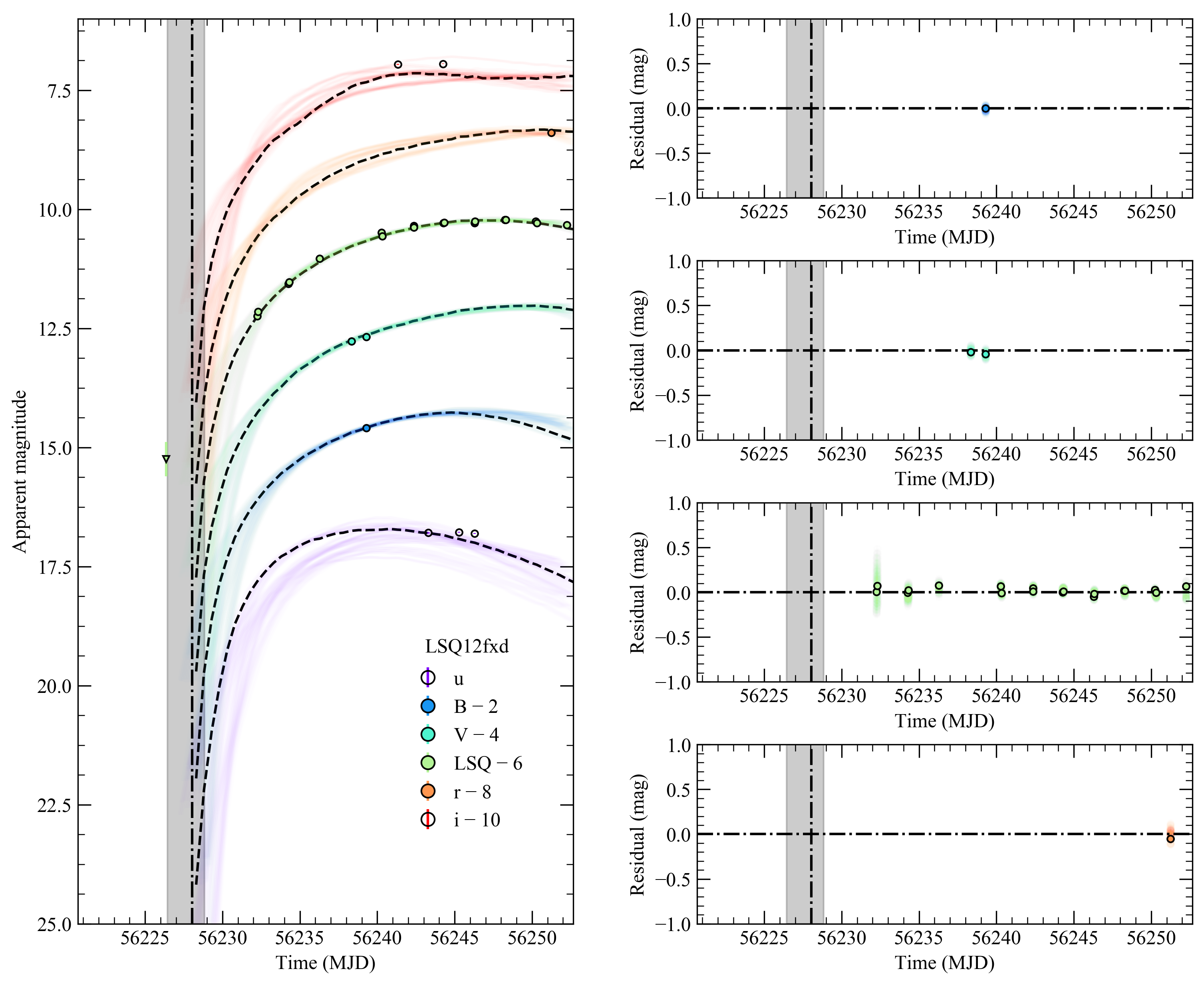}
        \caption{LSQ12fxd}
        \label{fig:LSQ12fxd_best}
    \end{subfigure}
    ~ 
    \caption{Same as in Fig.~\ref{fig:SN2012fr_best}.}
    \label{fig:KSN2012a_LSQ12fxd}
\end{figure}

\begin{figure}[h!]
    \centering
    \begin{subfigure}[b]{0.49\textwidth}
        \includegraphics[width=\textwidth]{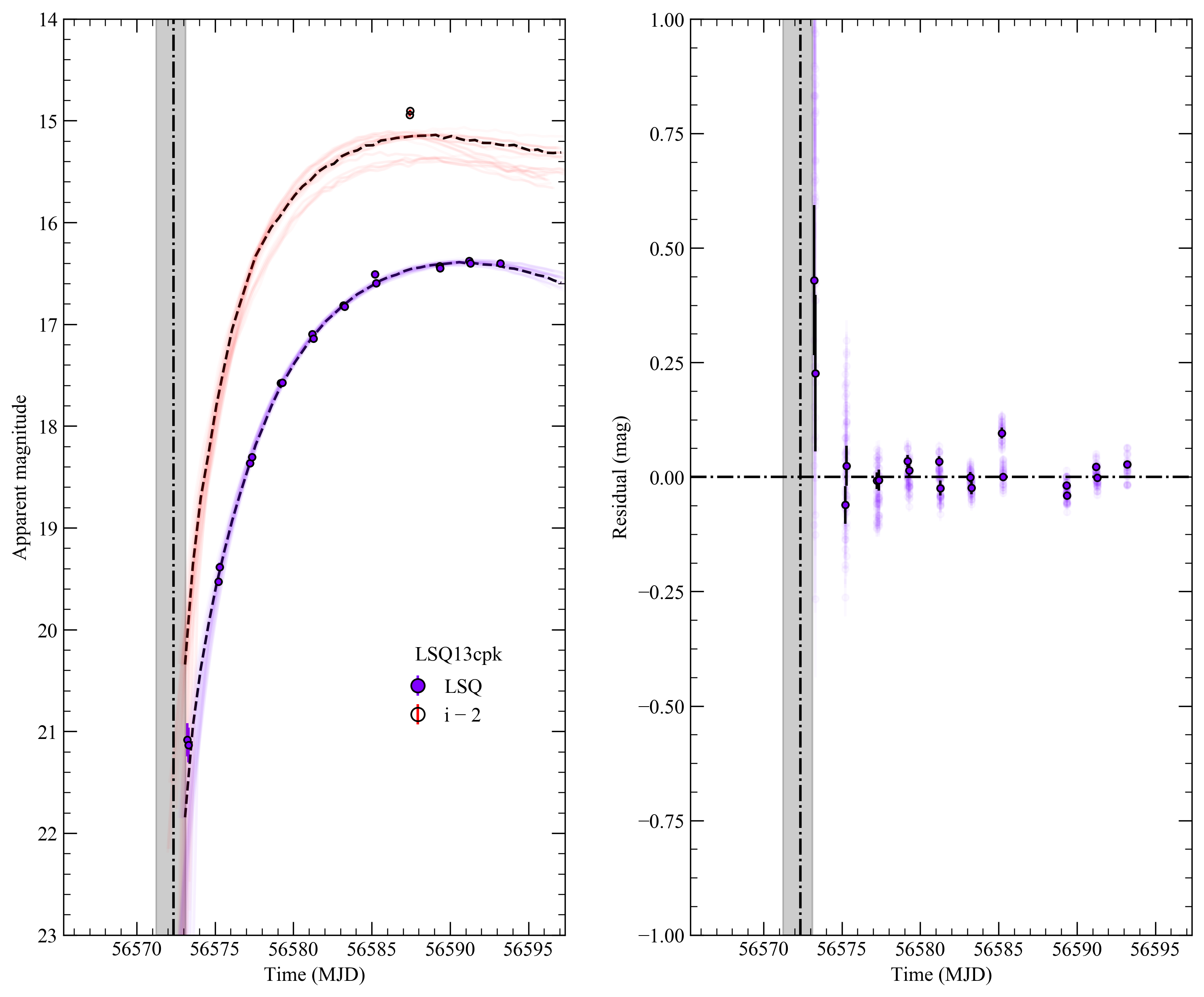}
        \caption{LSQ13cpk}
        \label{fig:LSQ13cpk_best}
    \end{subfigure}
    ~ 
    \begin{subfigure}[b]{0.49\textwidth}
        \includegraphics[width=\textwidth]{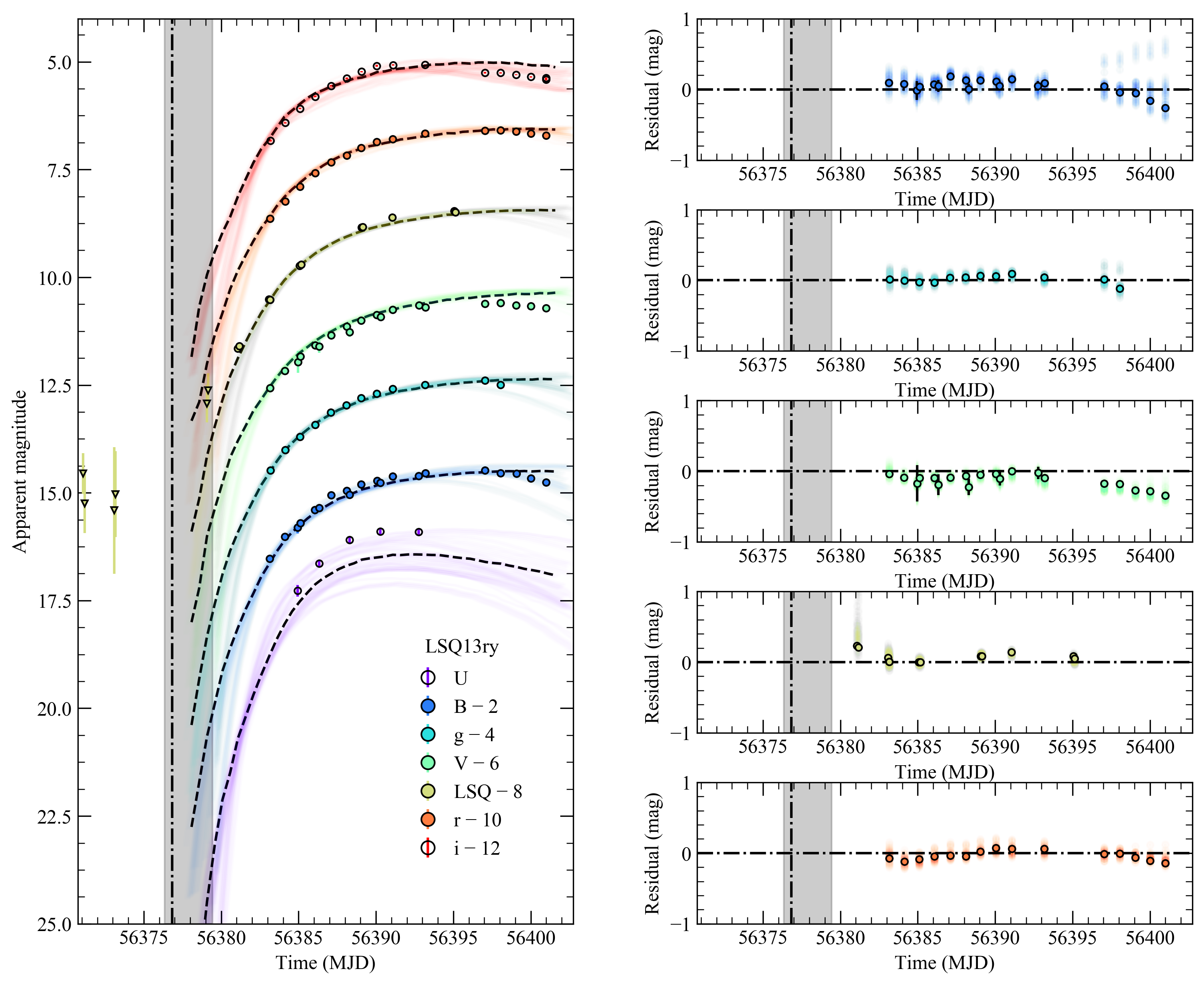}
        \caption{LSQ13ry}
        \label{fig:LSQ13ry_best}
    \end{subfigure}
    ~ 
    \caption{Same as in Fig.~\ref{fig:SN2012fr_best}.}
    \label{fig:LSQ13cpk_and_LSQ13ry}
\end{figure}

\begin{figure}[h!]
    \centering
    \begin{subfigure}[b]{0.49\textwidth}
        \includegraphics[width=\textwidth]{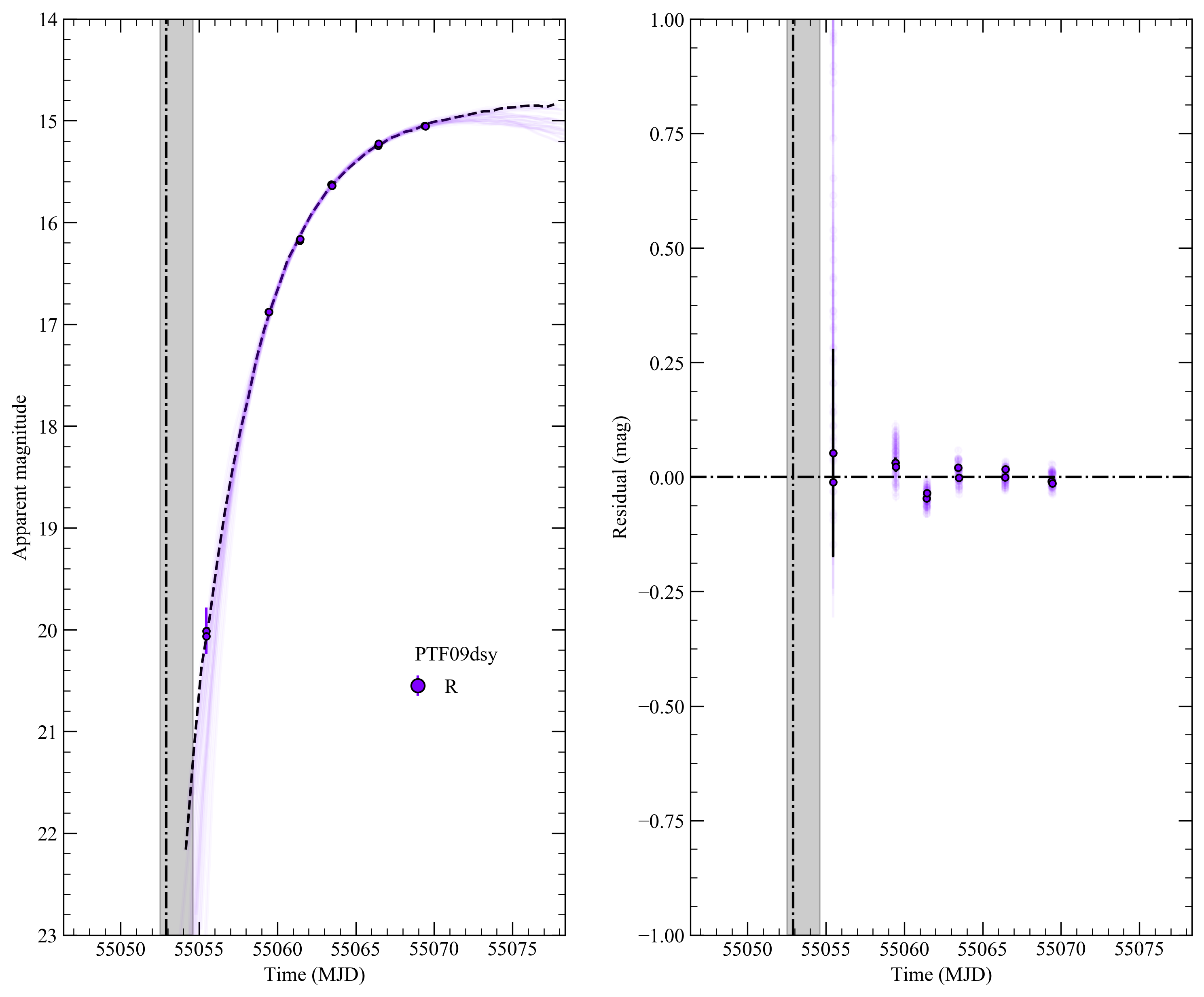}
        \caption{PTF09dsy}
        \label{fig:PTF09dsy_best}
    \end{subfigure}
    ~ 
    \begin{subfigure}[b]{0.49\textwidth}
        \includegraphics[width=\textwidth]{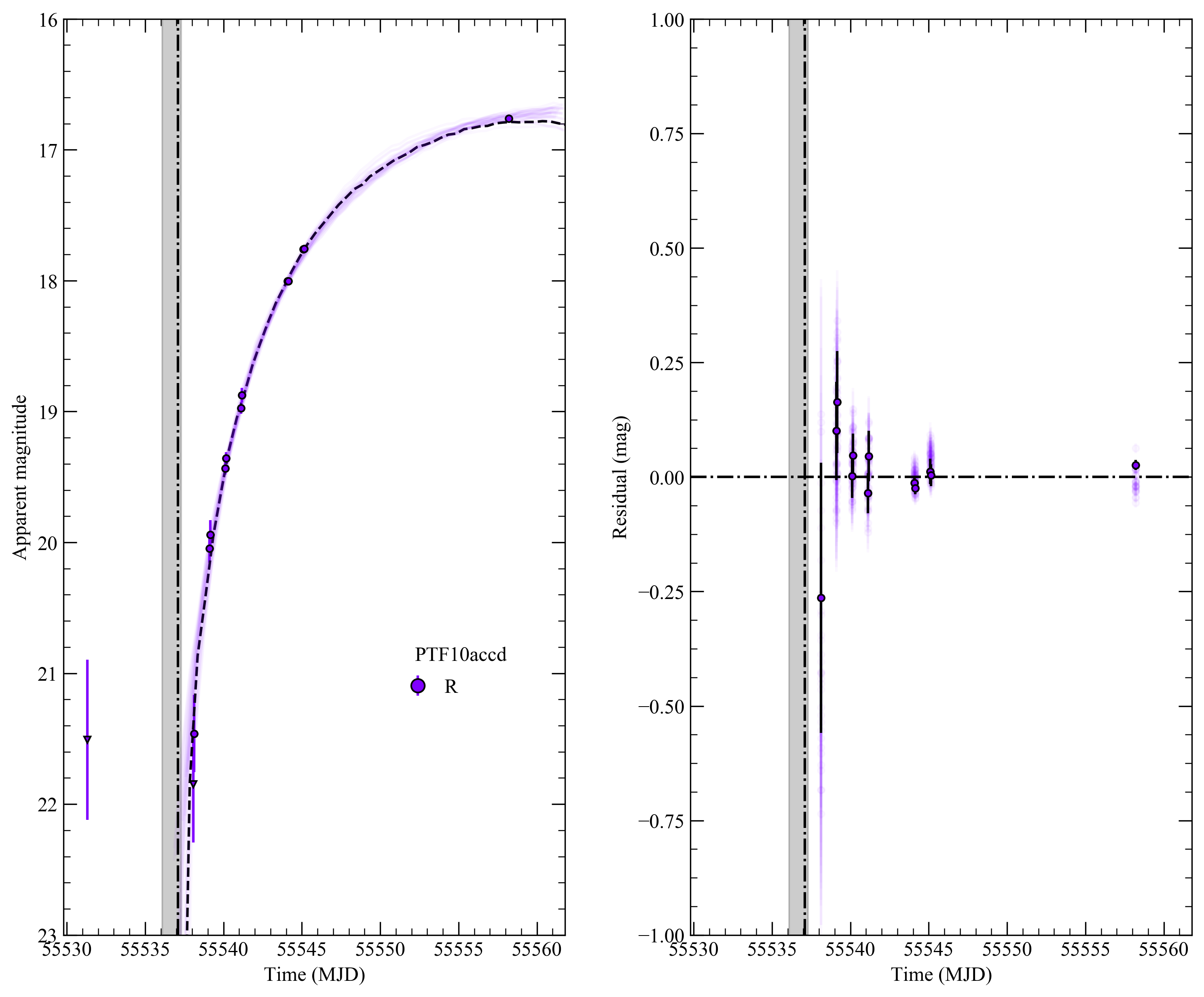}
        \caption{PTF10accd}
        \label{fig:PTF10accd_best}
    \end{subfigure}
    ~ 
    \caption{Same as in Fig.~\ref{fig:SN2012fr_best}.}
    \label{fig:PTF09dsy_and_PTF10accd}
\end{figure}

\begin{figure}[h!]
    \centering
    \begin{subfigure}[b]{0.49\textwidth}
        \includegraphics[width=\textwidth]{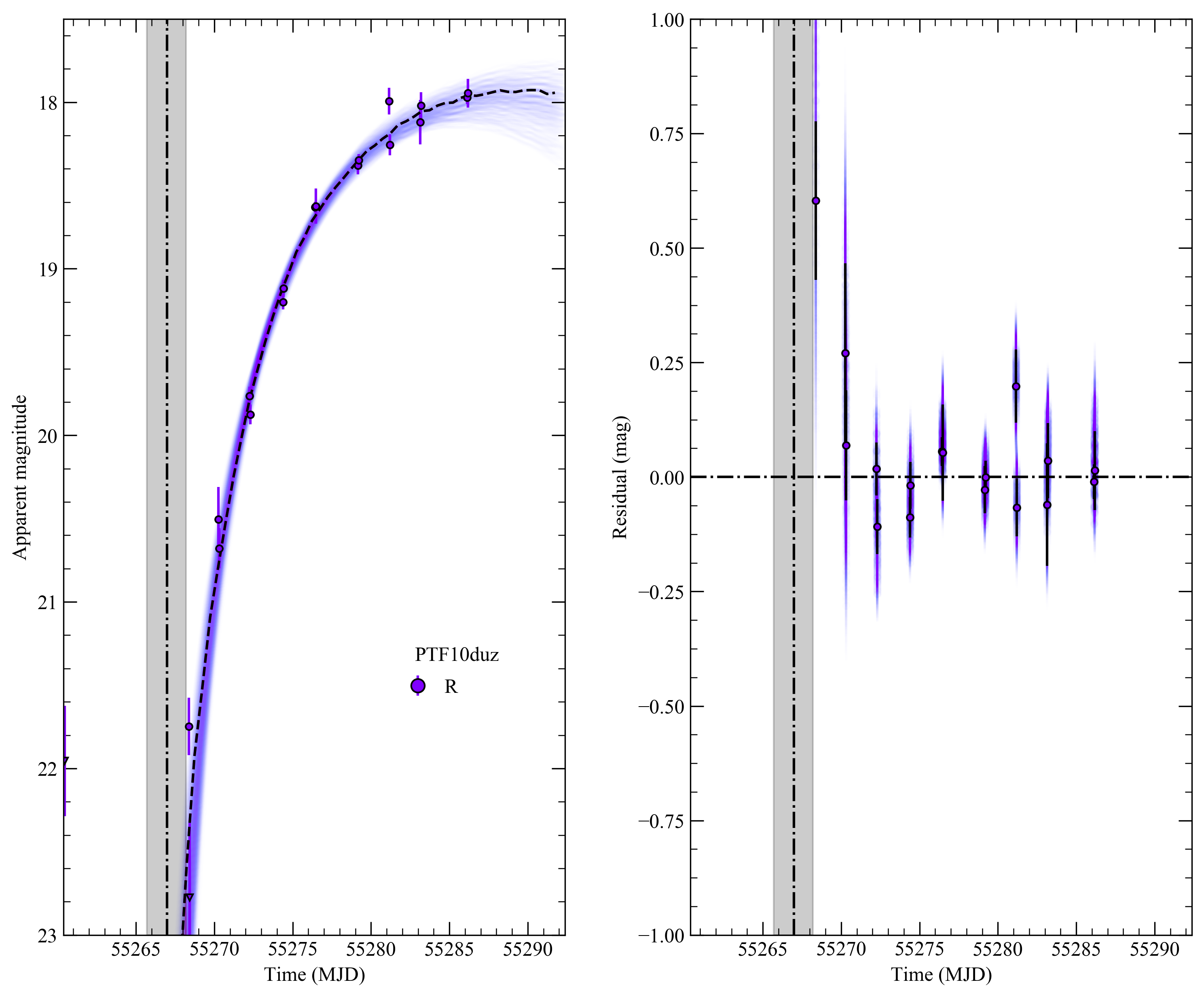}
        \caption{PTF10duz}
        \label{fig:PTF10duz_best}
    \end{subfigure}
    ~ 
    \begin{subfigure}[b]{0.49\textwidth}
        \includegraphics[width=\textwidth]{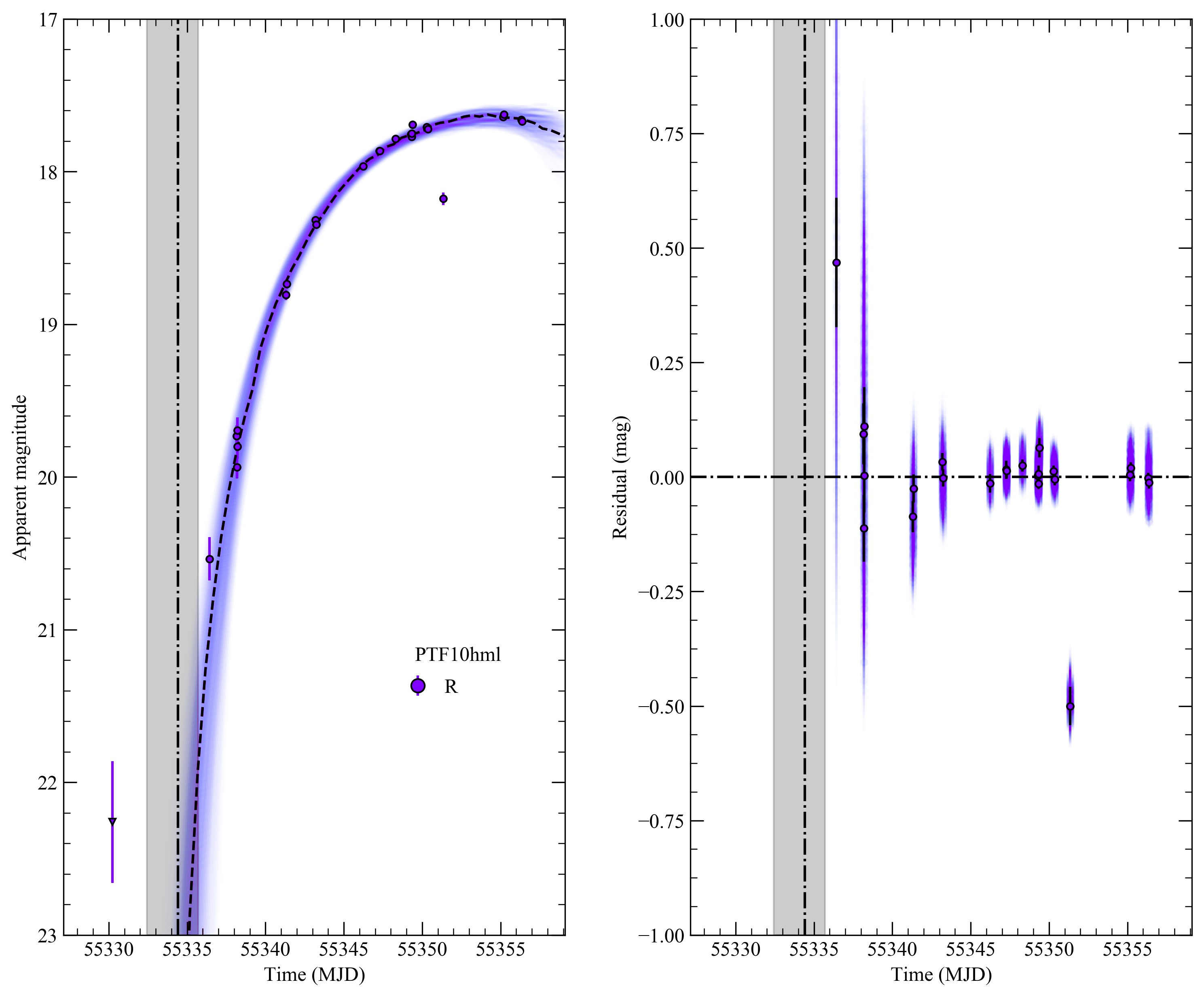}
        \caption{PTF10hml}
        \label{fig:PTF10hml_best}
    \end{subfigure}
    ~ 
    \caption{Same as in Fig.~\ref{fig:SN2012fr_best}.}
    \label{fig:PTF10duz_and_PTF10hml}
\end{figure}

\begin{figure}[h!]
    \centering
    \begin{subfigure}[b]{0.49\textwidth}
        \includegraphics[width=\textwidth]{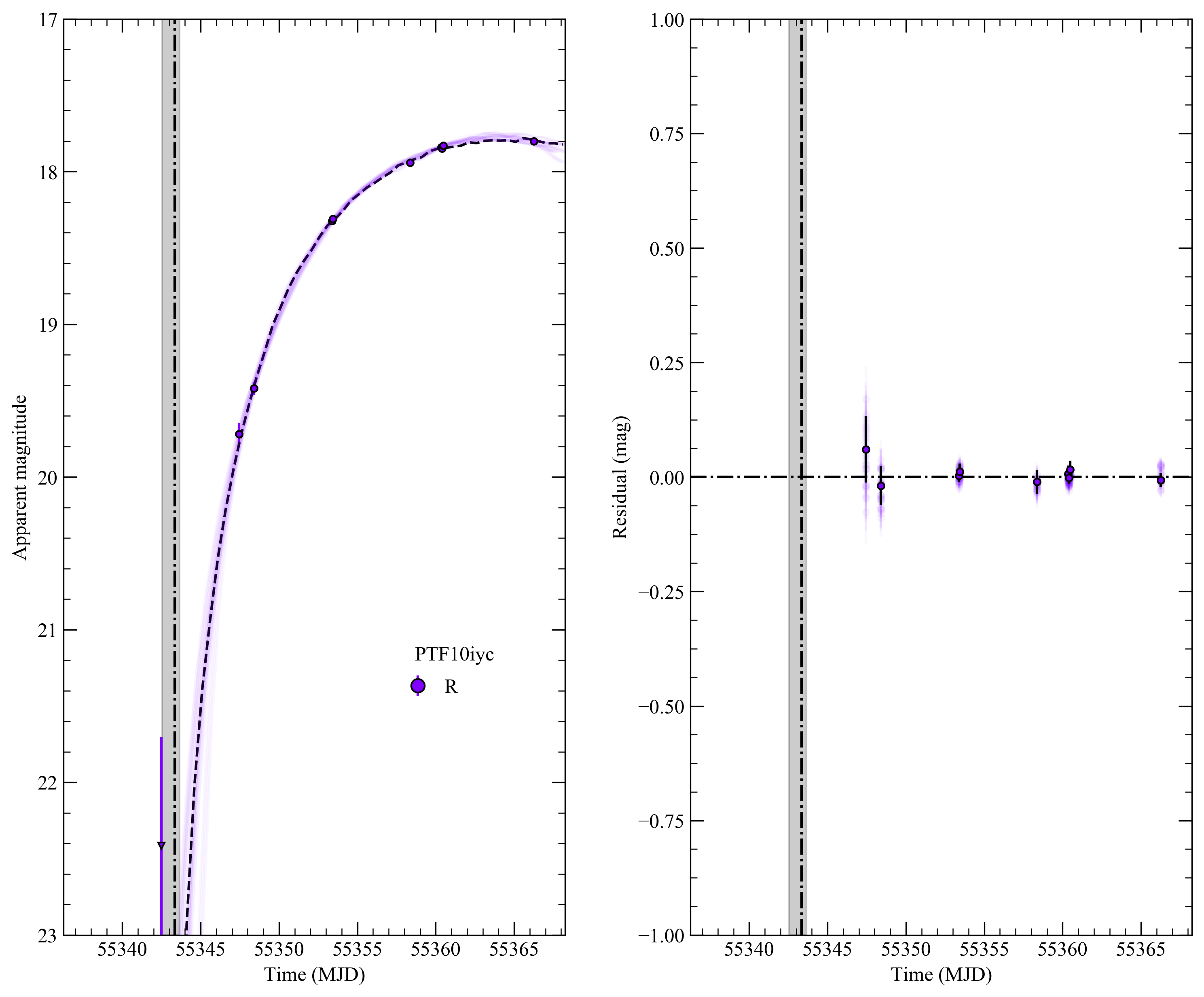}
        \caption{PTF10iyc}
        \label{fig:PTF10iyc_best}
    \end{subfigure}
    ~ 
    \begin{subfigure}[b]{0.49\textwidth}
        \includegraphics[width=\textwidth]{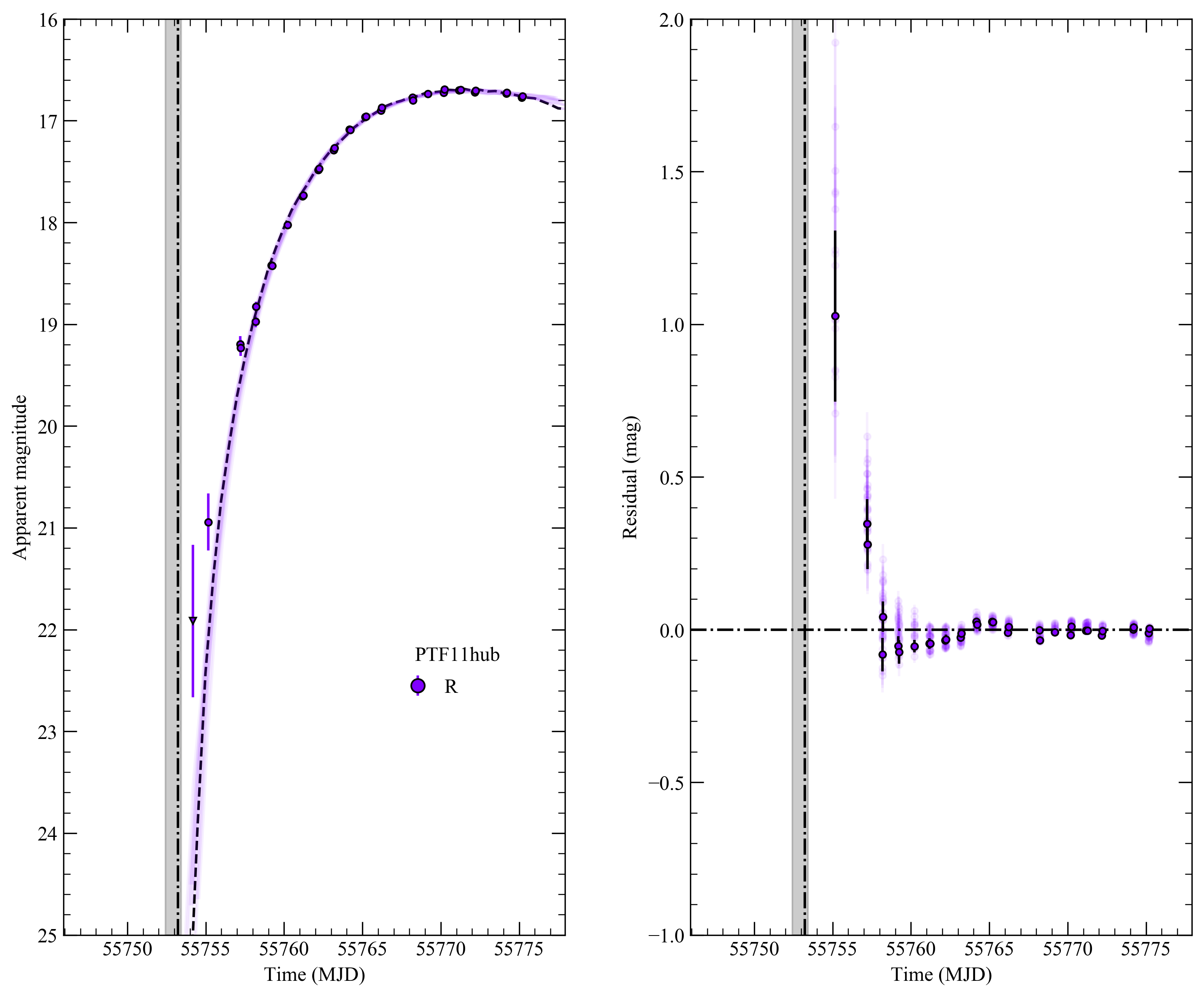}
        \caption{PTF11hub}
        \label{fig:PTF11hub_best}
    \end{subfigure}
    ~ 
    \caption{Same as in Fig.~\ref{fig:SN2012fr_best}.}
    \label{fig:PTF10iyc_and_PTF11hub}
\end{figure}

\begin{figure}[h!]
    \centering
    \begin{subfigure}[b]{0.49\textwidth}
        \includegraphics[width=\textwidth]{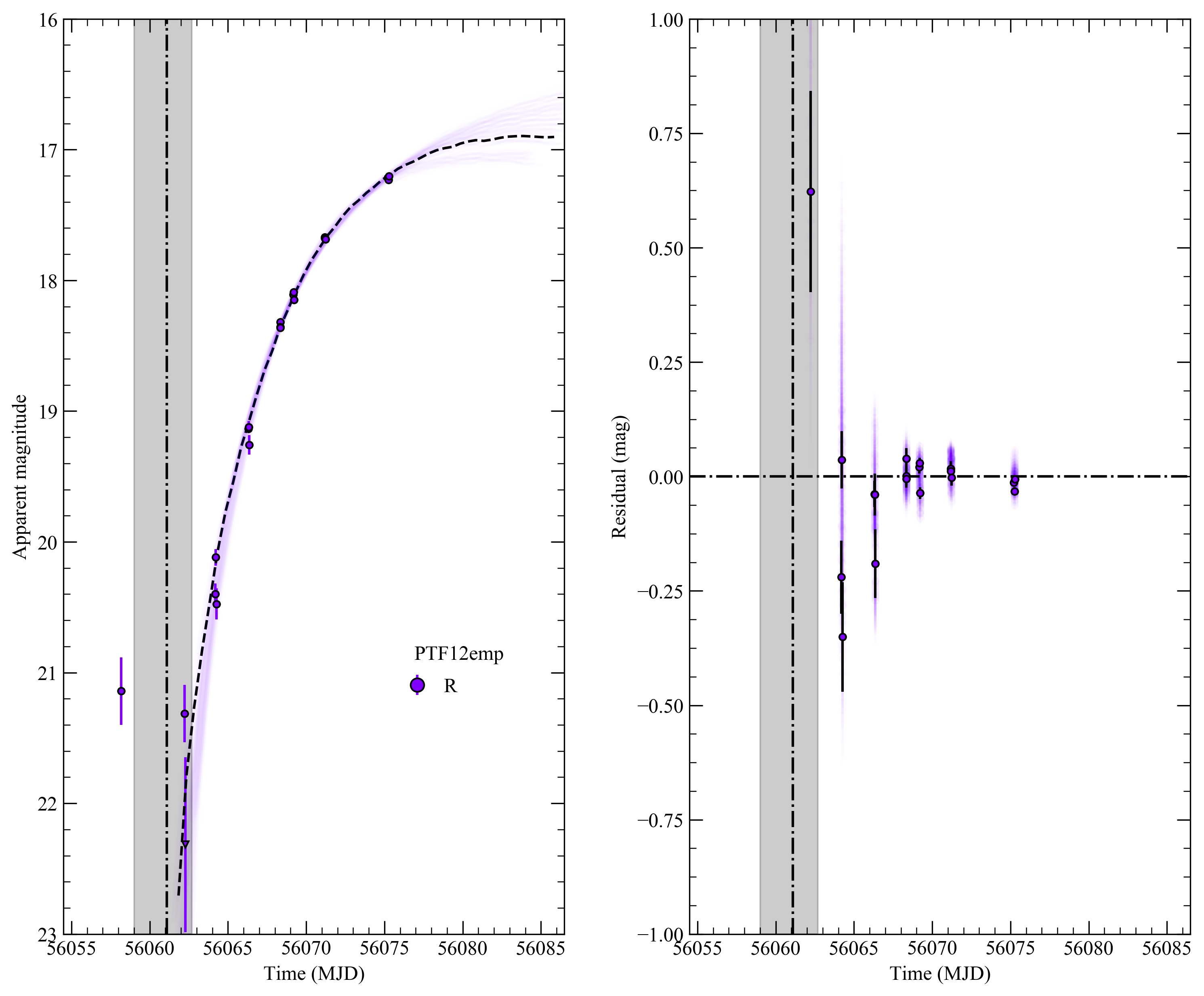}
        \caption{PTF12emp}
        \label{fig:PTF12emp_best}
    \end{subfigure}
    ~ 
    \begin{subfigure}[b]{0.49\textwidth}
        \includegraphics[width=\textwidth]{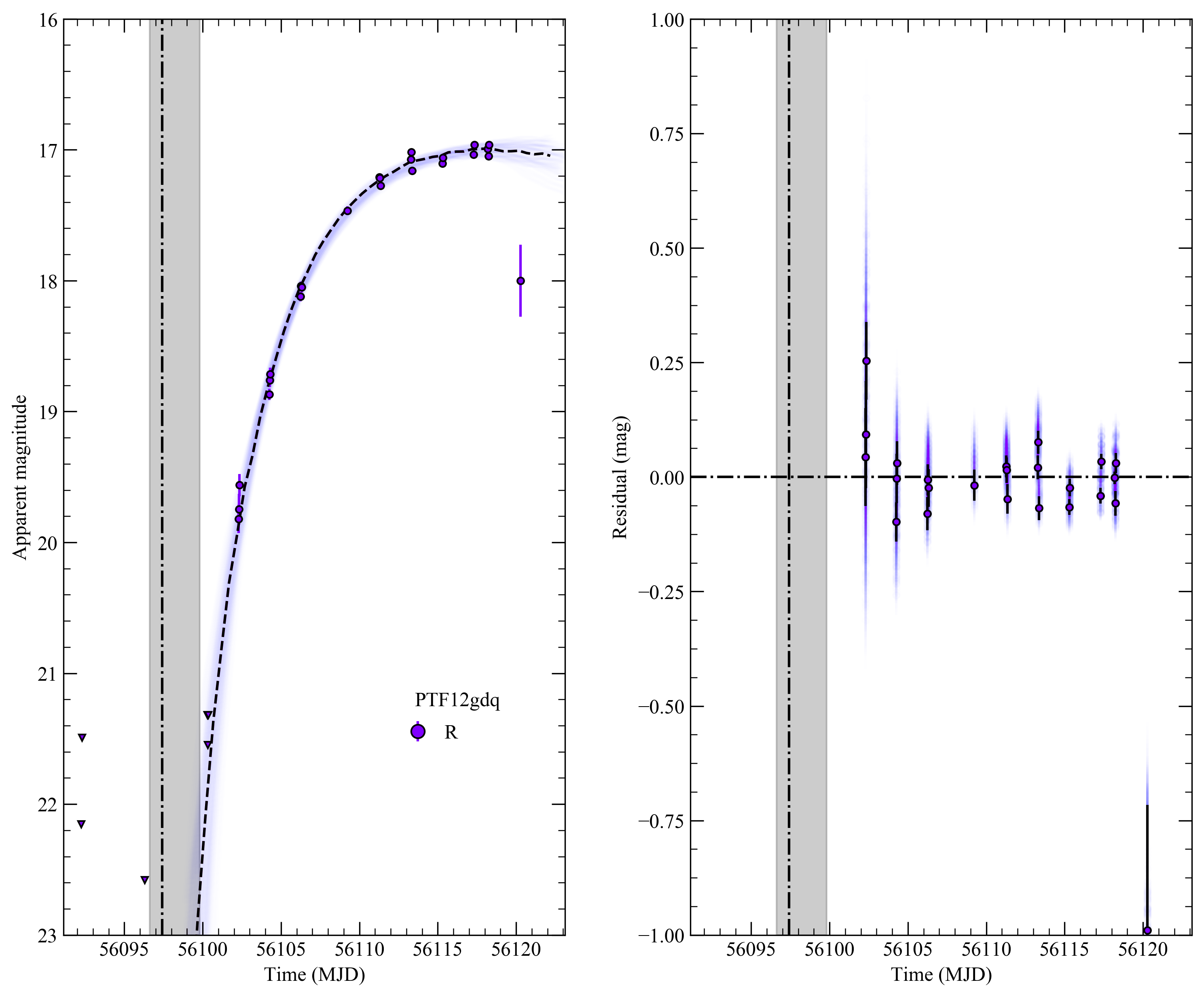}
        \caption{PTF12gdq}
        \label{fig:PTF12gdq_best}
    \end{subfigure}
    ~ 
    \caption{Same as in Fig.~\ref{fig:SN2012fr_best}.}
    \label{fig:PTF12emp_and_PTF12gdq}
\end{figure}

\begin{figure}[h!]
    \centering
    \begin{subfigure}[b]{0.49\textwidth}
        \includegraphics[width=\textwidth]{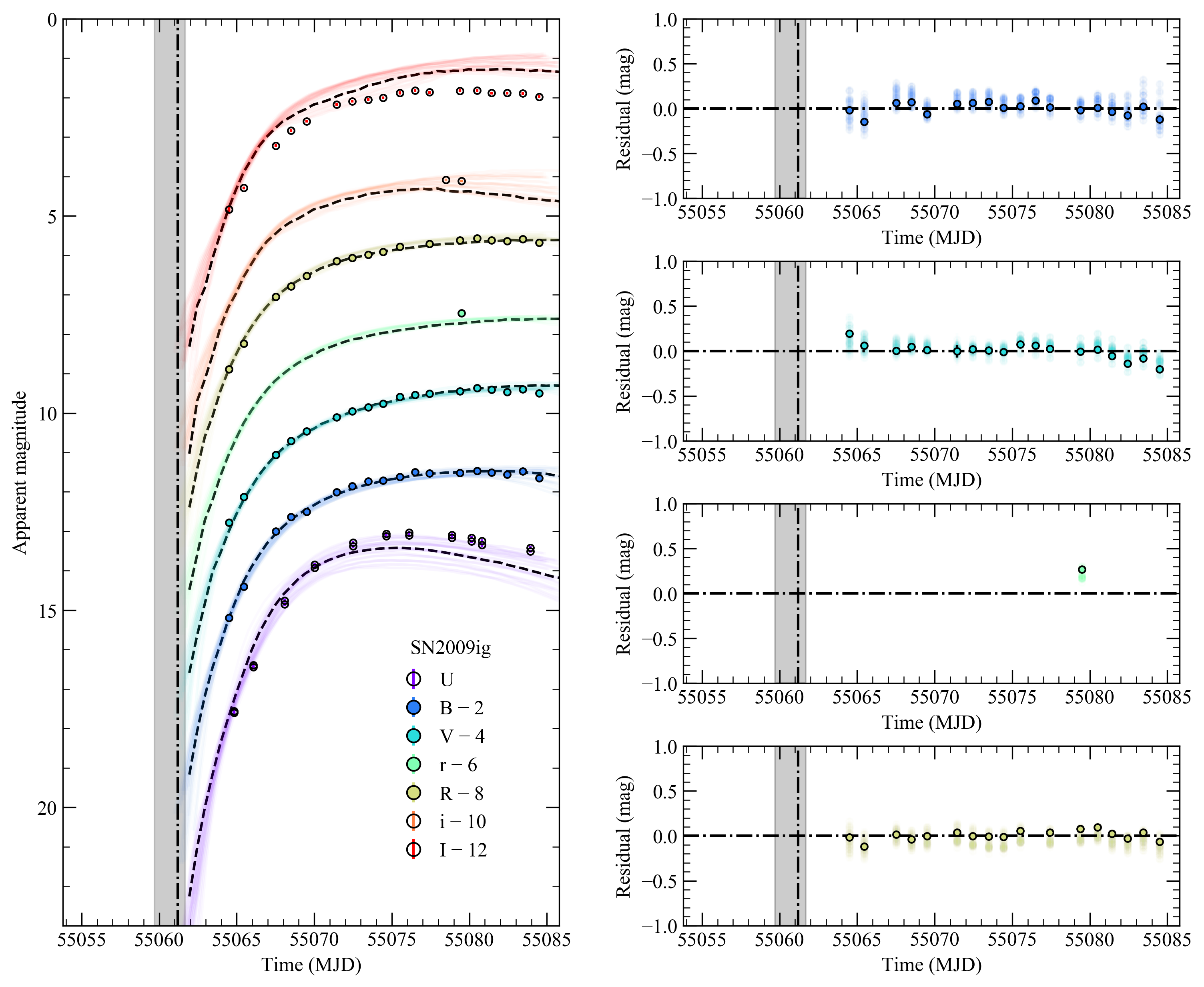}
        \caption{SN~2009ig}
        \label{fig:SN2009ig_best}
    \end{subfigure}
    ~ 
    \begin{subfigure}[b]{0.49\textwidth}
        \includegraphics[width=\textwidth]{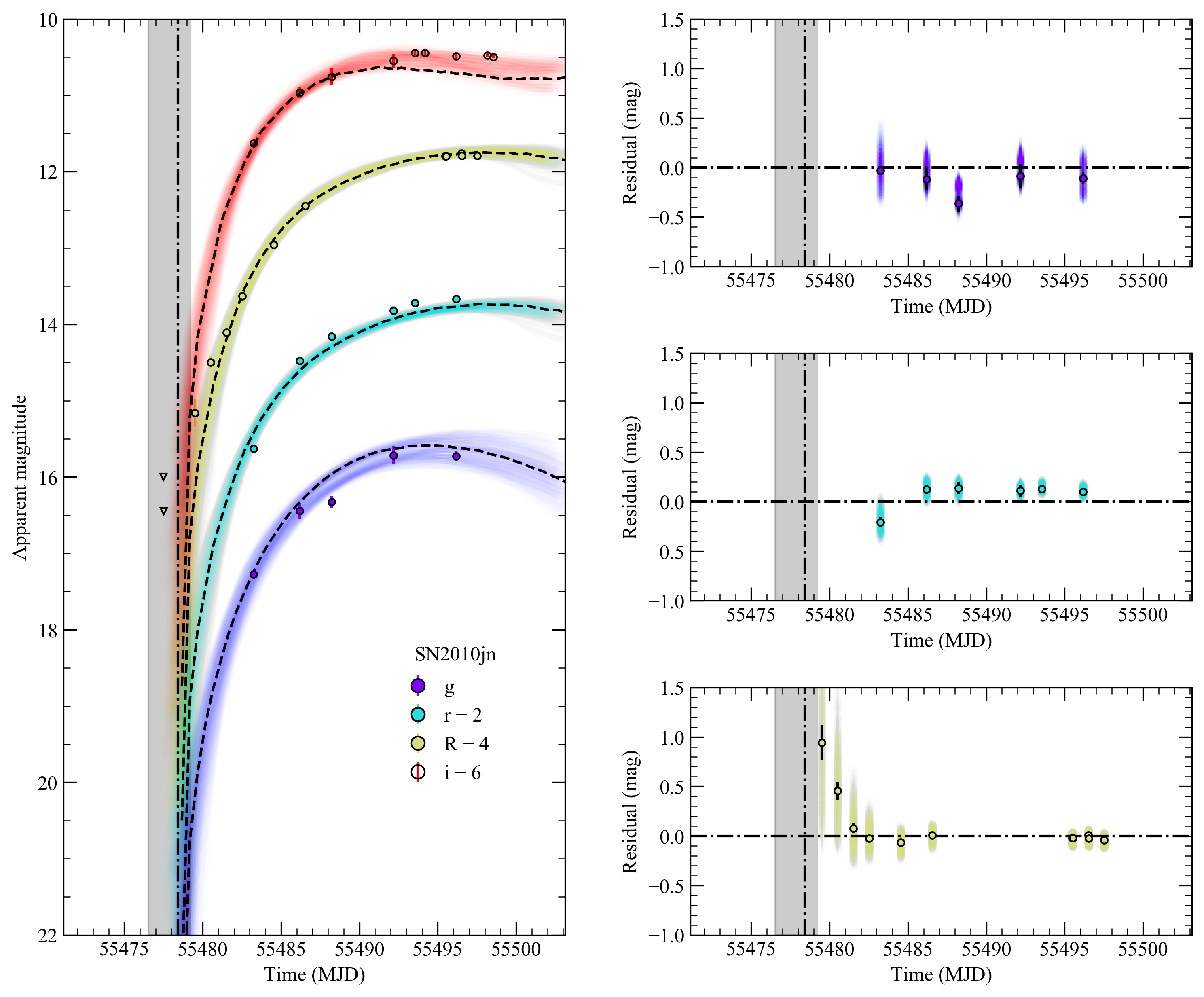}
        \caption{SN~2010jn}
        \label{fig:SN2010jn_best}
    \end{subfigure}
    ~ 
    \caption{Same as in Fig.~\ref{fig:SN2012fr_best}.}
    \label{fig:SN2009ig_and_SN2010jn}
\end{figure}

\begin{figure}[h!]
    \centering
    \begin{subfigure}[b]{0.49\textwidth}
        \includegraphics[width=\textwidth]{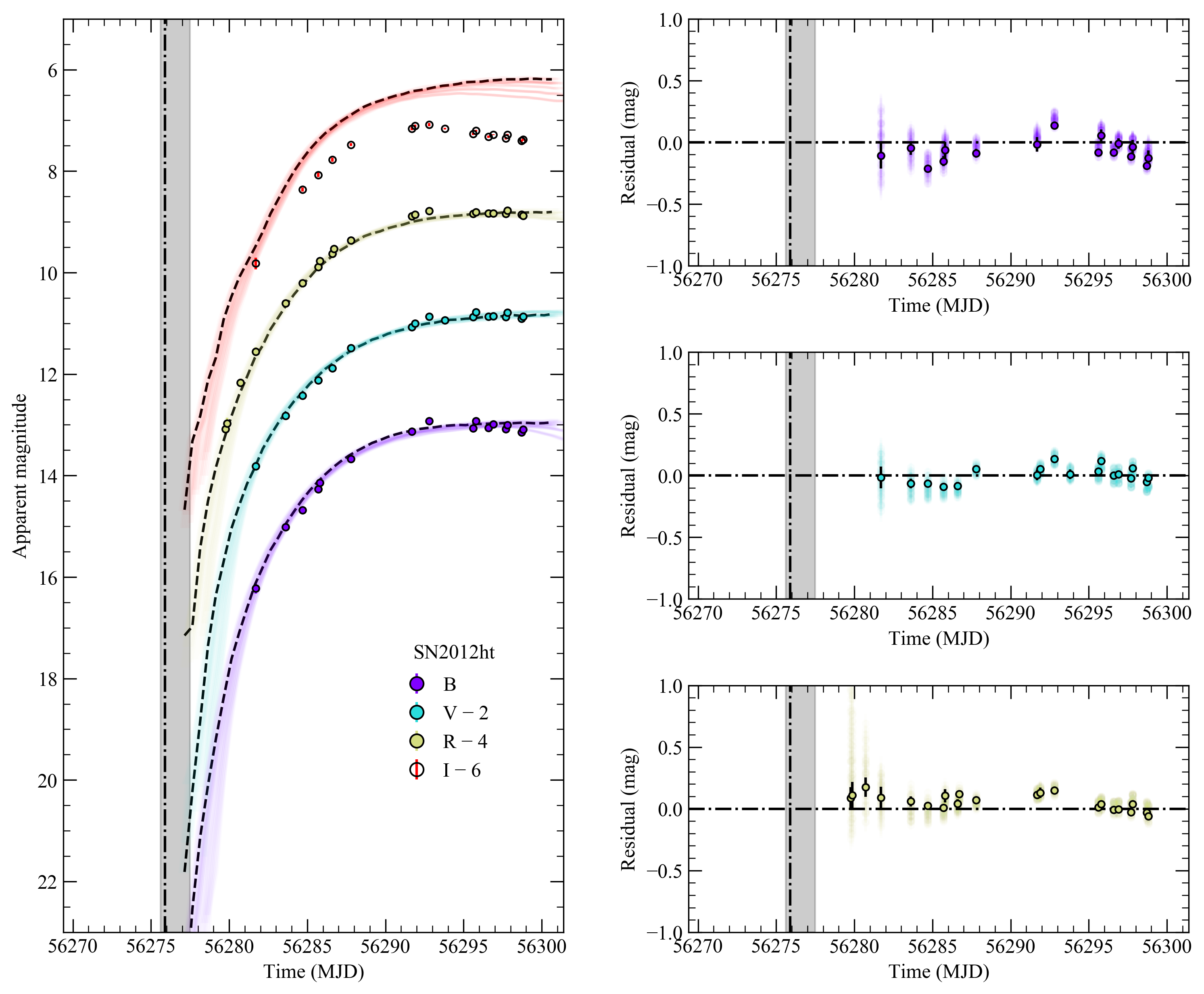}
        \caption{SN~2012ht}
        \label{fig:SN2012ht_best}
    \end{subfigure}
    ~ 
    \begin{subfigure}[b]{0.49\textwidth}
        \includegraphics[width=\textwidth]{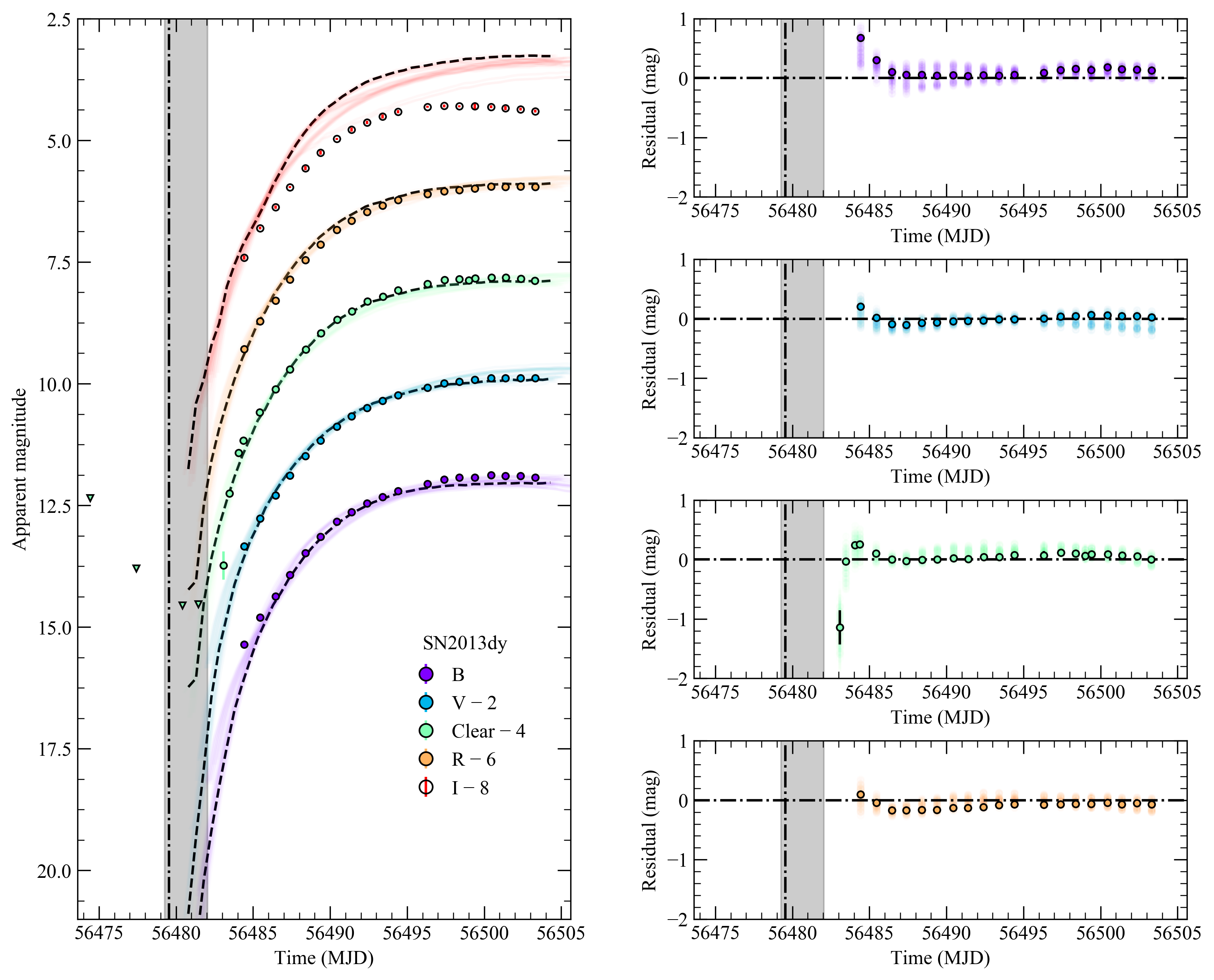}
        \caption{SN~2013dy}
        \label{fig:SN2013dy_best}
    \end{subfigure}
    ~ 
    \caption{Same as in Fig.~\ref{fig:SN2012fr_best}.}
    \label{fig:SN2012ht_and_SN2013dy}
\end{figure}

\begin{figure}[h!]
    \centering
    \begin{subfigure}[b]{0.49\textwidth}
        \includegraphics[width=\textwidth]{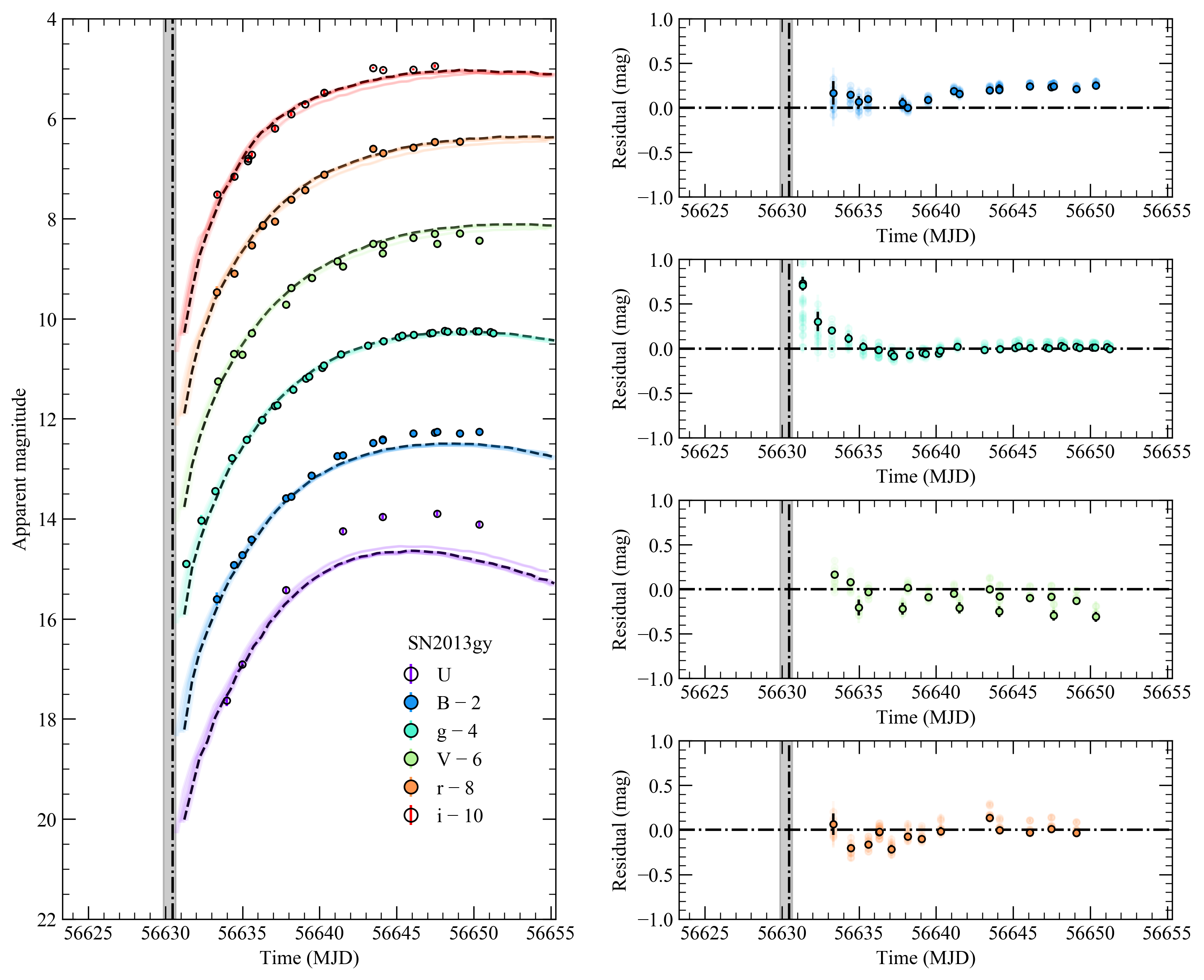}
        \caption{SN~2013gy}
        \label{fig:SN2013gy_best}
    \end{subfigure}
    ~ 
    \begin{subfigure}[b]{0.49\textwidth}
        \includegraphics[width=\textwidth]{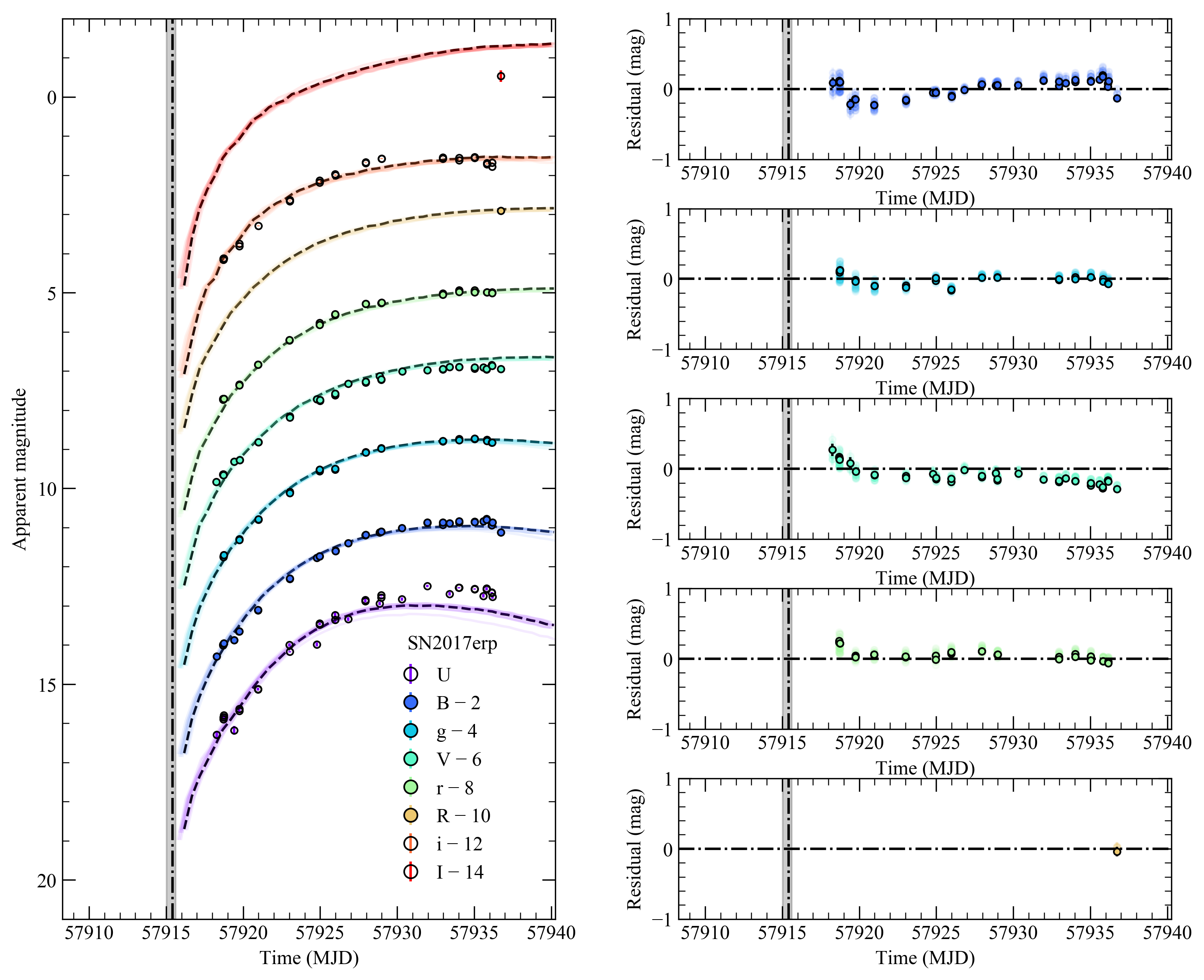}
        \caption{SN~2017erp}
        \label{fig:SN2017erp_best}
    \end{subfigure}
    ~ 
    \caption{Same as in Fig.~\ref{fig:SN2012fr_best}.}
    \label{fig:SN2013gy_and_SN2017erp}
\end{figure}

%

\clearpage
\section{SNe~Ia requiring additional tuning to $^{56}$Ni distribution}
\label{sect:apdx:good_but}
Figures are as in Fig.~\ref{fig:SN2012fr_best}. Objects included in this section are those for which additional tuning is required to improve agreement with our models.

\begin{figure}[h!]
    \centering
    \begin{subfigure}[b]{0.49\textwidth}
        \includegraphics[width=\textwidth]{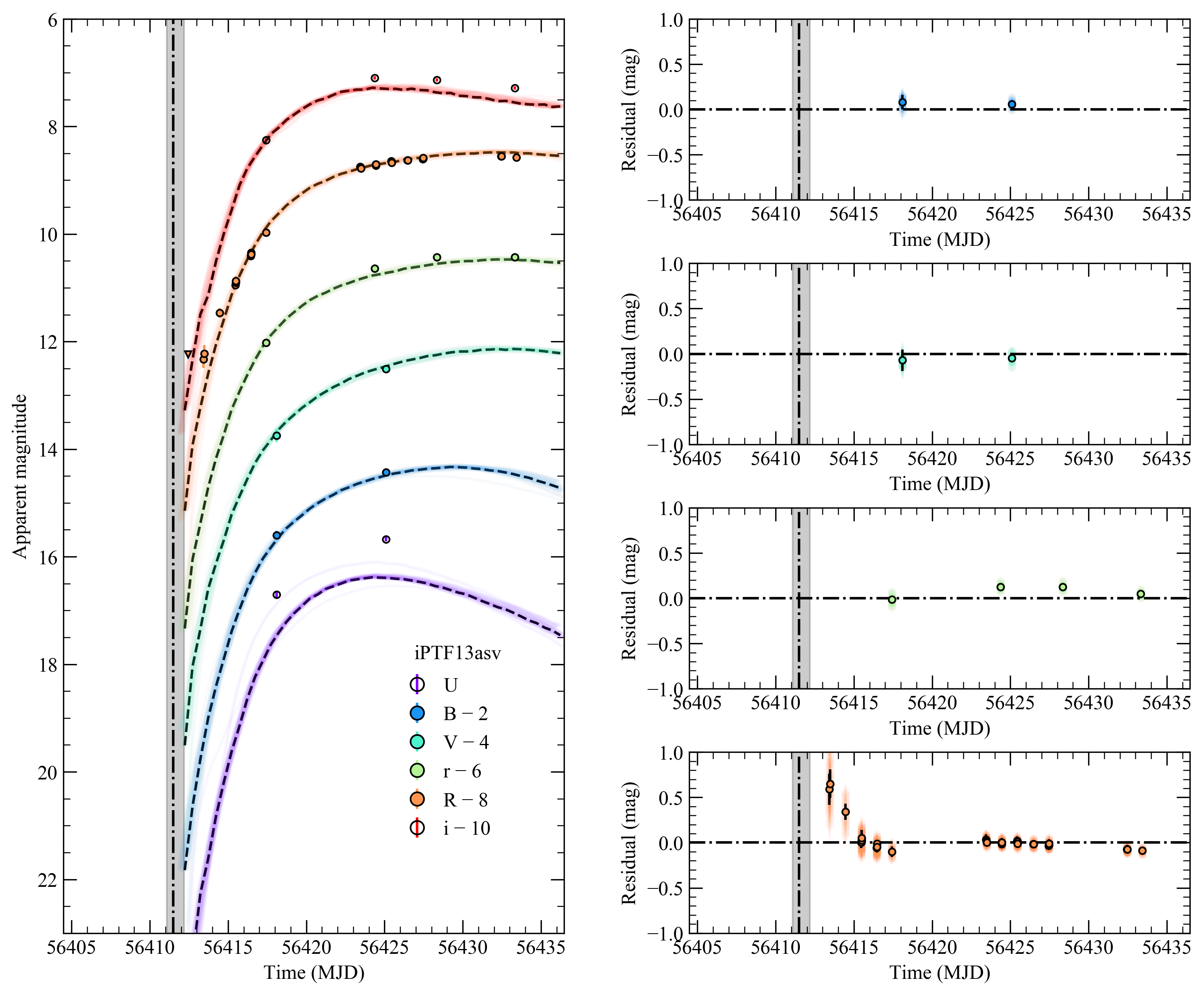}
        \caption{iPTF13asv}
        \label{fig:iPTF13asv_best}
    \end{subfigure}
    ~ 
    \begin{subfigure}[b]{0.49\textwidth}
        \includegraphics[width=\textwidth]{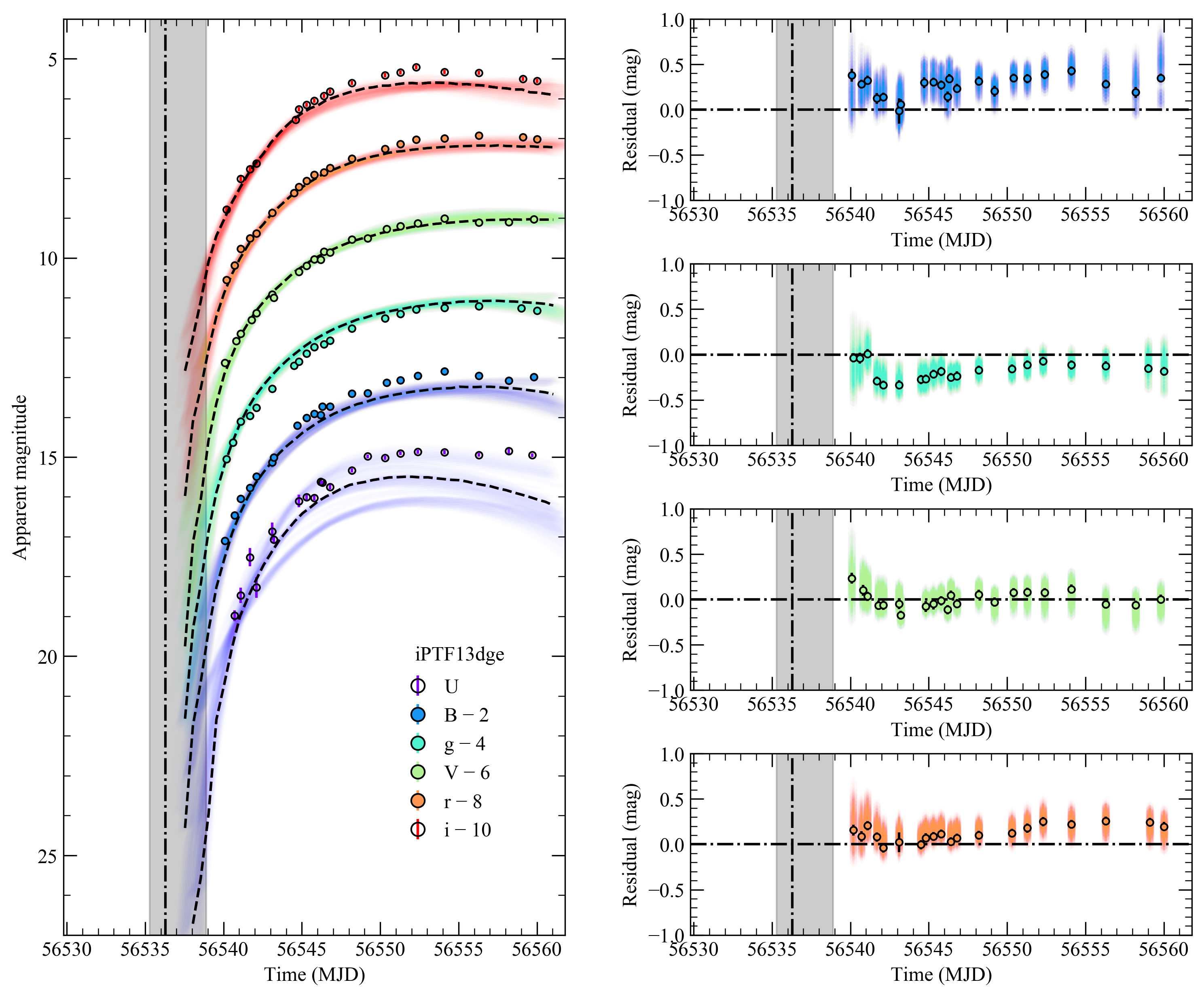}
        \caption{iPTF13dge}
        \label{fig:iPTF13dge_best}
    \end{subfigure}
    ~ 
    \caption{Same as in Fig.~\ref{fig:SN2012fr_best}.}
    \label{fig:iPTF13asv_and_iPTF13dge}
\end{figure}

\begin{figure}[h!]
    \centering
    \begin{subfigure}[b]{0.49\textwidth}
        \includegraphics[width=\textwidth]{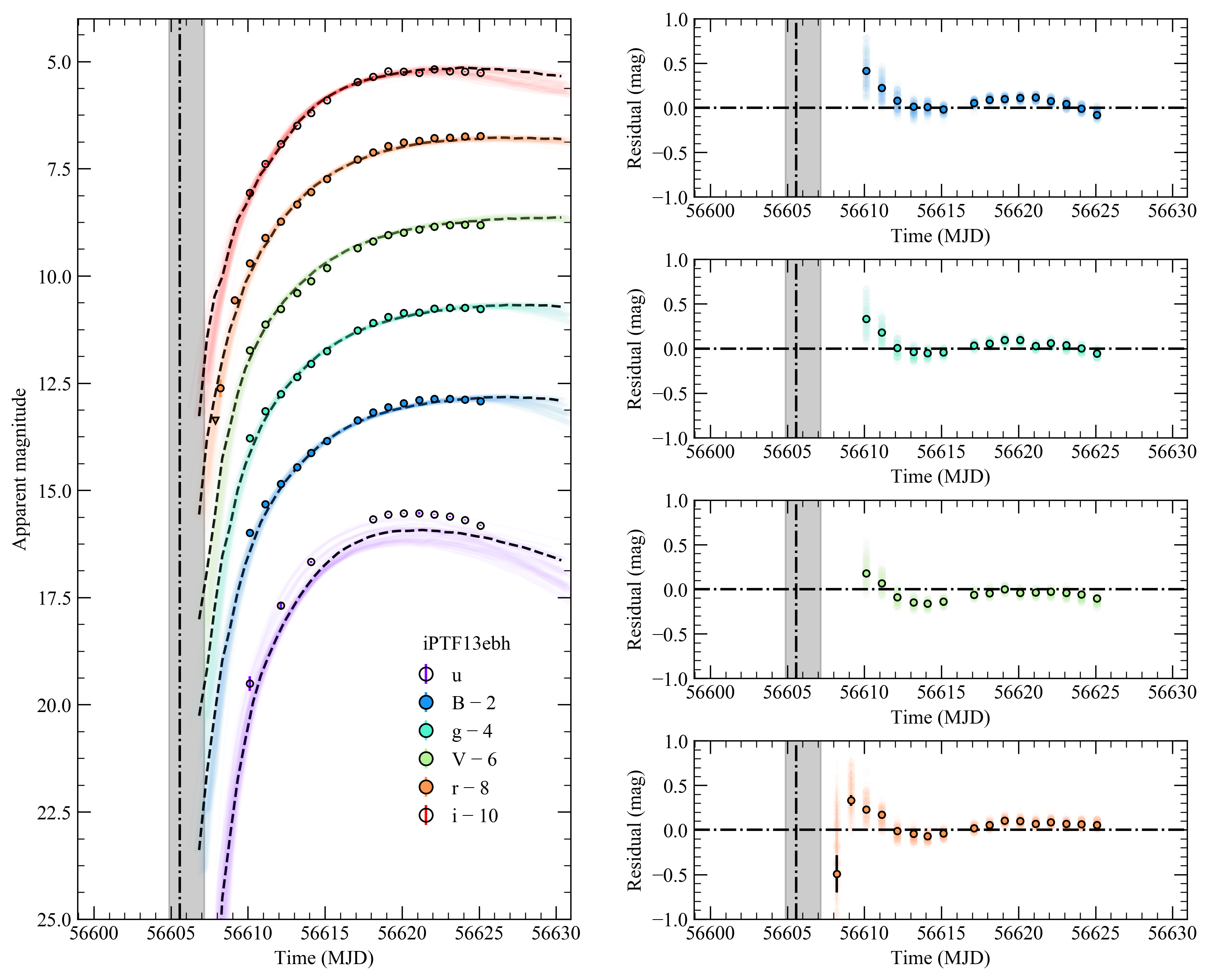}
        \caption{iPTF13ebh}
        \label{fig:iPTF13ebh_best}
    \end{subfigure}
    ~ 
    \begin{subfigure}[b]{0.49\textwidth}
        \includegraphics[width=\textwidth]{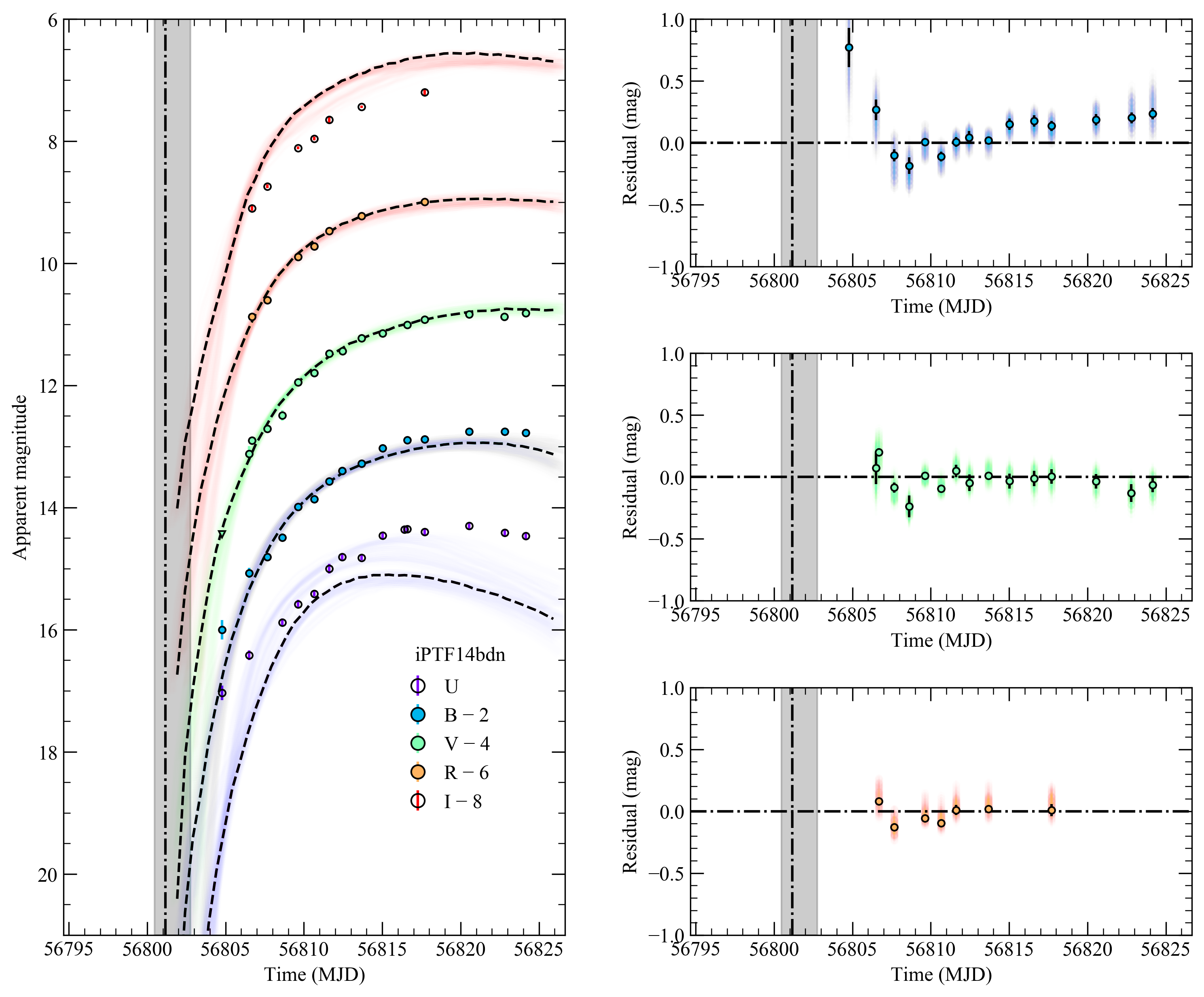}
        \caption{iPTF14bdn}
        \label{fig:iPTF14bdn_best}
    \end{subfigure}
    ~ 
    \caption{Same as in Fig.~\ref{fig:SN2012fr_best}.}
    \label{fig:iPTF13ebh_and_iPTF14bdn}
\end{figure}

\begin{figure}[h!]
    \centering
    \begin{subfigure}[b]{0.49\textwidth}
        \includegraphics[width=\textwidth]{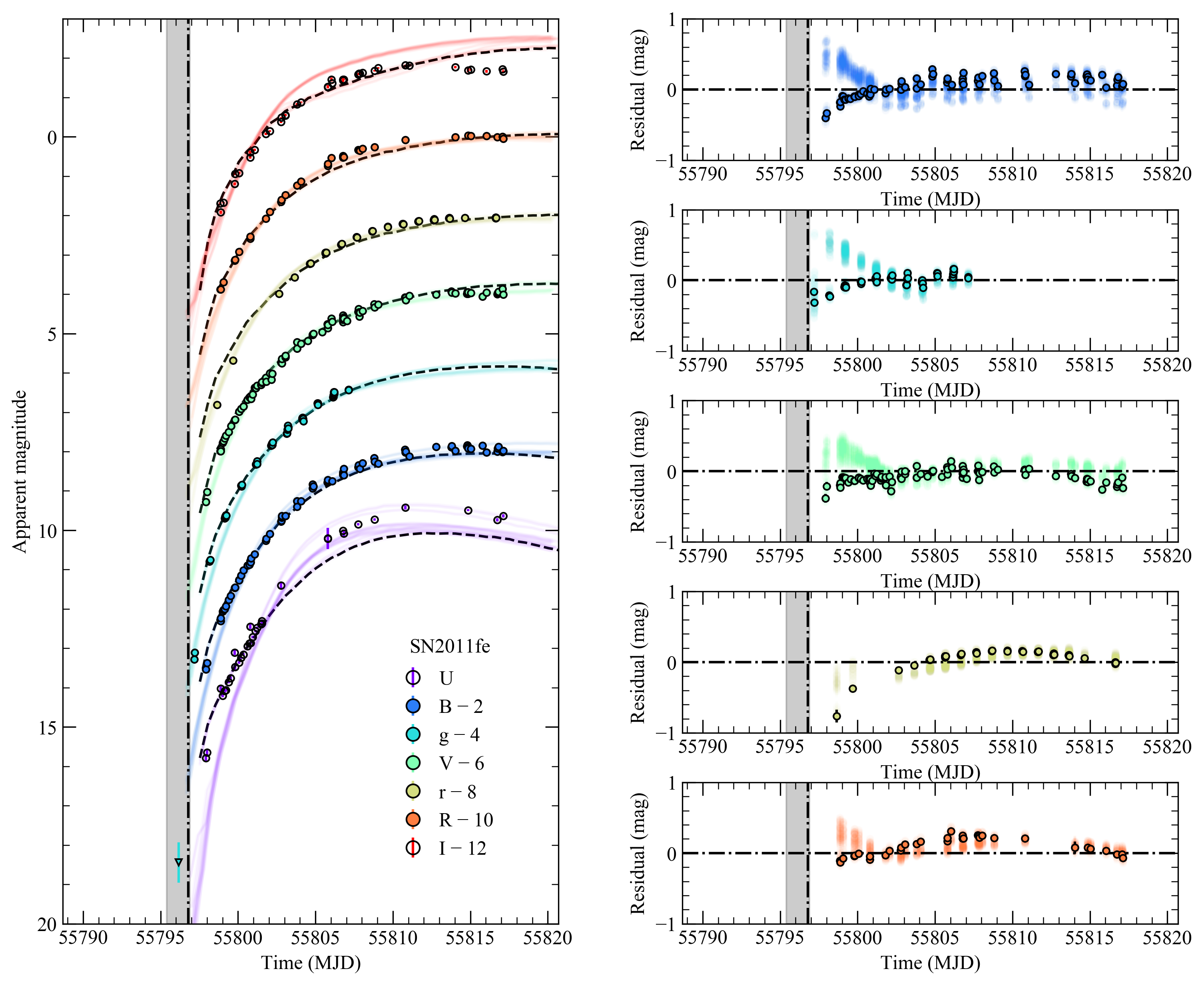}
        \caption{SN~2011fe}
        \label{fig:SN2011fe_best}
    \end{subfigure}
    ~ 
    \begin{subfigure}[b]{0.49\textwidth}
        \includegraphics[width=\textwidth]{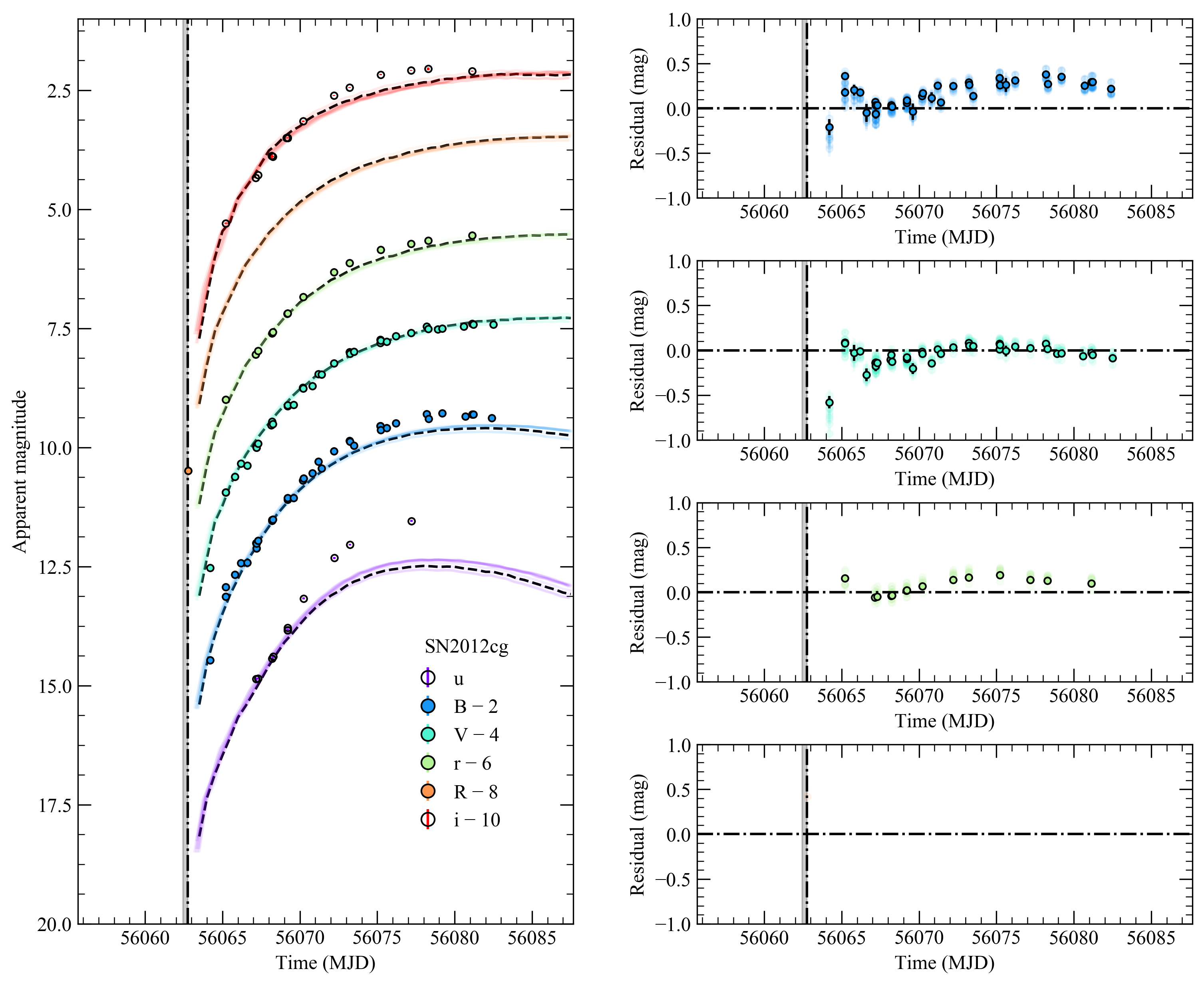}
        \caption{SN~2012cg}
        \label{fig:SN2012cg_best}
    \end{subfigure}
    ~ 
    \caption{Same as in Fig.~\ref{fig:SN2012fr_best}.}
    \label{fig:SN2011fe_and_SN2012cg}
\end{figure}

\begin{figure}[h!]
    \centering
    \begin{subfigure}[b]{0.49\textwidth}
        \includegraphics[width=\textwidth]{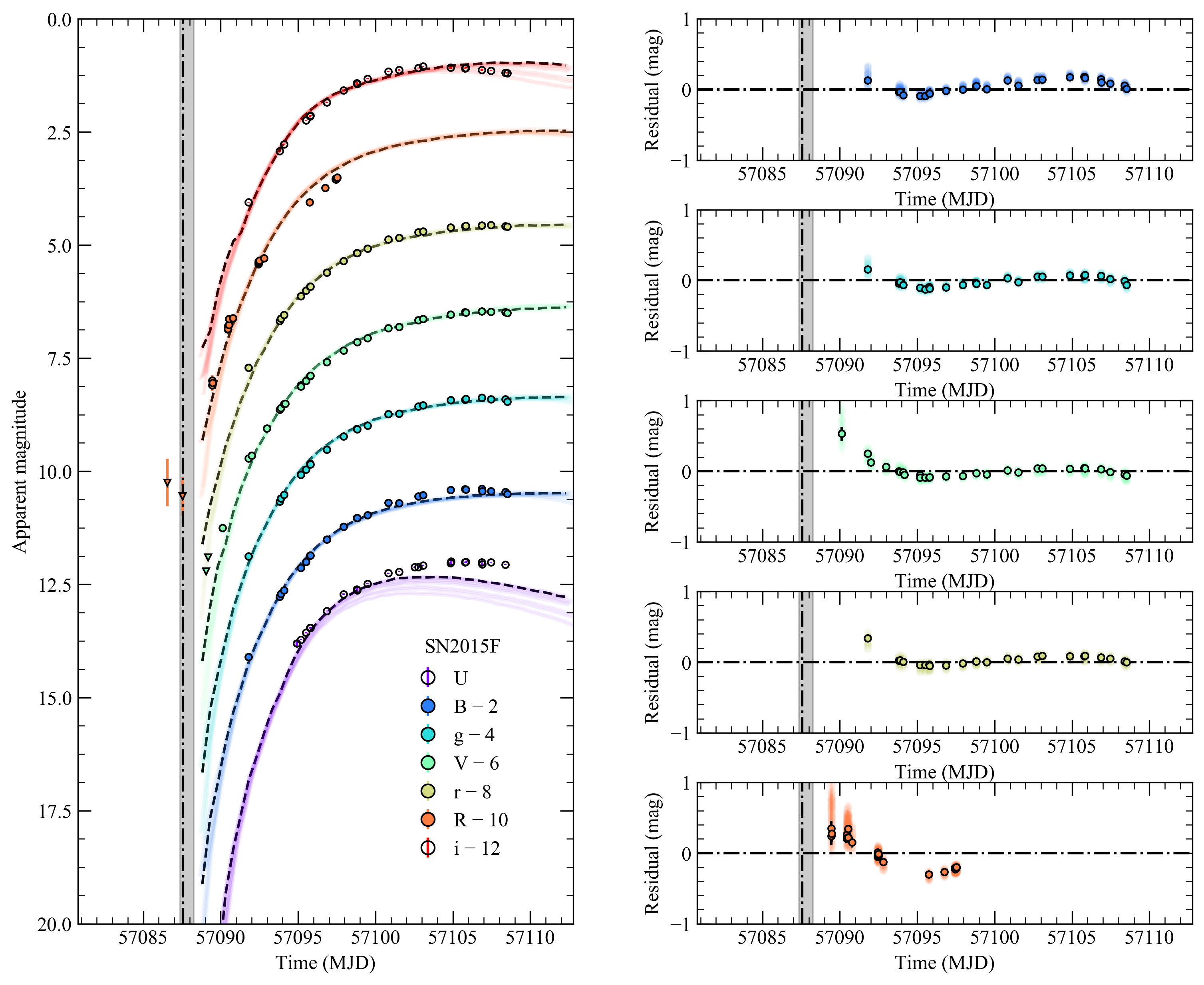}
        \caption{SN~2015F}
        \label{fig:SN2015F_best}
    \end{subfigure}
    ~ 
    ~ 
    \caption{Same as in Fig.~\ref{fig:SN2012fr_best}.}
    \label{fig:SN2015F}
\end{figure}

%

\clearpage
\section{SNe~Ia requiring an excess of flux at early times}
\label{sect:apdx:excess}
Figures are as in Fig.~\ref{fig:SN2012fr_best}. Objects included in this section are those for which our models do not produce good agreement with the full light curve and an excess of flux is required at early times.

\begin{figure}[h!]
    \centering
    \begin{subfigure}[b]{0.49\textwidth}
        \includegraphics[width=\textwidth]{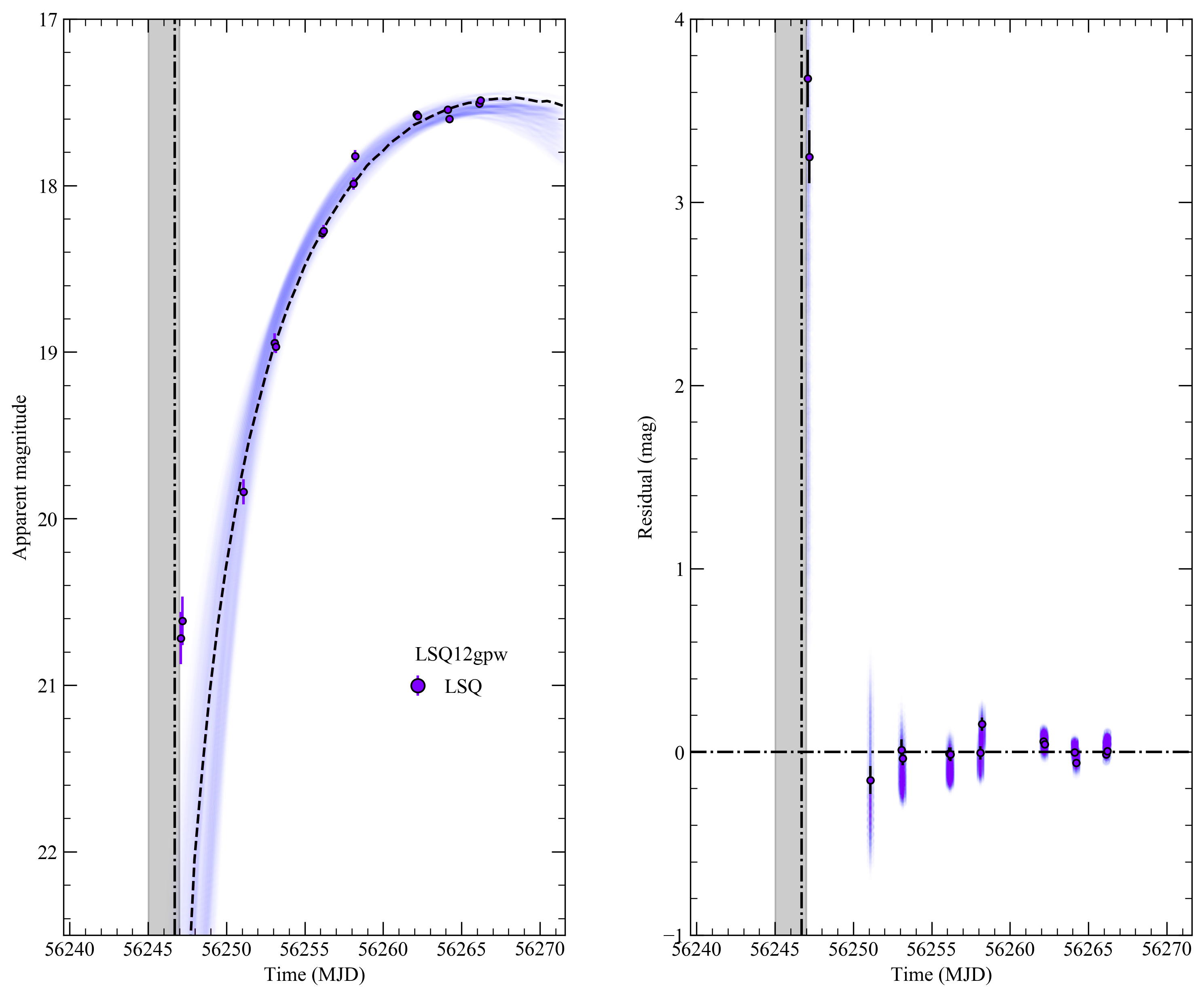}
        \caption{LSQ12gpw}
        \label{fig:LSQ12gpw_best}
    \end{subfigure}
    ~ 
    \begin{subfigure}[b]{0.49\textwidth}
        \includegraphics[width=\textwidth]{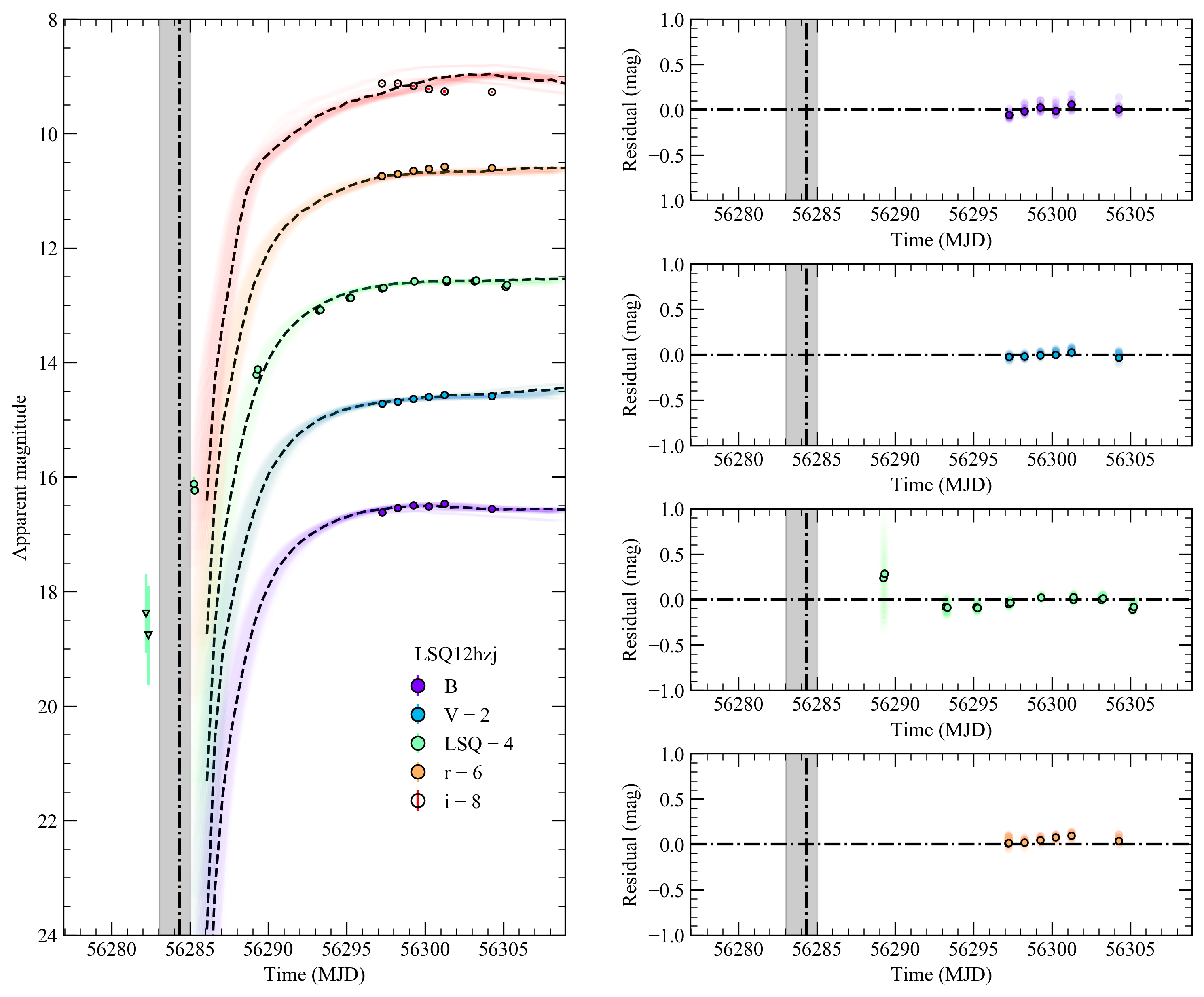}
        \caption{LSQ12hzj}
        \label{fig:LSQ12hzj_best}
    \end{subfigure}
    ~ 
    \caption{Same as in Fig.~\ref{fig:SN2012fr_best}.}
    \label{fig:LSQ12gpw_and_LSQ12hzj}
\end{figure}

\begin{figure}[h!]
    \centering
    \begin{subfigure}[b]{0.49\textwidth}
        \includegraphics[width=\textwidth]{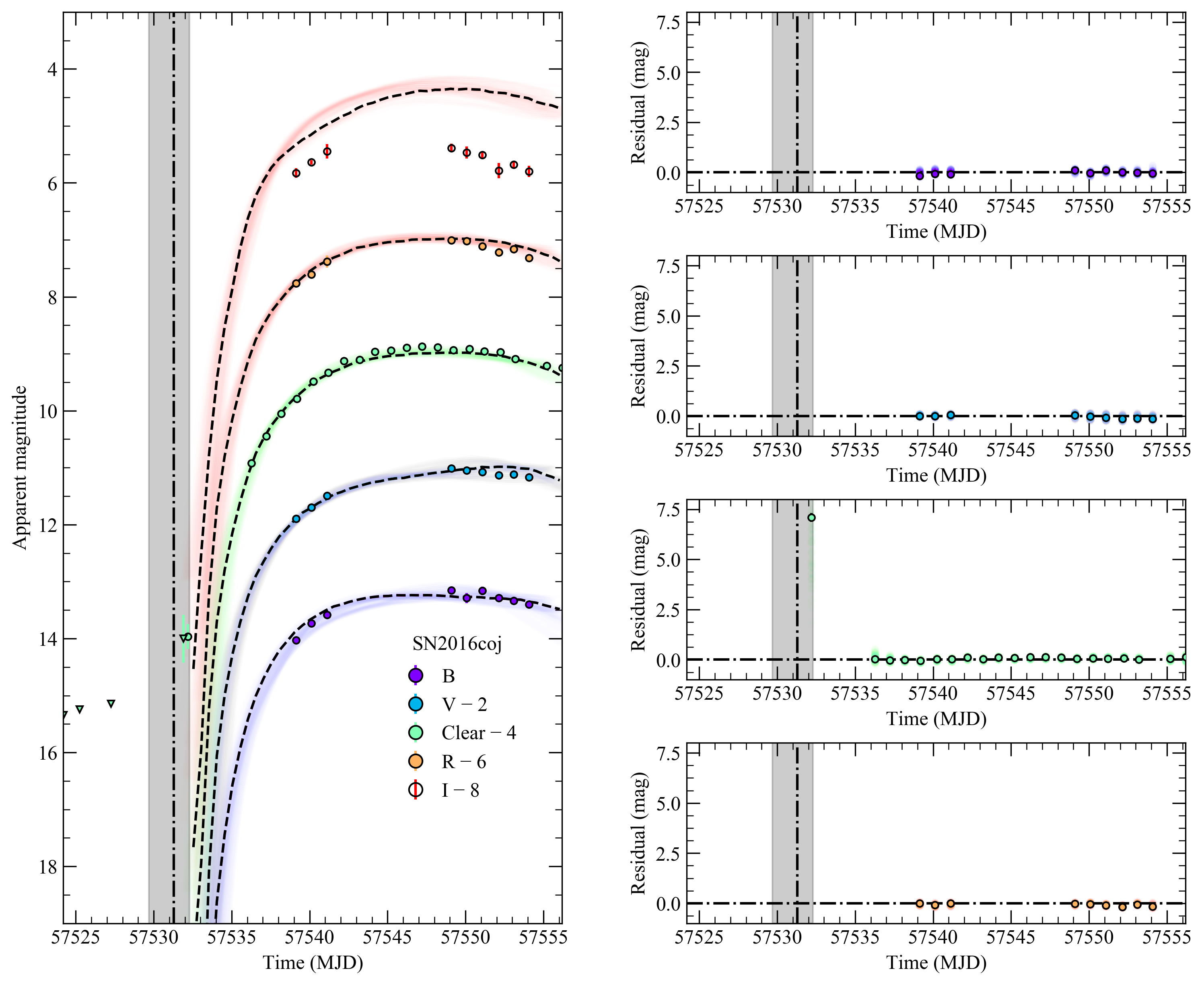}
        \caption{SN~2016coj}
        \label{fig:SN2016coj_best}
    \end{subfigure}
    ~ 
    \begin{subfigure}[b]{0.49\textwidth}
        \includegraphics[width=\textwidth]{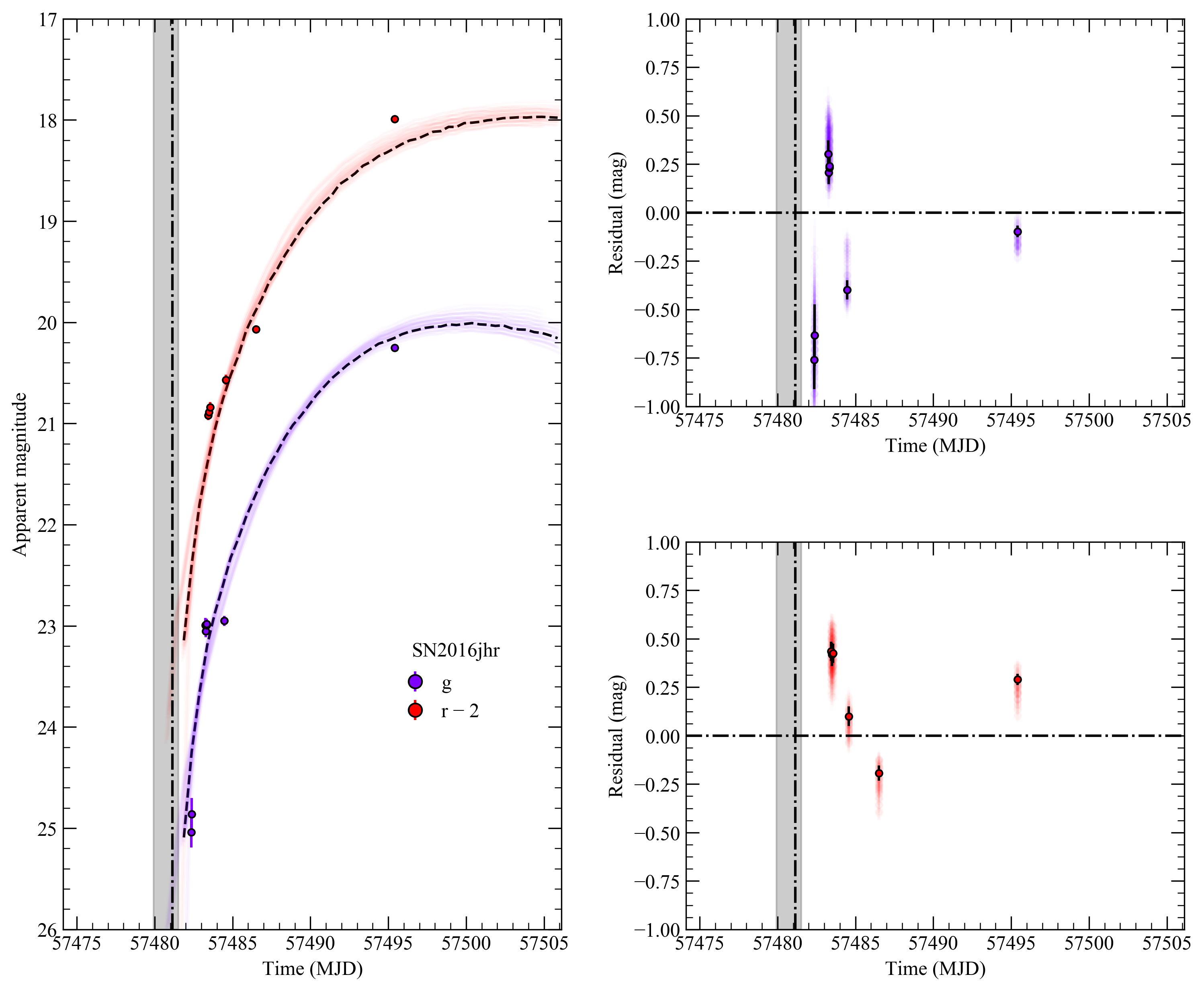}
        \caption{SN~2016jhr}
        \label{fig:SN2016jhr_best}
    \end{subfigure}
    ~ 
    \caption{Same as in Fig.~\ref{fig:SN2012fr_best}.}
    \label{fig:SN2016coj_and_SN2016jhr}
\end{figure}

%

\clearpage

\section{SNe~Ia with ambiguous results}
\label{sect:apdx:ambiguous}
Figures are as in Fig.~\ref{fig:SN2012fr_best}. Objects included in this section are those for which we find ambiguous results and are unable to distinguish whether or not an early flux excess is required.

\begin{figure}[h!]
    \centering
    \begin{subfigure}[b]{0.49\textwidth}
        \includegraphics[width=\textwidth]{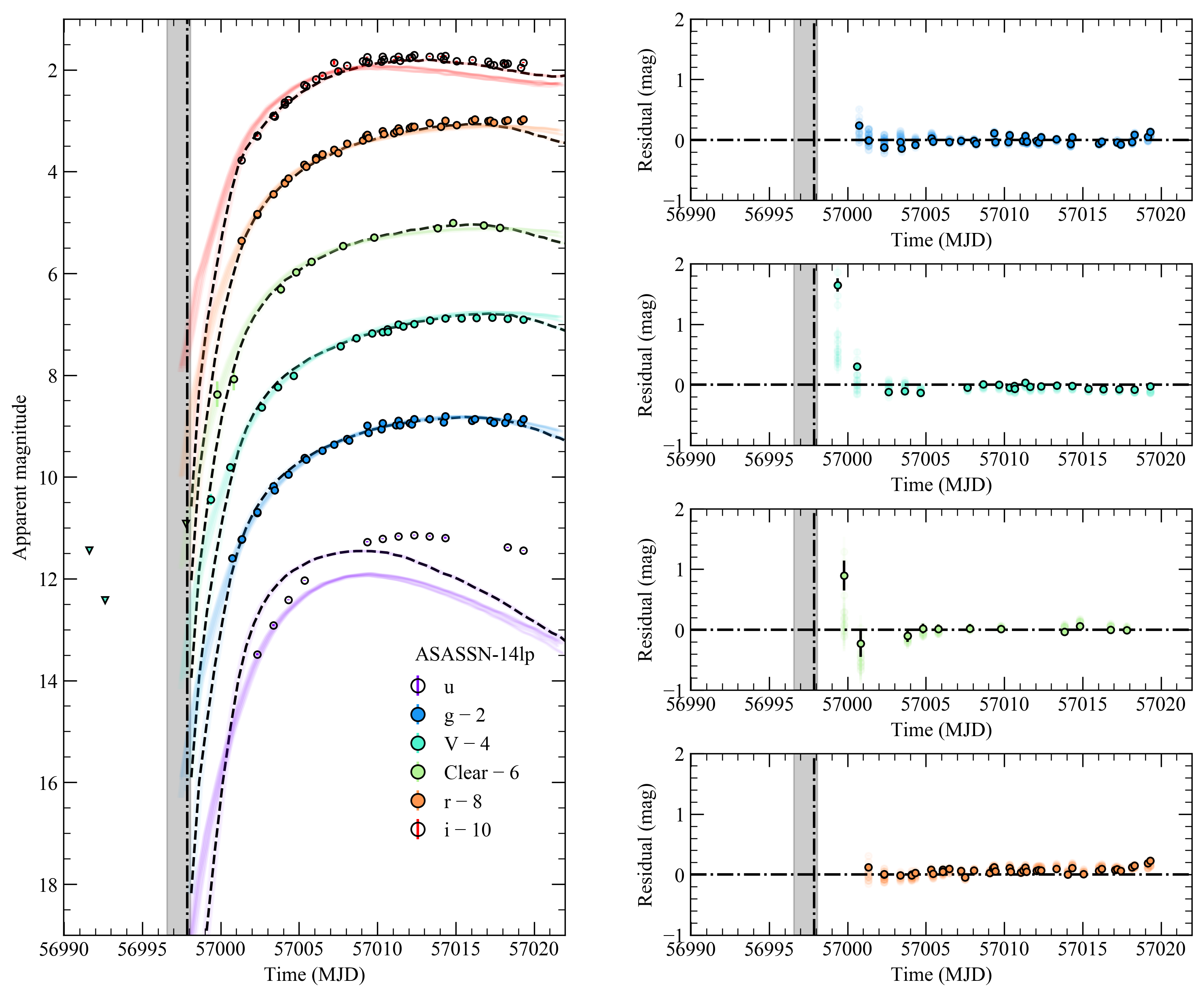}
        \caption{ASASSN-14lp}
        \label{fig:ASASSN-14lp_best}
    \end{subfigure}
    ~ 
    \begin{subfigure}[b]{0.49\textwidth}
        \includegraphics[width=\textwidth]{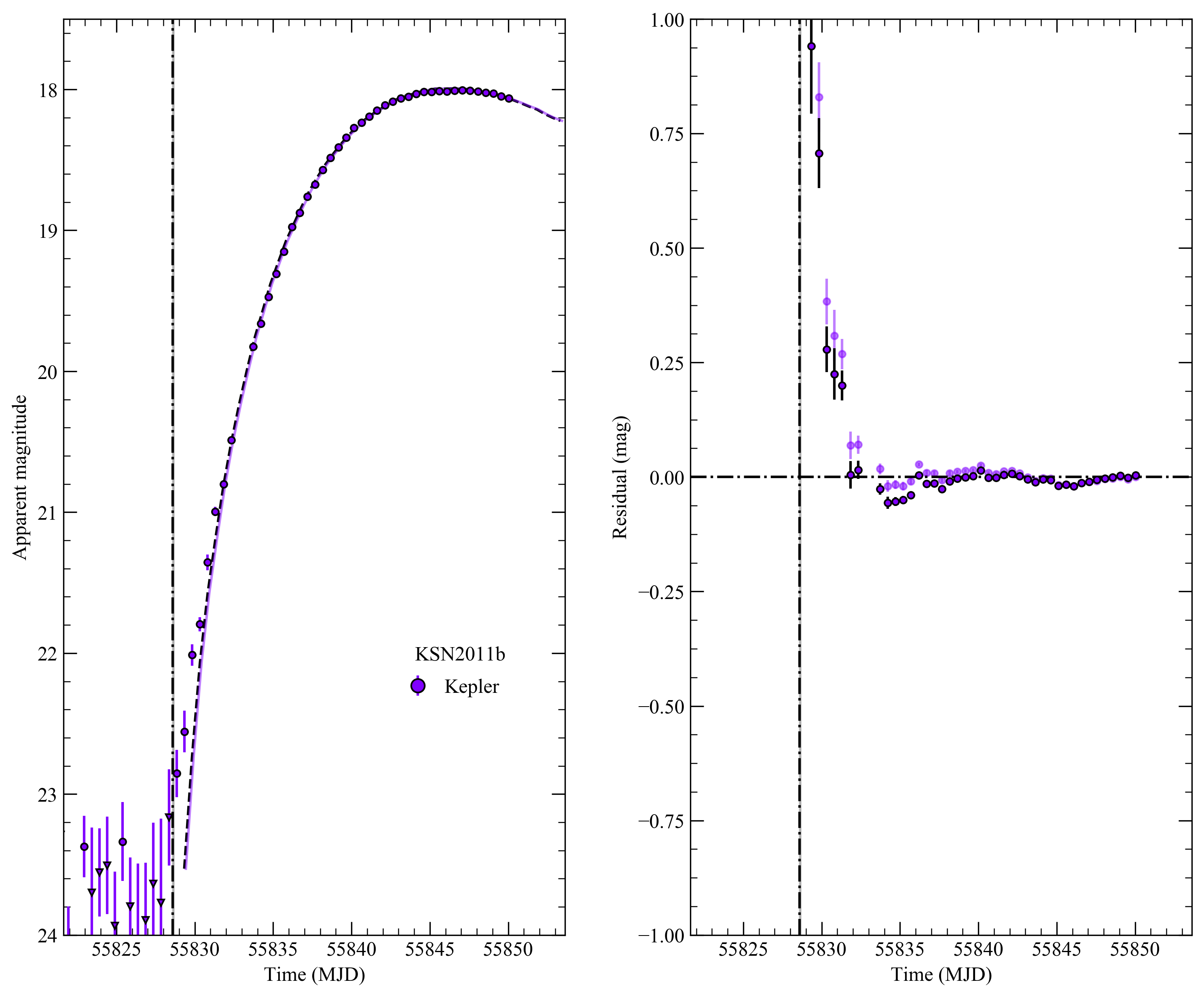}
        \caption{KSN2011b}
        \label{fig:KSN2011b_best}
    \end{subfigure}
    ~ 
    \caption{Same as in Fig.~\ref{fig:SN2012fr_best}.}
    \label{fig:ASASSN-14lp_and_KSN2011b}
\end{figure}

\end{appendix}
\end{document}